\documentclass[aps,prl,notitlepage,twocolumn,superscriptaddress, longbibliography,showpacs,psfig]{revtex4-2}

\makeatletter
\def\@hangfrom@section#1#2#3{\normalsize\@hangfrom{#1#2}#3}%\MakeTextUppercase{#3}}%
\def\@hangfroms@section#1#2{\normalsize#1#2}%\MakeTextUppercase{#2}}%
\makeatother

\RequirePackage[usenames, dvipsnames, x11names]{xcolor}
\usepackage{amsmath,amsfonts,amssymb,bm}
\usepackage{graphicx,color}
\usepackage{enumerate}
\definecolor{darkblue1}{rgb}{0.18,0.19,0.57}
\definecolor{darkblue2}{rgb}{0.27,0.4,0.6}
\usepackage[colorlinks=true, allcolors=darkblue1]{hyperref}

\usepackage{multirow}

\newcommand{\nc}{\newcommand}
\nc{\ket}[1]{|#1\rangle}
\nc{\bra}[1]{\langle#1|}
\nc{\ketbra}[2]{|#1\rangle\!\langle#2|}
\nc{\braket}[2]{\langle#1|#2\rangle}
\nc{\braoprket}[3]{\langle#1|#2|#3\rangle}
\nc{\opr}[1]{\operatorname{#1}}
\nc{\avg}[1]{\langle#1\rangle}
\nc{\ketbrasame}[1]{|#1\rangle\!\langle#1|}
\nc{\tr}{\opr{tr}}
\nc{\E}{\mathbb{E}}
\nc{\var}{\opr{Var}}
\nc{\up}{\uparrow}
\nc{\dn}{\downarrow}

\nc{\hk}[1]{\textcolor{violet}{\textbf{[hk: #1]}}}
\usepackage[normalem]{ulem}
\nc{\hknew}[1]{\textcolor{violet}{#1}}

\nc{\zjx}[1]{{\color{orange}{\textbf{[zjx: #1]}}}}
\nc{\qzjx}[1]{{\color{orange}{#1}}}
\nc{\weng}[1]{{\color{red}{[weng: #1]}}}

\begin{document}
% \graphicspath{{figure/}}
% \title{Quantum-interference-induced pairing in antiferromagnetic bosonic $t$-$J$ model}
\title{Quantum-interference-induced pairing in bosonic doped antiferromagnets}

\author{Hao-Kai Zhang}
\affiliation{Institute for Advanced Study, Tsinghua University, Beijing 100084, China}
\affiliation{Institute of Physics, Chinese Academy of Sciences, Beijing 100190, China}

\author{Jia-Xin Zhang}
\email{jiaxin\_zhang@ucsb.edu}
\affiliation{French American Center for Theoretical Science, CNRS, KITP, Santa Barbara, California 93106-4030, USA
}
\affiliation{Kavli Institute for Theoretical Physics, University of California, Santa Barbara, California 93106-4030, USA}
\affiliation{Institute for Advanced Study, Tsinghua University, Beijing 100084, China}

\author{Ji-Si Xu}
\affiliation{Institute for Advanced Study, Tsinghua University, Beijing 100084, China}

\author{Hong-Chen Jiang}
\affiliation{Stanford Institute for Materials and Energy Sciences, SLAC National Accelerator Laboratory, Menlo Park, California 94025, USA}

\author{Zheng-Yu Weng}
% \email{weng@mail.tsinghua.edu.cn}
\affiliation{Institute for Advanced Study, Tsinghua University, Beijing 100084, China}

\date{\today}
\begin{abstract}
The pairing mechanism in doped antiferromagnets is essential for understanding high-temperature superconductivity. In this work, we investigate the pairing mechanism in bosonic doped antiferromagnets via large-scale density matrix renormalization group calculations of the bosonic $t$-$J$ model. We discover a pair density wave (PDW) coexists with the antiferromagnetic (AFM) order forming a ``supersolid'' at small doping. The pairing is attributed to a hidden many-body Berry phase that introduces the sole ``sign problem'' into this bosonic model and imposes quantum phase frustration to the spin-charge interference pattern. Only via tightly pairing of doped holes, can such frustration be most effectively erased in an AFM background. By contrast, the pairing vanishes as the Berry phase is trivialized in the ferromagnetic condensate at larger doping or switched off into the Bose-Hubbard model at large $U$. The present pairing mechanism---distinct from the conventional mechanisms based on Fermi surface instabilities---may provide a different perspective and new insights for understanding the complex nature of doped Mott insulators and is promising to be probed on qudit-based quantum simulators such as ultracold Rydberg atom arrays.
\end{abstract}
\maketitle

% in contrast to the fermionic case,

\textit{Introduction.---}The pairing mechanism of unconventional superconductivity (SC) from doping antiferromagnetic (AFM) Mott insulators is one of the central puzzles in modern condensed matter physics~\cite{Keimer2015, Anderson1987a, Lee2006}. The two-dimensional Hubbard model and the related $t$-$J$ model are believed to capture many of the essential aspects of such systems~\cite{Zhang1988, Auerbach2012,  Arovas2022}. Despite intensive efforts and considerable progress over the past decades in analytical and numerical studies~\cite{Lee2006, Arovas2022, Anderson1987, Sheng1996, Weng1997, Wu2008, Fradkin2015, Zheng2017, Huang2017, Jiang2018, Jiang2019, Jiang2020a, Qin2020, Gong2021, Lu2023, Jiang2023, Jiang2021, Jiang2021SL, Chen2023, Xu2023, Wietek2020}, an undisputed systematic understanding of their ground and thermal states remains elusive due to the inherent complexity of them as paradigmatic models of strongly correlated systems.

In recent years, quantum simulators especially ultracold atoms have brought new opportunities for exploring unconventional SC mechanisms~\cite{Bohrdt2021, Mazurenko2017, Koepsell2021, Xu2023a, Lebrat2024} and other strongly correlated quantum matter~\cite{Sun2021, Browaeys2020, Jepsen2020, Semeghini2021, Chen2023a}. These simulators can reach regimes that are numerically difficult to access and are flexible in tuning and probing microscopic correlations. The Hubbard model and effectively the $t$-$J$ model can be artificially realized by cooling and trapping fermionic atoms into optical lattices~\cite{Bohrdt2021, Mazurenko2017, Koepsell2021, Xu2023a, Lebrat2024, Sun2021}. The lowest effective temperatures reached are about half of the superexchange energy scale, where extended-range AFM correlations have been observed~\cite{Mazurenko2017}, yet the potential SC phase still requires further lower temperatures.

On the other hand, various quantum spin (equivalently hard-core boson) models have been recently realized on Rydberg atom arrays~\cite{Browaeys2020, Jepsen2020, Semeghini2021, Chen2023a}, where the spin or boson states are encoded into the internal states of atoms. Measurements beyond diagonal density correlations are accessible by implementing programmable quantum gates~\cite{Bluvstein2022}. Experimental attempts have been reported to realize the AFM bosonic $t$-$J$ model on such platforms~\cite{Homeier2024, Qiao2025}. This model still describes doped Heisenberg antiferromagnets but with bosonic dopants. Comparative studies of the bosonic and fermionic $t$-$J$ models, alongside advanced cold atom simulations, are promising to offer deeper insights into the physics of doped Mott insulators.

\begin{figure*}
    \centering
    \includegraphics[width=0.9\linewidth]{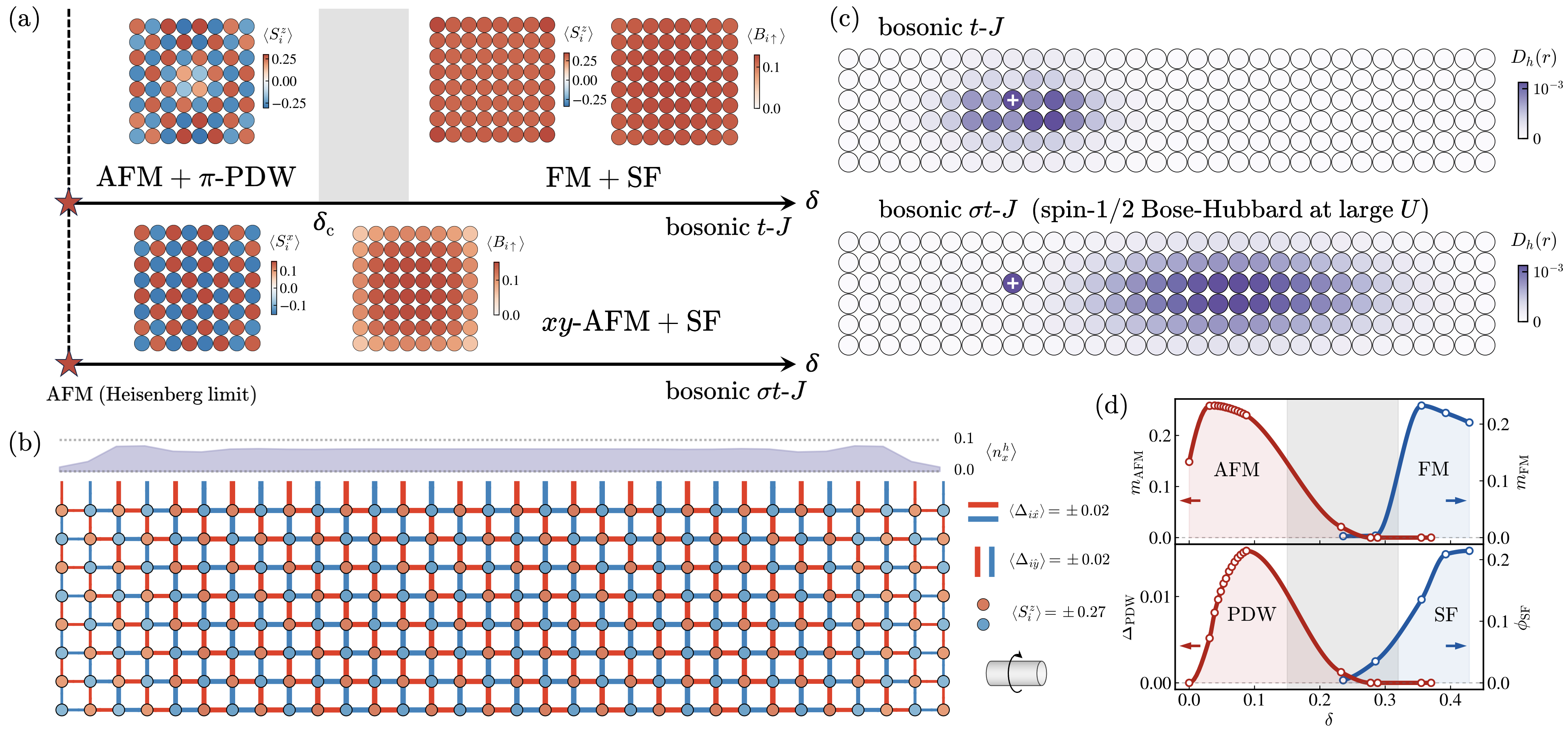}
    \caption{(a) The ground state phase diagram of the 2D bosonic $t$-$J$ model and $\sigma t$-$J$ model at $t/J=3$ and varying hole doping $\delta$ determined by simulations on width-$4,6,8$ square lattices. At half-filling, both models are reduced to the Heisenberg model. Upon doping, the bosonic $t$-$J$ model enters a pair density wave (PDW) phase on an antiferromagnetic (AFM) spin background until $\delta_{\mathrm{c}}\approx 0.1$ and goes into a ferromagnetic (FM) superfluid (SF) phase at larger doping. By contrast, the bosonic $\sigma t$-$J$ model directly enters an SF phase with an easy-plane AFM order. Representative local order parameters obtained from GCE simulations are shown accordingly. (b) A representative PDW pattern on a $32\times 8$ cylinder with the AFM order and hole density profile. The edge width is proportional to the magnitude of the singlet pairing order parameter $\avg{\Delta_{i\alpha}}$. (c) The hole-hole correlation $D_h(r)$ with two doped holes. The cross represents the reference site. (d) The global order parameters as functions of doping levels on a $32\times 8$ cylinder. The shadow regions in (a) and (d) indicate an intermediate phase.}
    \label{fig:btj_phase_diagram}
\end{figure*}

In this work, we identify the critical role of quantum interference in the bosonic $t$-$J$ model as a decisive factor that determines the emergence of a non-trivial pairing order on the AFM spin background. By performing large-scale density-matrix renormalization group (DMRG) simulations, we establish a ground-state phase diagram [cf.\,Fig.\,\ref{fig:btj_phase_diagram}(a)] for the bosonic $t$-$J$ model on a 2D square lattice with varying doping levels. Upon doping the antiferromagnet, a robust singlet pairing order emerges with a pair density wave (PDW) at momentum $(\pi,\pi)$, which coexists with the AFM order symbiotically resembling the feature of supersolids. The boson pairs are tightly bound in space and exhibit $p$-wave symmetry up to the PDW oscillation. Importantly, the single-particle propagation is strongly suppressed in this PDW phase, prohibiting the Bose-Einstein condensation (BEC) of individual bosons that occurs in common boson systems. Only after exceeding a critical doping level does the ground state gradually turn into a conventional superfluid (SF) with the spins polarized into a Nagaoka ferromagnet (FM) concomitantly.

Therefore, we find a robust phase at small doping where the doped holes can propagate coherently on the AFM background only if they are tightly bound in pairs. This pairing can be well understood by the intrinsic quantum frustration in the model, where the hopping of a doped hole can accumulate a $\mathbb{Z}_2$ Berry phase. Such a Berry phase will result in destructive interference for the single-hole motion, whereas by canceling the frustration phases, the tightly bound pairs can realize a constructive coherent propagation. We further demonstrate that if such frustration is artificially removed, the pairing order will disappear and be replaced by a conventional SF. 

% Finally, we discuss the analogy to the flat-band SC, the implication to the fermionic case, and the potential experimental observations in cold atom simulations.

\textit{Model and phase diagram.---}We first present the numerical results and then discuss the physical interpretation. We consider the spin-$1/2$ hard-core bosonic $t$-$J$ model on a 2D square lattice of size $N=L_x\times L_y$. The Hamiltonian reads $H_{t\text{-}J} = \mathcal{P}_s (H_t + H_J) \mathcal{P}_s$ composed of a hopping term and an isotropic AFM spin-exchange term
\begin{equation}\label{eq:bosonic_tj}
\begin{aligned}
    H_t &= - t\sum_{\avg{ij} \sigma} \left( B^{\dagger}_{i\sigma} B_{j\sigma} + \text{H.c.}\right), \\
    H_J &= J \sum_{\avg{ij}} \left (\mathbf{S}_i\cdot \mathbf{S}_j - \frac{1}{4} n_i n_j \right).
\end{aligned}
\end{equation}
Here $B_{i\sigma}^\dagger$ and $B_{i\sigma}$ are the creation and annihilation operators of a hard-core boson of spin $\sigma\in\{\up,\dn\}$ at site $i=(x,y)$. $\mathbf{S}_i$ and $n_{i}=\sum_\sigma n_{i\sigma}$ are the spin and particle number operator. $\avg{ij}$ denotes nearest-neighbor (NN) links. The Hilbert space is restricted by the no-double occupancy constraint $n_i \leq 1$ via the projector $\mathcal{P}_s$. We set $J=1$ as the energy unit with $t/J=3$ and vary the hole doping level $\delta=\sum_i \avg{n_i^h}/N$ where $n_i^h=1-n_i$. We implement DMRG simulations using canonical ensembles (CE)~\cite{Jiang2018, Jiang2019, Jiang2020a, Qin2020, Gong2021, Lu2023, Jiang2023} for $L_y=4,6,8$ and grand canonical ensembles (GCE)~\cite{Jiang2021, Chen2023} for $L_y=6,8$ independently to obtain reliable results. We keep up to $M=25000$ SU(2) multiplets (equivalent to $D\approx 90000$ U(1) bond dimensions) in CE simulations with truncation error $\epsilon \lesssim 5\times 10^{-6}$, while up to $D=4000$ in GCE simulations only to measure local properties on the natural ground states~\cite{Tasaki2019}. The methodological details can be found in the Supplemental Material~\footnote{See Supplemental Material for more technical details and additional supporting data. The Supplemental Material also contains Refs. \cite{Sandvik2010, Chen2011, Chen2024, Yang2024, Trugman1988, Iglovikov2015, Liao2017, Liu2025}}.

\begin{figure}[t]
    \centering
    \includegraphics[width=0.97\linewidth]{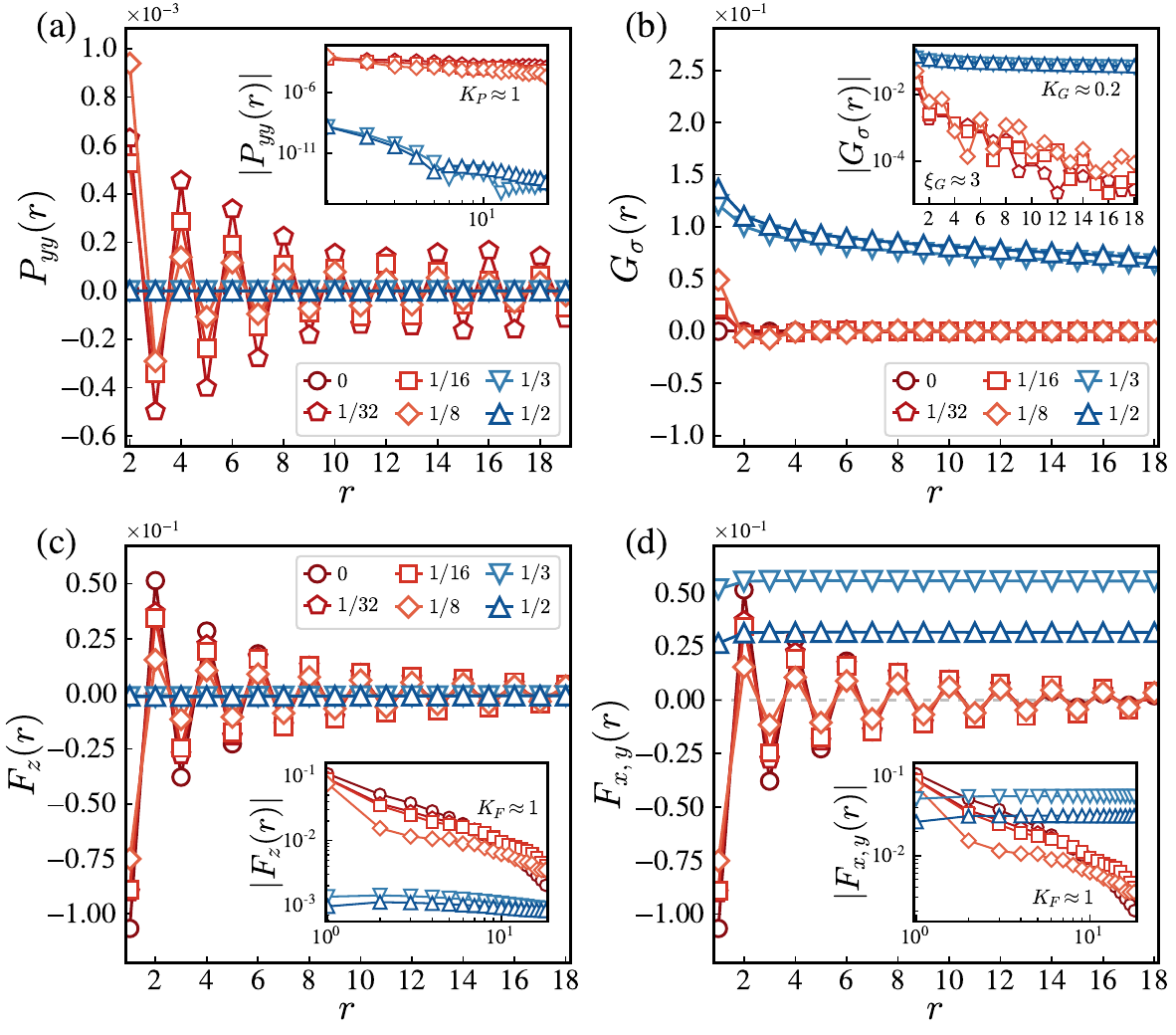}
    \caption{Correlation functions in the ground state of the bosonic $t$-$J$ model on cylinders of size $32\times 4$ or $36\times 4$ with $t/J=3$ and varying doping $\delta$ obtained from CE simulations. (a) The singlet pair-pair correlation. (b) The single-particle Green's function. (c) and (d) The spin-spin correlation along the $S^z$ and $S^{x,y}$ directions. The inset shows the logarithmic or semi-logarithmic plot with fitted power exponent $K$ or correlation length $\xi$. The red and blue markers indicate the AFM+PDW and FM+SF phases, respectively.}
    \label{fig:bosonic_sc_green_spin}
\end{figure}

\begin{figure}[b]
    \centering
    \includegraphics[width=0.97\linewidth]{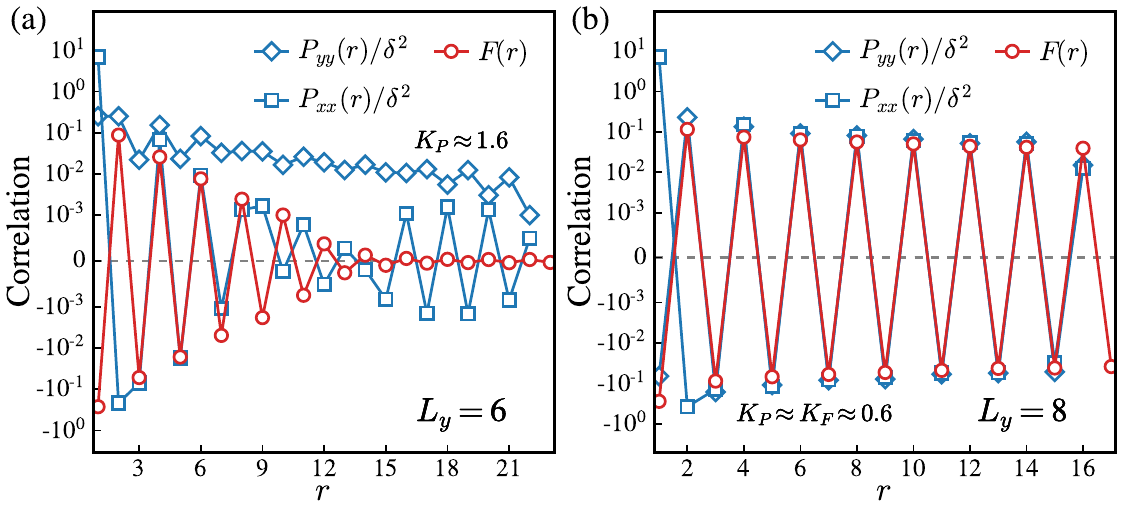}
    \caption{The pairing and spin correlations in the ground state of the bosonic $t$-$J$ model on cylinders of sizes (a) $32\times 6$ and (b) $24\times 8$ with $t/J=3$ and $\delta=1/24$ obtained from CE simulations, where the symmetric semi-logarithmic scale is used and $F(r)=\sum_{\gamma}F_\gamma(r)$. The fitted power exponents $K$ are labeled accordingly.}
    \label{fig:bosonic_CE_sc_spin_Ly=6_8}
\end{figure}

The long-range AFM order at half-filling persists to the finite-doping regime until a critical doping $\delta_{\mathrm{c}} \approx 0.1$, as evidenced by the spontaneous broken local magnetization $\avg{S_i^z}$ in the N\'{e}el order from GCE simulations in Fig.\,\ref{fig:btj_phase_diagram}(a), and the quasi-long-range spin-spin correlations $F_\gamma(r) = \avg{S_{i_0}^\gamma S_{i_0+r}^\gamma}$ from CE simulations in Figs.\,\ref{fig:bosonic_sc_green_spin}(c), (d) and Fig.\,\ref{fig:bosonic_CE_sc_spin_Ly=6_8}. Here $\gamma\in\{x,y,z\}$ denotes the spin component and $i_0=(x_0,y_0)$ with $x_0=L_x/4$ denotes the reference site. The doped holes even enhance the AFM order as evidenced by the slower decay of $F_\gamma(r)$ at finite doping and the larger order parameter $m_{\mathrm{AFM}}=\sum_i (-1)^{x+y} \avg{S_i^z}/N$ in Fig.\,\ref{fig:btj_phase_diagram}(d).

A robust singlet pairing order $\avg{\Delta_{i\alpha}}$ emerges at small doping with the sign staggered in both spatial directions, giving rise to a PDW order~\cite{Agterberg2020, Yue2024, Zhang2022a} at momentum $(\pi, \pi)$, coexisting with the AFM order, as shown by the spatial pattern in Fig.\,\ref{fig:btj_phase_diagram}(b) and the order parameter $\Delta_{\mathrm{PDW}}=\sum_{i\alpha} (-1)^{x+y} \avg{\Delta_{i\alpha}}/(2N-L_y)$ in Fig.\,\ref{fig:btj_phase_diagram}(d). Here $\Delta_{i\alpha} = \frac{1}{\sqrt{2}}\sum_{\sigma}\sigma B_{i,\sigma} B_{i+\alpha,-\sigma}$ is the spin-singlet pair operator. $\alpha\in \{\hat{x},\hat{y}\}$ is the NN link. The coefficient $\sigma$ takes $\{+1,-1\}$ for $\{\up,\dn\}$. The PDW is also observed by the quasi-long-range pair-pair correlation $P_{\alpha\beta}(r) = \langle \Delta_{i_0\alpha}^{\dagger} \Delta_{(i_0+r),\beta}\rangle$ with staggered oscillations in Fig.\,\ref{fig:bosonic_sc_green_spin}(a) and Fig.\,\ref{fig:bosonic_CE_sc_spin_Ly=6_8}. The case of $L_y=6$ exhibits a slightly different PDW pattern, on which we elaborate in the Supplemental Material with additional supporting data~\cite{Note1}.

The tight binding of hole pairs also manifests in the hole-hole correlation $D_{h}(r)= \avg{n^h_{i_0} n^h_{(i_0+r)}}$ in a two-hole-doped system, as shown in Fig.\,\ref{fig:btj_phase_diagram}(c). It decays exponentially with a correlation length of about one lattice constant, implying an extremely small pair size. The single-particle Green's function $G_{\sigma}(r)= \avg{B_{i_0,\sigma}^\dagger B_{(i_0+r),\sigma}}$ decays exponentially at small doping, as shown in Fig.\,\ref{fig:bosonic_sc_green_spin}(b), indicating that $\avg{\Delta_{i\alpha}}$ is a primary pairing order instead of a secondary result from single-particle condensate. 

% Moreover, the hole density distribution tends to be uniform in the bulks at moderately low doping, as shown by the profile $\avg{n_x^h}=\sum_{y}\avg{n_{x,y}^h}/L_y$ in Fig.\,\ref{fig:btj_phase_diagram}(b), compared to the distinct charge and spin stripes observed in the fermionic case that compete with the SC order~\cite{Zheng2017, Huang2017, Jiang2018, Jiang2019, Jiang2020a, Qin2020, Gong2021, Lu2023, Jiang2023, Jiang2021, Chen2023}.

% e.g., the fluctuation is of magnitude around $10^{-3}$ for width-$8$ cylinders

The AFM and PDW orders break down simultaneously at large doping and are replaced by a single-particle condensate as in common boson systems. This is evidenced by the prominent order parameter $\avg{B_{i\sigma}}$ in Fig.\,\ref{fig:btj_phase_diagram}(a) and the extremely slow decay of $G_\sigma(r)$ at large doping in Fig.\,\ref{fig:bosonic_sc_green_spin}(b). Compatible with the single-particle's unimpeded motion, the spin background is polarized to an FM order like the Nagaoka ferromagnetism~\cite{Auerbach2012}, as indicated by $\avg{S_i^z}$ in Fig.\,\ref{fig:btj_phase_diagram}(a) and $F_\gamma(r)$ in Figs.\,\ref{fig:bosonic_sc_green_spin}(c) and (d), respectively. The two phases can be seen clearly from the order parameter evolutions in Fig.\,\ref{fig:btj_phase_diagram}(d), with an intermediate phase in between~\cite{Note1}. $m_{\mathrm{FM}}=\sum_i \avg{S_i^z}/N$ and $\phi_{\mathrm{SF}}=\sum_{i\sigma} \avg{B_{i\sigma}}/N$ represent the FM and SF order parameters, respectively.

\textit{Quantum frustration from an exact Berry phase.---}The usual fate for bosons at low temperatures is BEC, once the single-particle phase coherence is established throughout the system, e.g., the Bose-Hubbard model at fractional filling exhibits an SF phase all the time~\cite{PhysRevB.40.546}. A significant feature revealed by our simulations is the absence of a single-boson BEC phase at zero temperature at small doping. A robust pairing order emerges instead. This suggests that the bosonic $t$-$J$ model may possess some hidden intrinsic phase frustration for the single-boson motion.

\begin{figure}[t]
    \centering
    \includegraphics[width=\linewidth]{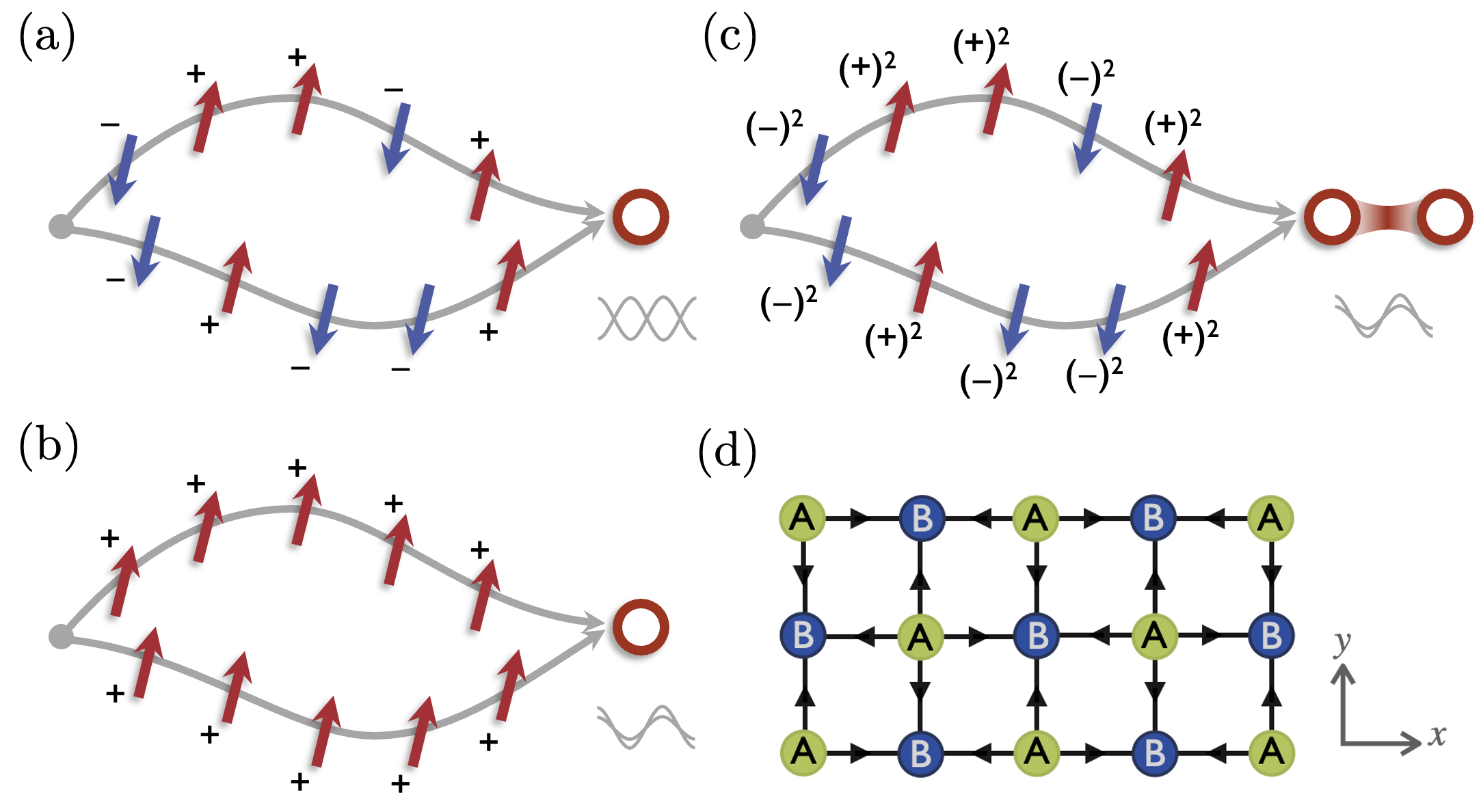}
    \caption{(a)-(c) Schematic illustration of the interference effect in the bosonic $t$-$J$ model. The gray curves represent the worldlines of holes (hollow circles) passing through the spin background (red and blue arrows). (d) The sign pattern $(-1)^{i+\alpha}$ for the directed singlet pair operator $\Delta_{i\alpha}$ of bosons under the Marshall basis. $\alpha$ taking the same direction as the arrow on the link leads to a positive sign.
}
    \label{fig:berry}
\end{figure}

We first consider an undoped Mott insulator described by the Heisenberg model $H_J$ on a square lattice. $H_J$ is stoquastic under the Marshall basis~\cite{Auerbach2012}, i.e., all the matrix elements of $(-H_J)$ are non-negative after performing a gauge transformation that attaches a negative sign for all the $\dn$ spins on sublattice A, $\left|\downarrow\right\rangle_\text{A} \rightarrow - \left|\downarrow\right\rangle_\text{A}$~\cite{Note1}. Hence, $H_J$ itself is sign-problem-free. However, upon doping, the hopping of doped holes will bring a nontrivial sign structure $\tau_C$ in the partition function
\begin{equation}\label{eq:Z_tJ}
    Z_{t\text{-}J}\equiv \opr{Tr} e^{-\beta H_{t\text{-}J}}=\sum_C \tau_C W_{t\text{-}J}[C],
\end{equation}
where $W_{t\text{-}J}[C] \geqslant 0$ is the non-negative weight corresponding to each closed imaginary-time evolution path $C$. The phase factor $\tau_{C}$ has the following exact form
\begin{equation}\label{eq:tau}
  \tau_{C} = (-1)^{N^h_{\dn}} ,
\end{equation}
which is also known as the phase string~\cite{Sheng1996, Weng1997, Wu2008}. $N^h_{\downarrow}$ is the total number of mutual exchanges between holes and $\downarrow$ spins in a closed loop $C$. This indicates that the single hole hopping can accumulate a $\mathbb{Z}_2$ Berry phase, as illustrated in Fig.\,\ref{fig:berry}(a), with the sign determined by the orientation of the spin exchanged with the hole—positive for $\up$ spin and negative for $\dn$ spin. In fact, this finding is broadly applicable to other strongly correlated systems, such as the fermionic $t$-$J$ and Hubbard model~\cite{Wu2008, Zhang2023c, Zhang2014}, where this Berry phase is a direct consequence of the Mottness independent of model details. In particular, for the bosonic $t$-$J$ model, $(-1)^{N^h_{\dn}}$ is the \emph{only} nontrivial sign structure, whereas, for the fermionic case, the additional fermionic sign of dopants is also present. The phase factor $\tau_C$ captures the possible quantum interference among different evolution paths and is an ``adiabatic'' phase accumulated during an evolution loop that cannot be gauged away. Interestingly, this frustration phase can be regarded as a generalized discrete version of the Berry phase action in the continuous path integral formalism of conventional models~\cite{Altland2010, Auerbach2012}.

Consequently, the coherence of a single hole is strongly disrupted by destructive interference among different paths due to the frustration phase that depends on the fluctuating spin configuration in the AFM background, as illustrated in Fig.\,\ref{fig:berry}(a). This theoretical insight is corroborated by the DMRG results in Fig.\,\ref{fig:bosonic_sc_green_spin}(b), which show a rapid decay in the single-particle Green's function $G_{\sigma}(r)$ at small doping. However, the phase frustration can be significantly canceled by the formation of tightly bound hole pairs in an ``in-tandem'' motion, where one hole always follows the other, as depicted in Fig.\,\ref{fig:berry}(c). Then, the string of phases created by the front hole can always be compensated by the identical phases created by the back hole, resulting in constructive interference among different paths of in-tandem hole-pair propagation, giving rise to a paired BEC. Moreover, given the dominant positive contribution from the constructive interference among extensive paths, the sign of the pairing order parameter $\avg{\Delta_{i\alpha}}$ is just determined by the residual static Marshall sign in the pair operator $\Delta_{i\alpha}$, which takes a staggered form of $(-1)^{i+\alpha}$ as depicted in Fig.\,\ref{fig:berry}(d). Here $(-1)^{i}$ takes $(\mp 1)$ for sublattice A and B respectively. This static sign exactly gives rise to the $\pi$-PDW compatible with the AFM order.

% as the fluctuating Berry phases can be effectively canceled out by forming hole pairs

% $B_{i\dn} \rightarrow (-1)^i B_{i\dn}$

Alternatively, if the spin background is fully polarized to an FM order as shown in Fig.\,\ref{fig:berry}(b), the single-hole hopping will accumulate identical phases, so all paths contribute constructively and hence the phase coherence of the single-hole motion can be fully restored to realize a superfluid. Thus, at larger doping, the hopping energy becomes dominant and forces the spins to polarize, which in turn allows individual holes to condense, leading to the FM+SF phase~\cite{Note1}.

\textit{Sign-problem-free: $\sigma t$-$J$ model.---}To further showcase the critical phase frustration effect, we turn it off artificially by modifying the hopping term $H_t$ to $H_{\sigma t} = - t \sum_{\langle i j\rangle \sigma} \sigma \left(B_{i \sigma}^{\dagger} B_{j \sigma} + \text {H.c.}\right)$. The resulting bosonic $\sigma t$-$J$ model, $H_{\sigma t\text{-}J}=\mathcal{P}_s(H_{\sigma t}+H_J)\mathcal{P}_s$, have the same partition function in Eq.\,\eqref{eq:Z_tJ} except for $\tau_C\equiv 1$, i.e., with the sign problem completely removed. It is interesting to note that such a sign-problem-free model is equivalent to the effective model of the spin-$1/2$ hard-core Bose-Hubbard model in the large-$U$ limit~\cite{Note1}.

\begin{figure}[t]
    \centering
    \includegraphics[width=0.95\linewidth]{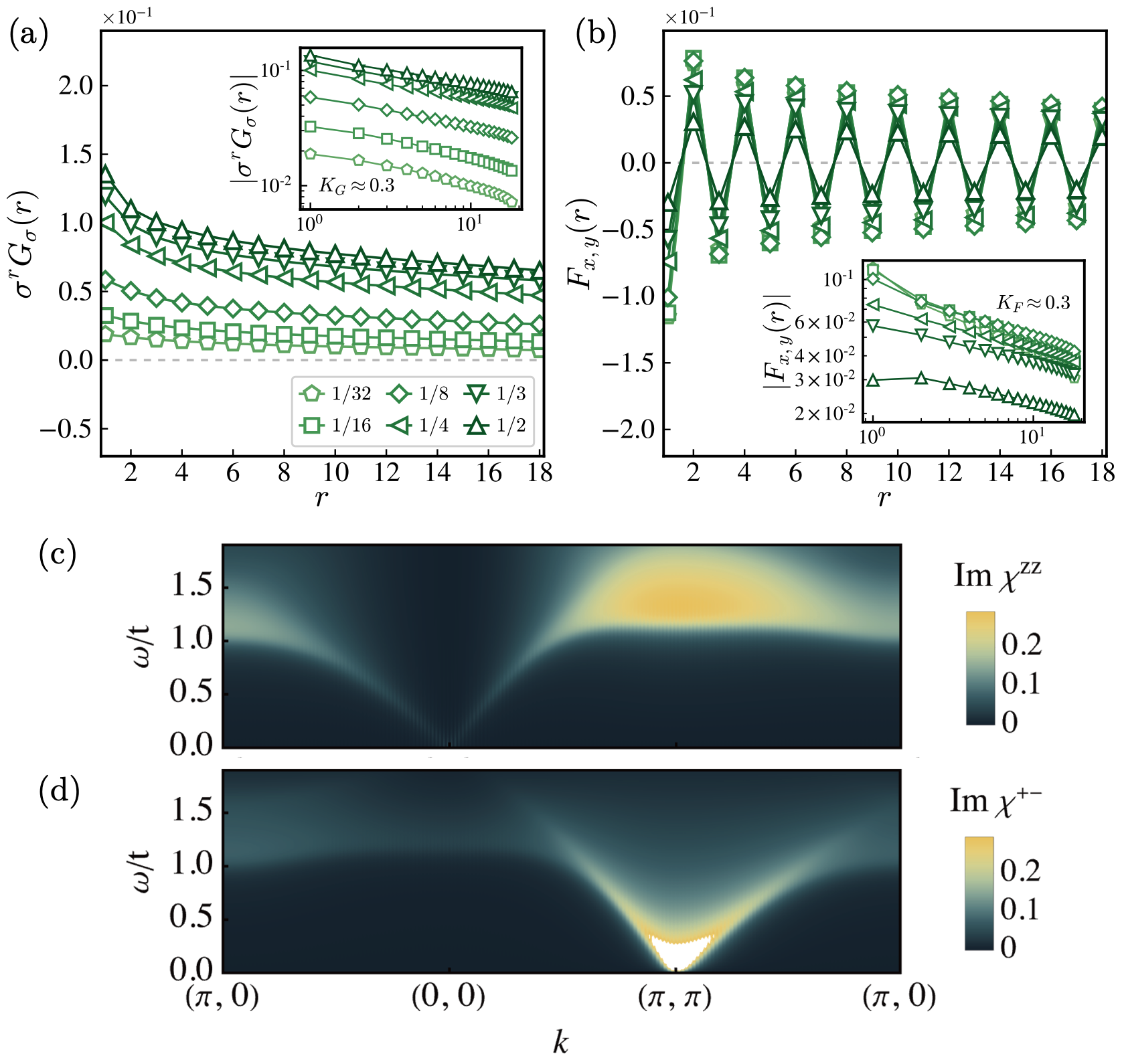}
    \caption{(a-b) The single-particle Green's function and the spin correlation for the $S^{x,y}$ components in the ground state of the bosonic $\sigma t$-$J$ model on cylinders of size $32\times 4$ or $36\times 4$ with $t/J=3$ and varying doping $\delta$ obtained from CE simulations. (c-d) The spin dynamic susceptibility for the $S^z$ and $S^{x,y}$ components at $\delta=1/12$ by parton mean-field calculations.}
    \label{fig:b_stj}
\end{figure}

Without the phase frustration, the dopants in the bosonic $\sigma t$-$J$ model can propagate coherently and hence undergo a BEC, as evidenced by the DMRG results in Fig.\,\ref{fig:b_stj}(a)~\cite{Boninsegni2001, Boninsegni2002, Boninsegni2008, Smakov2004}, which is similar to the case of the FM+SF phase in the bosonic $t$-$J$ model. In contrast to the latter, however, the AFM order here remains robust within the $S^{x,y}$ plane, as shown in Fig.\,\ref{fig:b_stj}(b). Namely, the mutual entanglement between the spin background and the dopants vanishes with $\tau_{C}=1$, leading to an effective decoupling between them. The no-double-occupancy constraint is still present but can be effectively addressed using a conventional parton mean-field scheme~\cite{Note1}, by which we show that at arbitrary non-zero doping, the bosons with opposite spins undergo a BEC at zero temperature separately at the $(0,0)$ and $(\pi,\pi)$ momenta, respectively. Figs.\,\ref{fig:b_stj}(c) and (d) show the spin spectrums at $\delta = 1/12$, indicating an AFM order in the $S^{x,y}$ plane in good agreement with the DMRG calculations. 

In contrast to the success of the slave-particle description in the $\sigma t$-$J$ model, the bosonic $t$-$J$ model encounters inherent difficulty: if both holons and spinons are bosonic, they tend to condense individually at zero temperature at the mean-field level, resulting in the single-particle condensation rather than the PDW order observed here; if both are fermionic, the AFM order can not be easily captured by the fermionic spinons at the mean-field level~\cite{PhysRevX.7.031051}. This contradiction highlights the crucial role of the quantum interference effect between charge and spin degrees of freedom in a correct low-energy effective theory for the bosonic $t$-$J$ model.

%\hk{$S^z$-string, reparable by the spin-flip process}

\textit{Discussion.---}In this work, we show that a quantum interference effect governs the physics in the bosonic doped antiferromagnets. Specifically, the destructive interference caused by the $\mathbb{Z}_2$ Berry phase can significantly hinder the propagation of individual bosonic dopants on an AFM background. However, the doped holes in pairs can effectively eliminate such quantum frustration to restore phase coherence, giving rise to a pairing order. This leads to an exotic pairing mechanism rooted in strong correlation effects, in stark contrast to those conventional pairing mechanisms based on Fermi surface instabilities. At small doping, this mechanism predicts an AFM+PDW supersolid. At larger doping, the same mechanism enforces the spin background to transit to an FM order such that the single bosonic dopants become coherent to condense into a superfluid. These novel phases of matter are promising to be observed on ultracold atom platforms~\cite{Koepsell2021, Xu2023a, Lebrat2024, Sun2021, Homeier2024, Note1}. We point out that, although related, the pairing mechanism discussed here differs from the mechanism purely based on the spin mismatch caused by the hopping of doped holes~\cite{Note1}, which is supported by the observation that the ground state of the bosonic $t$-$J_z$ model is a single-particle superfluid~\cite{Smakov2004} instead of the pairing order found here. Interestingly, due to the exact mapping between the hard-core spin-1/2 bosonic system and spin-1 system~\cite{Note1}, the framework presented here also provides a novel perspective for exploring exotic magnetic orders in high-spin systems, such as quadrupolar or nematic spin supersolids.

Intrinsic physics from interference also manifests in other systems, such as Moir\'{e} structures, exotic lattices, and the Hofstadter model. In these cases, however, the frustrated interference patterns are static, dictated by lattice geometry or external field, allowing destructive interference to be effectively neutralized by expanding the unit cell, rendering it imperceptible at long wavelengths. By contrast, the quantum interference discussed here depends on the fluctuating background spins and hence is inherently dynamic, cannot be effectively removed, and remains significant even in the infrared limit.

The present bosonic model only differs from the well-known fermionic $t$-$J$ model by the statistics of dopants. An additional fermionic sign structure will appear in the partition function of the latter. Two models are expected to bear certain similarities in the presence of a few holes, but predict dramatically different phases at finite doping. The fermionic statistics is expected to play an important role when involving the overlap of hole pairs in a fashion of the BEC-BCS crossover, with a partial recurrence of the phase frustration giving rise to competing orders such as stripes~\cite{Fradkin2015} or short-range AFM with local pair binding. Given a similar structure, the bosonic model thus provides a different perspective and new insights into the nature of the complex phase diagram emerging in the fermionic doped Mott insulators that may be closely related to the high-$T_c$ cuprate superconductors.

\begin{acknowledgments}
\textit{Note added.---}After we finished this manuscript, we became aware of a closely related work by Harris et al.~\cite{Harris2024}, in which they performed DMRG simulations on width-$4$ and $6$ cylinders with an additional density-density repulsion term compared to the model used here, with a special focus on the magnetic transition and signatures of stripes. A detailed discussion on this difference is provided in the Supplemental Material~\cite{Note1}.

We acknowledge the stimulating discussions with Zhi-Jian Song, Jing-Yu Zhao, Boyan Shen, Yu-Peng Wang, Lukas Homeier and Leon Balents. H.-K.Z, J.-S.X., and Z.-Y.W. are supported by MOST of China (Grant No. 2021YFA1402101) and NSF of China (Grant No. 12347107); H.-K.Z is supported by the Postdoctoral Fellowship Program and China Postdoctoral Science Foundation under Grant No. BX20250169; J.-X.Z was funded by the European Research Council (ERC) under the European Union’s Horizon 2020 research and innovation program (Grant Agreement No. 853116, acronym TRANSPORT). This research was supported in part by grant NSF PHY-2309135 to the Kavli Institute for Theoretical Physics (KITP). H.-C.J. was supported by the Department of Energy (DOE), Office of Sciences, Basic Energy Sciences, Materials Sciences and Engineering Division, under Contract No. DE-AC02-76SF00515.
\end{acknowledgments}

\bibliographystyle{apsreve}
\bibliography{main}

\clearpage
\newpage
% \appendix
\widetext

\begin{center}
\textbf{\large Supplementary Information for \\``Quantum-interference-induced pairing in bosonic doped antiferromagnets''}
\end{center}
% \tableofcontents
\addtocontents{toc}{\protect\setcounter{tocdepth}{0}}
{
\tableofcontents
}

\renewcommand{\thefigure}{S\arabic{figure}}
\setcounter{figure}{0}
\renewcommand{\theequation}{S\arabic{equation}}
\setcounter{equation}{0}
\renewcommand{\thesection}{\Roman{section}}
\setcounter{section}{0}
\renewcommand{\thetable}{S\arabic{table}}
\setcounter{table}{0}
\setcounter{secnumdepth}{4}

\section{The effective model of the spinful hard-core Bose-Hubbard model}
In this section, we will derive the effective model of the spinful Bose-Hubbard model in the limit of $U\gg t$ and show that it is equivalent to the bosonic $\sigma t$-$J$ model investigated in this work under an on-site unitary (phase) transformation on bipartite lattices, similar to the derivation from the Fermi-Hubbard model to the standard fermionic $t$-$J$ model via the second-order perturbation in the large $U$ limit. The Hamiltonian of the one-band spinful hard-core Bose-Hubbard model reads
\begin{equation}\label{eq:spinful_hard_core_bose_hubbard}
    H_{t\text{-}U} = H_t + H_U = - t \sum_{\avg{ij} \sigma} \left( B^{\dagger}_{i\sigma} B_{j\sigma} + \text{H.c.}\right) + U \sum_{i} n_{i\up}n_{i\dn},
\end{equation}
where $B_{i\sigma}^\dagger$ and $B_{i\sigma}$ are the creation and annihilation operators of a hard-core boson of spin $\sigma\in\{\up,\dn\}$ at site $i$, respectively. ``$\text{H.c.}$'' means the Hermitian conjugation. The corresponding commutation relation is
\begin{equation}
\begin{aligned}
    &[B_{i\sigma}, B_{j\sigma'}^\dagger] = 0, ~~[B_{i\sigma}, B_{j\sigma'}] = 0, \quad \text{if } i\neq j \text{ or } \sigma\neq\sigma',\\
    &\{B_{i\sigma}, B_{j\sigma'}^\dagger \} = 1, \quad \text{if } i = j \text{ and }  \sigma=\sigma',
\end{aligned}
\end{equation}
or equivalently,
\begin{equation}
    [B_{i\sigma}, B^\dagger_{j\sigma'}] = \delta_{ij}\delta_{\sigma\sigma'} \left(1 - 2 B^\dagger_{j\sigma'} B_{i\sigma} \right)  ,~~[B_{i\sigma}, B_{j\sigma'}] = 0,
\end{equation}
where $\delta_{ij}$ denotes the Kronecker delta function and the identity operator is omitted for simplicity and denoted as $1$. $n_{i\sigma}=B_{i\sigma}^\dagger B_{i\sigma}$ is the particle density operator of spin $\sigma$ at site $i$. We use $n_{i}=\sum_\sigma n_{i\sigma}$ to denote the total particle density at site $i$. $t\geq 0$ is the hopping integral between the nearest-neighbor (NN) sites $\avg{ij}$ and $U\geq 0$ is the on-site Hubbard repulsion.

Suppose that the total particle number is below half-filling. In the limit of $U/t\rightarrow +\infty$, the ground states are highly degenerate, forming a subspace spanned by all possible no-double-occupied configurations. Denote the projector to this subspace as $\mathcal{P}_s$. According to the Brillouin-Wigner perturbation theory, the contributions to the effective Hamiltonian up to the second order are
\begin{equation}
\begin{aligned}
    & H_{\text{eff}}^{(0)} = \mathcal{P}_s H_U \mathcal{P}_s = 0, \\
    & H_{\text{eff}}^{(1)} = \mathcal{P}_s H_t \mathcal{P}_s, \\
    & H_{\text{eff}}^{(2)} = \mathcal{P}_s H_t \frac{1-\mathcal{P}_s}{0-H_U} H_t \mathcal{P}_s = -\frac{1}{U} \mathcal{P}_s H_t (1-\mathcal{P}_s) H_t \mathcal{P}_s \\
    &\quad\quad= -\frac{t^2}{U}\sum_{\avg{ij}\avg{kl}\sigma\sigma'} \mathcal{P}_s \left(  B_{i\sigma}^\dagger B_{j\sigma} + \text{H.c.} \right) (1-\mathcal{P}_s) \left(  B_{k\sigma'}^\dagger B_{l\sigma'} + \text{H.c.} \right) \mathcal{P}_s.
\end{aligned}
\end{equation}
In the second-order contribution, the successive action of the projectors $\mathcal{P}_s$, $(1-\mathcal{P}_s)$, and $\mathcal{P}_s$ results in the surviving terms in $H_t$ must map a no-double-occupied configuration to a double-occupied configuration and then back to a no-double-occupied configuration. Hence, $H_U^{-1}$ directly becomes $1/U$ and the two links $\avg{ij}$ and $\avg{kl}$ must have overlaps. In the following, we only consider the cases where the two links completely coincide $\avg{ij}=\avg{kl}$, i.e., the cases where the two links overlap by only one site are omitted, as in the derivation of the standard fermionic $t$-$J$ model. Hence, for each link $\avg{ij}$, the effect of the projector $(1-\mathcal{P}_s)$ can be equivalently represented by a factor $n_in_j$ just after the first (rightmost) $\mathcal{P}_s$, which means that only the configuration with singly-occupied sites $i,j$ can survive and then naturally leads to a double-occupied configuration after the action of $H_t$. Thus, the second-order contribution becomes
\begin{equation}
\begin{aligned}
    H_{\text{eff}}^{(2)\prime} = -\frac{t^2}{U}\sum_{\avg{ij}\sigma\sigma'} \mathcal{P}_s \left(  B_{i\sigma}^\dagger B_{j\sigma} + \text{H.c.} \right) \left(  B_{i\sigma'}^\dagger B_{j\sigma'} + \text{H.c.} \right) n_i n_j \mathcal{P}_s.
\end{aligned}
\end{equation}
The terms in the summation on a single link $\avg{ij}$ can be expressed as
\begin{equation}
\begin{aligned}
    &\mathcal{P}_s\left( \sum_{\sigma} B_{i\sigma}^\dagger B_{j\sigma} + \text{H.c.} \right)^2 n_i n_j \mathcal{P}_s \\
    &= \sum_{\sigma\sigma'} \mathcal{P}_s\left(  B_{i\sigma}^\dagger B_{j\sigma} B_{i\sigma'}^\dagger B_{j\sigma'} + B_{i\sigma}^\dagger B_{j\sigma} B_{j\sigma'}^\dagger B_{i\sigma'} + B_{j\sigma}^\dagger B_{i\sigma} B_{i\sigma'}^\dagger B_{j\sigma'} + B_{j\sigma}^\dagger B_{i\sigma} B_{j\sigma'}^\dagger B_{i\sigma'} \right) n_i n_j \mathcal{P}_s \\
    &= \sum_{\sigma\sigma'} \mathcal{P}_s \left( B_{i\sigma}^\dagger B_{i\sigma'} ( \delta_{\sigma\sigma'} - (2\delta_{\sigma\sigma'}-1) B_{j\sigma'}^\dagger B_{j\sigma} ) + ( \delta_{\sigma\sigma'} - (2\delta_{\sigma\sigma'}-1) B_{i\sigma'}^\dagger B_{i\sigma} ) B_{j\sigma}^\dagger B_{j\sigma'} \right) n_i n_j \mathcal{P}_s, \\
    &= \mathcal{P}_s \left(n_i + n_j + 2 \sum_{\sigma\sigma'} B_{i\sigma}^\dagger B_{i\sigma'} B_{j\sigma'}^\dagger B_{j\sigma} - 4\sum_{\sigma} n_{i\sigma}n_{j\sigma}\right) n_i n_j \mathcal{P}_s.
\end{aligned}
\end{equation}
Note that the first and last terms in the second line vanish due to the projector $\mathcal{P}_s$. Compared to the fermionic case, the sign of the exchange term $2\sum_{\sigma\sigma'} B_{i\sigma}^\dagger B_{i\sigma'} B_{j\sigma'}^\dagger B_{j\sigma}$ is reversed and there is an additional term $-4\sum_{\sigma} n_{i\sigma}n_{j\sigma}$. The exchange term can be rewritten via the Schwinger boson representation of the spin operator 
\begin{equation}
    \mathbf{S}_i=(S^x_i, S^y_i, S^z_i)=\frac{1}{2}\sum_{\sigma\sigma'} B_{i\sigma}^\dagger \boldsymbol{\tau}_{\sigma\sigma'} B_{i\sigma'},
\end{equation}
as
\begin{equation}
\begin{aligned}
    2\sum_{\sigma\sigma'} B_{i\sigma}^\dagger B_{i\sigma'} B_{j\sigma'}^\dagger B_{j\sigma} = 4 \mathbf{S}_i\cdot \mathbf{S}_j + n_{i} n_{j}.
\end{aligned}
\end{equation}
by use of the Pauli decomposition of the swap operation
\begin{equation}
    2 \delta_{\sigma\rho'} \delta_{\sigma'\rho} = \boldsymbol{\tau}_{\sigma\sigma'} \cdot \boldsymbol{\tau}_{\rho\rho'} + \delta_{\sigma\sigma'}\delta_{\rho\rho'} ,
\end{equation}
where $\boldsymbol{\tau}_{\sigma\sigma'}$ denote the vector composed of the three Pauli matrices. Hence, the second-order contribution to the effective Hamiltonian is
\begin{equation}\label{eq:Heff_2_spinful}
    H_{\text{eff}}^{(2)\prime} = - J \mathcal{P}_s \left( \mathbf{S}_i\cdot \mathbf{S}_j + \frac{3}{4} n_i n_j - \sum_{\sigma} n_{i\sigma}n_{j\sigma} \right) \mathcal{P}_s,
\end{equation}
where the spin-exchange coupling constant is still $J=4t^2/U$. However, the coupling term $\mathbf{S}_i\cdot \mathbf{S}_j$ becomes ferromagnetic, which is completely different from the fermionic case. Recombining the terms in Eq.\,\eqref{eq:Heff_2_spinful} using the identity
\begin{equation}
    4 S_i^z S_j^z + n_i n_j = \sum_{\sigma\sigma'} (\sigma\sigma'+1) n_{i\sigma} n_{j\sigma'} = \sum_{\sigma\sigma'} 2\delta_{\sigma\sigma'} n_{i\sigma} n_{j\sigma'} = 2\sum_{\sigma} n_{i\sigma} n_{j\sigma},
\end{equation}
will obtain
\begin{equation}
\begin{aligned}
    \mathbf{S}_i\cdot \mathbf{S}_j + \frac{3}{4} n_i n_j - \sum_{\sigma} n_{i\sigma}n_{j\sigma} &= \mathbf{S}_i\cdot \mathbf{S}_j + \frac{3}{4} n_i n_j - \frac{1}{2}\left(4 S_i^z S_j^z + n_i n_j\right) \\
    &= S_i^x S_j^x + S_i^y S_j^y - \left(S_i^z S_j^z - \frac{1}{4} n_i n_j\right),
\end{aligned}
\end{equation}
where $\sigma\in\{+1,-1\}$ for $\{\up,\dn\}$ respectively when serving as a coefficient. Therefore, the effective Hamiltonian of the spinful hard-core Bose-Hubbard model at $U\gg t$ up to the second order is 
\begin{equation}\label{eq:btj_minus}
    H_{\text{eff}} = \mathcal{P}_s \left(H_t + H_J^{-} \right) \mathcal{P}_s,
\end{equation}
where
\begin{equation}\label{eq:eff_tj_ferro_xy}
\begin{aligned}
    H_t &= - t\sum_{\avg{ij} \sigma} \left( B^{\dagger}_{i\sigma} B_{j\sigma} + \text{H.c.}\right), \\
    H_J^{-} &= J \sum_{\avg{ij}} \left( - S_i^x S_j^x - S_i^y S_j^y + S_i^z S_j^z - \frac{1}{4} n_i n_j \right).
\end{aligned}
\end{equation}
This effective Hamiltonian is quite similar to the bosonic $t$-$J$ model except that the in-plane $(S^x, S^y)$ spin-exchange is ferromagnetic (FM). The root cause is that exchanging two fermions of different spin states yields a negative sign while that for spinful hard-core bosons does not. This is a significant difference compared to the fermionic case where the effective model of the Fermi-Hubbard model at $U\gg t$ is just the fermionic $t$-$J$ model. Note that the spin-exchange coupling along $S^z$ in Eq.\,\eqref{eq:eff_tj_ferro_xy} is still antiferromagnetic, in line with the common argument about the Hubbard model, i.e., the nearest-neighboring (NN) spins tend to be antiparallel because NN parallel spins cannot gain kinetic energy from the second-order virtual hopping process due to the Pauli exclusion principle or hard-core property while NN antiparallel spins can. Note that if the following identity is applied
\begin{equation}
    S_i^z S_j^z - \frac{1}{4} n_i n_j = \frac{1}{4} \sum_{\sigma\sigma'} (\sigma\sigma'-1) n_{i\sigma} n_{j\sigma'} = \frac{1}{4} \sum_{\sigma\sigma'} (-2\delta_{\sigma,-\sigma'}) n_{i\sigma} n_{j\sigma'} = -\frac{1}{2} \sum_{\sigma} n_{i\sigma} n_{j,-\sigma},
\end{equation}
the anisotropic bosonic $t$-$J$ model in Eq.\,\eqref{eq:eff_tj_ferro_xy} will have exactly the same form as in Ref.\,\cite{Boninsegni2001, Boninsegni2002, Boninsegni2008}.

We remark that the term $H_J^{-}$ in Eq.\,\eqref{eq:eff_tj_ferro_xy} explicitly breaks the spin-$\mathrm{SU}(2)$ rotation symmetry. The reason can be actually traced back to the fact that the hopping term in the original spinful hard-core Bose-Hubbard model in Eq.\,\eqref{eq:spinful_hard_core_bose_hubbard} has already broken the spin-$\mathrm{SU}(2)$ rotation symmetry implicitly as the on-site fermion sign between different $\up$ or $\dn$ spins has been removed and replaced by the ``artificial'' spinful hard-core boson statistics. Instead, the bosonic $t$-$J$ model indeed has the spin-$\mathrm{SU}(2)$ rotation symmetry since there is no double-occupied state that needs to be considered in the hopping term. Moreover, as we will discuss in the next section, the effective model in Eq.\,\eqref{eq:eff_tj_ferro_xy} is sign-problem-free or say sign-free, consistent with the fact that the original spinful hard-core Bose-Hubbard model in Eq.\,\eqref{eq:spinful_hard_core_bose_hubbard} is also a sign-free model.

From above one can see that the effective model of the spinful hard-core Bose-Hubbard model at half-filling is not the Heisenberg model anymore. But it is possible to recover the Heisenberg limit by introducing an on-site unitary (phase) transformation on bipartite lattices like square lattices that changes the sign of the in-plane spin-exchange interaction, i.e., the so-called Marshall transformation
\begin{equation}\label{eq:marshall}
    U_{\text{Ms}} = (-1)^{N_A^{\dn}} = \prod_{i\in A} (-1)^{n_{i\dn}} = \prod_{i\in A} e^{-i S_i^z \pi } e^{i\pi/2},
\end{equation}
where $N_A^{\dn}$ denotes the total number of $\dn$ spins on sublattice $A$. The choices of $A$ or $B$ sublattice and $\dn$ or $\up$ spins in this transformation are arbitrary. Here we just choose to count $\dn$ spins on sublattice $A$. This Marshall transformation can be seen as a $\mathrm{U}(1)$ gauge transformation via the mapping between spins and hard-core bosons.

For a real-space Fock state, here referring to a product state composed of single-site spin-up, spin-down, and unoccupied states $\{\left|\up\right\rangle, \left|\dn\right\rangle, \ket{0}\}$ (a hole-spin configuration), if the state on site $i$ of sublattice $A$ is $\left|\dn\right\rangle$, $U_{\text{Ms}}$ will attach a negative sign to the state $\left|\dn\right\rangle \rightarrow -\left|\dn\right\rangle$. The transformed basis is called the ``Marshall basis''. In such a way, $S_i^x$ and $S_i^y$ on sublattice $A$ can obtain a negative sign while $S_i^z$ on sublattice $A$ keeps unchanged, i.e.,
\begin{equation}
    U_{\text{Ms}}^\dagger S_i^x U_{\text{Ms}} = -S_i^x,\quad U_{\text{Ms}}^\dagger S_i^y U_{\text{Ms}} = -S_i^y,\quad
    U_{\text{Ms}}^\dagger S_i^z U_{\text{Ms}} = S_i^z,\quad i\in A.
\end{equation}
Hence, the transformed spin-exchange term becomes isotropically antiferromagnetic (AFM). At the same time, the hopping term should also change under this unitary transformation according to
\begin{equation}
    U_{\text{Ms}}^\dagger B_{i\sigma} U_{\text{Ms}} = \sigma B_{i\sigma},\quad  U_{\text{Ms}}^\dagger B_{i\sigma}^\dagger U_{\text{Ms}} = \sigma B_{i\sigma}^\dagger, \quad i\in A.
\end{equation}
Namely, a spin-dependent sign $\sigma$ will be attached to the NN hopping term. Assuming $i$ is an odd number for each $i\in A$, the expression above can be simplified to 
\begin{equation}
    U_{\text{Ms}}^\dagger B_{i\sigma} U_{\text{Ms}} = \sigma^i B_{i\sigma}.
\end{equation}
Therefore, the transformed effective Hamiltonian becomes exactly the bosonic $\sigma t$-$J$ model
\begin{equation}\label{eq:bosonic_sigma_tj}
    H_{\sigma t\text{-}J} = U_{\text{Ms}}^\dagger H_{\text{eff}} U_{\text{Ms}} = \mathcal{P}_s( H_{\sigma t} + H_{J} ) \mathcal{P}_s,
\end{equation}
where
\begin{equation}
\begin{aligned}
    H_{\sigma t} &= - t\sum_{\avg{ij} \sigma} \sigma \left( B^{\dagger}_{i\sigma} B_{j\sigma} + \text{H.c.}\right), \\
    H_J &= J \sum_{\avg{ij}} \left( \mathbf{S}_i\cdot \mathbf{S}_j - \frac{1}{4} n_i n_j \right).
\end{aligned}
\end{equation}
Remember that $\sigma\in\{+1,-1\}$ for $\{\up,\dn\}$ respectively when serving as a coefficient. In summary, in the limit of $U\gg t$, the spinful hard-core Bose-Hubbard model should effectively have the same nature as the bosonic $\sigma t$-$J$ model, up to the Marshall transformation that maps in-plane ferromagnetism to in-plane antiferromagnetism. That is to say, the correlation functions on the ground states of the two models are closely related to each other by applying $U_{\text{Ms}}$ to the observables. Especially, the pure ``charge'' properties are exactly the same, such as the hole-hole density correlation $D_h(r)$ in Fig.\,\textcolor{darkblue1}{1}(c).

\section{Quantum frustration analysis}
In this section, we analyze the quantum frustrations or say the sign structures of the models mentioned in the main text based on explicit power series expansions of the partition functions. Here the term ``quantum frustration'' or ``sign structure'' refers to the form of the sign (phase) factor in the path-integral-type summation of the partition function.

\subsection{General introduction of quantum frustration}
The partition function of a generic quantum system $Z=\opr{Tr} e^{-\beta H}$ can be expanded as a summation over closed paths of imaginary-time evolution~\cite{Sandvik2010}
\begin{equation}\label{eq:Z_expansion}
    Z=\sum_{n=0}^{\infty} \sum_{\left\{\alpha\right\}_n} \frac{\beta^{n}}{n !} \prod_{k=0}^{n-1} \left\langle\alpha_{k+1}\right| (-H) \left| \alpha_{k}\right\rangle,
\end{equation} 
where each $\ket{\alpha_i}$ runs over a complete basis of the Hilbert space, e.g., the real-space Fock basis. Each time series of states $\{\alpha\}_n$ should satisfy the temporal periodic boundary condition $\ket{\alpha_{n}}=\ket{\alpha_{0}}$. Each non-zero matrix element $\left\langle\alpha_{k+1}|(-H)| \alpha_{k}\right\rangle$ describes a single nontrivial evolution step. By decomposing matrix elements into magnitudes and phases (signs) and multiplying them respectively along each evolution path $C$, the partition function can be written as a path-integral-type summation
\begin{equation}\label{eq:Z_tau_W}
    Z=\sum_C \tau_C W[C],
\end{equation}
where $\tau_C$ and $W[C]$ represent the accumulated phase factor and the non-negative weight, respectively. The expectation values of observables, including various correlation functions, can also be expanded as in Eq.\,\eqref{eq:Z_expansion} but with extra operators inserted. There are other expansion schemes besides the direct series expansion used here, such as those in the auxiliary-field quantum Monte Carlo algorithm for interacting fermion systems, but they can still be written in the form of amplitude-phase summation like in Eq.\,\eqref{eq:Z_tau_W}. The phase factor $\tau_C$ captures the possible quantum interference among different evolution paths, or say the ``quantumness'' of the system under the expansion scheme. If $\tau_C$ is positive-definite, the quantum system is effectively mapped to a classical system and can be solved by Monte Carlo sampling methods. Otherwise, we say there is quantum frustration or a ``sign (phase) problem'', especially when the phase strongly fluctuates. Quantum frustration is not only a crucial algorithmic problem in quantum many-body physics but also plays a critical role in determining the low-energy physics of corresponding models. Interestingly, this frustration phase can be regarded as a generalized discrete version of the Berry phase action in the continuous path integral formalism of conventional models~\cite{Altland2010, Auerbach2012}. It is an ``adiabatic'' phase accumulated during an evolution loop that cannot be gauged away and is independent of the evolution speed, in contrast to the dynamical phase stemming from the exponentiated imaginary unit $i$ in the real-time evolution.

For the undoped AFM Heisenberg model $H_J$ on bipartite lattices, the phase factor $\tau_C$ is positive-definite because despite that the non-zero off-diagonal matrix element $\left\langle \alpha_{k+1}|(-S^+_iS^-_j)| \alpha_{k}\right\rangle$ is always negative, an even number of spin flips is required in every allowed evolution path to satisfy the temporal periodicity. This can be seen more explicitly and locally by transforming the Hamiltonian into a stoquastic form according to the so-called Marshall sign rule~\cite{Auerbach2012}. Namely, we perform the following gauge transformation
\begin{equation}\label{eq:U_Ms}
    U_{\text{Ms}} = (-1)^{N_{\text{A}}^{\dn}} = \prod_{i\in \text{A}} e^{i\pi n_{i\dn}},
\end{equation}
which counts the parity of the total number of $\dn$ spins $N_\text{A}^{\dn}$ on sublattice $\text{A}$ (here the choices of sublattice and $\dn$ or $\up$ spins are arbitrary). The transformed Heisenberg interaction carries an extra minus sign for the spin-flip process
\begin{equation}\label{eq:ferro_xy}
\begin{aligned}
    &H_J^{-} = U_{\text{Ms}} H_J U_{\text{Ms}}^\dagger \\
    &= J \sum_{\avg{ij}} \left( - S_i^x S_j^x - S_i^y S_j^y + S_i^z S_j^z - \frac{1}{4} n_i n_j \right),
\end{aligned}
\end{equation}
so that all non-zero matrix elements in Eq.\,\eqref{eq:Z_expansion} become positive (``stoquastic''), resulting in a positive-definite $\tau_C$. Note that each contribution in Eq.\,\eqref{eq:Z_tau_W} is invariant under a phase transformation like $U_{\text{Ms}}$.

For the doped AFM bosonic $t$-$J$ model, quantum frustration emerges upon doping holes because the number of spin exchanges is not necessarily even anymore due to the additional hole-spin exchange process. If one removes the negative sign in the spin-flip process by $U_{\text{Ms}}$, another negative sign will arise in the transformed hopping term $H_{\sigma t} = U_{\text{Ms}} H_{t} U_{\text{Ms}}^\dagger$ correspondingly, where the phase factor takes the form of $(-1)^{N_{\dn}^h}$ as the exchange of a hole and a $\dn$ spin contributes a negative sign. If one focuses on the ``worldline'' of a single hole, a string of $\mathbb{Z}_2$ phases $( \pm 1) \times( \pm 1) \times \cdots \times( \pm 1)$ will be picked up by the moving hole from the spin background depending on the orientation of the spin exchanged with the hole~\cite{Sheng1996, Weng1997, Wu2008, Chen2011}. We emphasize that the key advantage of the Marshall transformation is that it allows us to focus on the motion of holes by rendering the spin-flip process positive-definite, thereby eliminating the need to account for the phases arising from the realignment of the spin background. In other words, the frustration phase in the form of $(-1)^{N_{\dn}^h}$ treats doped holes as fundamental objects.

\subsection{Bosonic $t$-$J$ model}
We first analyze the sign structure of the bosonic $t$-$J$ model on bipartite lattices. After the Marshall transformation in Eq.\,\eqref{eq:marshall}, the Hamiltonian of the bosonic $t$-$J$ model can be rewritten as
\begin{equation}
\begin{aligned}
    U_{\text{Ms}}^\dagger H_{t\text{-}J} U_{\text{Ms}} &= H_{\sigma t} + H_{J}^{(-)} \\
    &= -t\left(T_{o \up}-T_{o \dn}\right) - \frac{J}{2} \left(T_{\up \dn} + V_{\up \dn}\right),
\end{aligned}
\end{equation}
where
\begin{eqnarray}
    T_{o \sigma} &=&\sum_{\avg{ij}} B_{i \sigma}^{\dagger} B_{j \sigma} + \text{H.c.}, \\
    T_{\up \dn} &=&\sum_{\langle i j\rangle} B_{i \up}^{\dagger} B_{i \dn} B_{j \dn}^{\dagger} B_{j \up} + \text{H.c.}, \\
    V_{\up \dn} &=&\sum_{\avg{ij}} \left(n_{i \up} n_{j \dn}+n_{i \dn} n_{j \up}\right).
\end{eqnarray}
Here we omit to write down the no-double-occupancy projector $\mathcal{P}_s$ explicitly and absorb it into each single-particle operator for simplicity (the same below). $T_{o \sigma}$ describes the NN exchange process of holes and $\sigma$ spins, $T_{\up \dn}$ describes the NN exchange process of $\up$ spins and $\dn$ spins, and $V_{\up \dn}$ describes the potential energy of NN antiparallel spins. The series expansion of the corresponding partition function up to infinite orders reads~\cite{Wu2008}
\begin{equation}\label{eq:expand_tj}
\begin{aligned}
    Z_{t \text{-}J}&= \opr{Tr} e^{-\beta  H_{t\text{-}J}} = \opr{Tr} e^{-\beta U_{\text{Ms}}^\dagger H_{t\text{-}J} U_{\text{Ms}}} = \opr{Tr} \sum_{n=0}^{\infty} \frac{\beta^{n}}{n!}\left(- U_{\text{Ms}}^\dagger H_{t\text{-}J} U_{\text{Ms}} \right)^{n} \\
    &= \sum_{n=0}^{\infty}  \frac{\beta^{n}}{n !} \opr{Tr}\left[\sum \cdots \left(t T_{o \up}\right) \cdots \left(\frac{J}{2} T_{\up \dn}\right) \cdots\left(-t T_{o \dn}\right) \cdots \left(\frac{J}{2} V_{\up \dn} \right) \cdots\right]_{n} \\
    &= \sum_{n=0}^{\infty} \frac{\beta^{n}}{n !} \opr{Tr}\left[\sum (-1)^{N_{ \dn}^h} \cdots\left(t T_{o \up}\right) \cdots \left(\frac{J}{2} T_{\up \dn}\right) \cdots\left(t T_{o \dn}\right) \cdots \left(\frac{J}{2} V_{\up \dn}\right) \cdots\right]_{n}.
\end{aligned}
\end{equation}
The formal notation $[\sum \cdots]_n$ indicates the summation over all length-$n$ process combination of $T_{o\up}$, $T_{o\dn}$, $T_{\up \dn}$, and $V_{\up \dn}$. Here $N_{\dn}^h$ denotes the total number of NN exchanges between holes and $\dn$ spins $T_{o\dn}$. We remark that the remaining part in Eq.\,\eqref{eq:expand_tj} except $(-1)^{N_{ \dn}^h}$ is always non-negative because each matrix element in $T_{o\up}$, $T_{o\dn}$, $T_{\up \dn}$, and $V_{\up \dn}$ is non-negative.

By further expanding $T_{o\up}$, $T_{o\dn}$, $T_{\up \dn}$, and $V_{\up \dn}$ into elementary local terms and writing the trace as the sum of expectations over the complete Fock basis (or equivalently by inserting complete bases between each ``time slice''), the partition function can be expressed by a huge summation of real numbers with each number indexed by a discrete evolution of hole-spin configurations, where in each step, one of the four events $T_{o\up}$, $T_{o\dn}$, $T_{\up \dn}$, and $V_{\up \dn}$ occurs at a certain link $\avg{ij}$. Note that due to the trace operation, the initial and final hole-spin configurations in the evolution should be the same, so in each possible evolution path $C$, the motion of holes and spins must form closed loops. Namely, the partition function can be expressed by a summation that runs over all possible closed evolution paths $C$, i.e.,
\begin{equation}\label{eq:partition_sign_tj}
    Z_{t\text{-}J} = \sum_{C} \tau_{C} W_{t\text{-}J} [C],
\end{equation}
where
\begin{equation}\label{eq:sign_tj}
    \tau_{C} = (-1)^{N_{\dn}^h},
\end{equation}
is the sign structure of the bosonic $t$-$J$ model and $W_{t\text{-}J}[C]\geq 0$ denotes a non-negative weight corresponding to the evolution path $C$ with its explicit form omitted. Note that the summation over evolution length $n$ in Eq.\,\eqref{eq:expand_tj} has been included in the summation over evolution path $C$ in Eq.\,\eqref{eq:partition_sign_tj}. Importantly, we find that the sign structure $(-1)^{N_{\dn}^h}$ can be interpreted as mutual statistics of holes and $\dn$ spins. Namely, an exchange of a hole and a $\dn$-spin will give rise to a minus sign during the evolution determined by the bosonic $t$-$J$ model.

Here we elaborate on why we focus on the sign structure after applying the Marshall transformation. Of course, one can write down the sign structure without applying the Marshall transformation, which is just $(-1)^{N_{\up\dn}}$ with $N_{\up\dn}$ denoting the number of NN exchanges between $\up$ spins and $\dn$ spins $T_{\up\dn}$. This can be seen as a type of ``fermion'' statistics between $\up$ spins and $\dn$ spins. It is equivalent to Eq.\,\eqref{eq:sign_tj} because phase transformations cannot change the overall sign of an evolution loop. However, at half-filling where the model is reduced to the Heisenberg model in which the Marshall transformation is first introduced, the sign $(-1)^{N_{\dn}^h}$ naturally disappears due to the absence of holes while $(-1)^{N_{\up\dn}}$ is still present, in spite that eventually $(-1)^{N_{\up\dn}}$ will cancel with each other due to the periodic boundary condition along the time direction. This implies that at low doping levels, it is natural and beneficial to exploit the sign structure $(-1)^{N_{\dn}^h}$ which treats doped holes as an essential object. Physically, at half-filling, treating spinons as bosons can naturally and correctly give rise to AFM long-range order via the Schwinger boson mean-field theory. This suggests that one shall continuously treat spinons as bosons at least when slightly deviating from half-filling, which also justifies the fundamental role of the sign structure $(-1)^{N_{\dn}^h}$ at low doping levels.

\subsection{Bosonic $\sigma t$-$J$ model}
For the bosonic $\sigma t$-$J$ model in Eq.\,\eqref{eq:bosonic_sigma_tj}, an extra spin-dependent sign $\sigma$ is attached to the NN hopping term of the bosonic $t$-$J$ model. As a result, after the Marshall transformation $U_{\text{Ms}}$, this spin-dependent sign can exactly cancel the negative sign in front of $T_{o \dn}$ that arises from the transformation, i.e.,
\begin{equation}\label{eq:sigma_tj_marshall}
    U_{\text{Ms}}^\dagger H_{\sigma t\text{-}J} U_{\text{Ms}} = -t \left(T_{o \up} + T_{o \dn}\right)- \frac{J}{2}\left(T_{\up \dn} + V_{\up \dn}\right).
\end{equation}
Similar to Eq.\,\eqref{eq:expand_tj}, one can perform series expansions for the bosonic $\sigma t\text{-}J$ model and find that everything is the same as the expression of the bosonic $t$-$J$ model except that the sign structure $(-1)^{N_{\dn}^h}$ disappears (or say simply becomes $1$). The non-negative weight $W[C]$ corresponding to each evolution path $C$ is exactly the same as that of the bosonic $t$-$J$ model. That is to say, the bosonic $\sigma t\text{-}J$ model is a ``sign-free'' model, i.e., it can be efficiently simulated by quantum Monte Carlo methods without any sign (phase) problem, such as the stochastic series expansion (SSE) method. This can be easily seen from Eq.\,\eqref{eq:sigma_tj_marshall} because each matrix element of $(-U_{\text{Ms}}^\dagger H_{\sigma t\text{-}J} U_{\text{Ms}})$ under the real-space Fock basis is non-negative.

\subsection{Spinful hard-core Bose-Hubbard model}
Similar to the derivation above, it is straightforward to prove that the spinful hard-core Bose-Hubbard model defined in Eq.\,\eqref{eq:spinful_hard_core_bose_hubbard} is also sign-free because the matrix elements of the minus hopping term $(-H_t)$ (off-diagonal) are all non-negative, i.e., equal $t$ or $0$, and the matrix elements of the minus on-site Hubbard term $(-H_U)$ (diagonal) can also be shifted to non-negative values by adding a constant to each local term to make $(-U)$ or $0$ shifted to $0$ or $U$. Note that one can always add a constant to the Hamiltonian without changing the expectation values of any other observables, and add it back when calculating the energy. The sign-free property of the spinful hard-core Bose-Hubbard model is consistent with the fact that its effective model in the limit of $U\gg t$, i.e., the bosonic $\sigma t$-$J$ model, is also sign-free. In other word, the sign-problem-free property of the bosonic $\sigma t$-$J$ model is hence a natural result from the sign-problem-free Bose-Hubbard model.

\subsection{Gauge invariance of the frustration phase}

Here, we further clarify the basis dependency of the sign problem. Whether the sign problem is basis-dependent relies on the allowed basis transformations. 

Firstly, if the basis transformation is allowed to be an arbitrary unitary transformation without any constraint, the sign problem is indeed trivially basis-dependent. For example, under the eigen-basis of the Hamiltonian, all weights in the partition function can be positive, and hence there is no apparent sign problem (though the diagonalization unitary is impractical to calculate for large systems). However, this does not imply the sign problem is just a technical challenge. Instead, the sign problem may encode profound physical insights into the model. For example, numerical calculations~\cite{Iglovikov2015} reveal that the sign problem of the Hubbard model becomes the most severe precisely near the optimal doping for cuprate high-temperature superconductors. As another example, the quantum frustration in AFM spin models on non-bipartite lattices such as Kagome lattices signifies potential quantum-spin-liquid ground states~\cite{Liao2017}.

\begin{figure}
    \centering
    \includegraphics[width=0.38\linewidth]{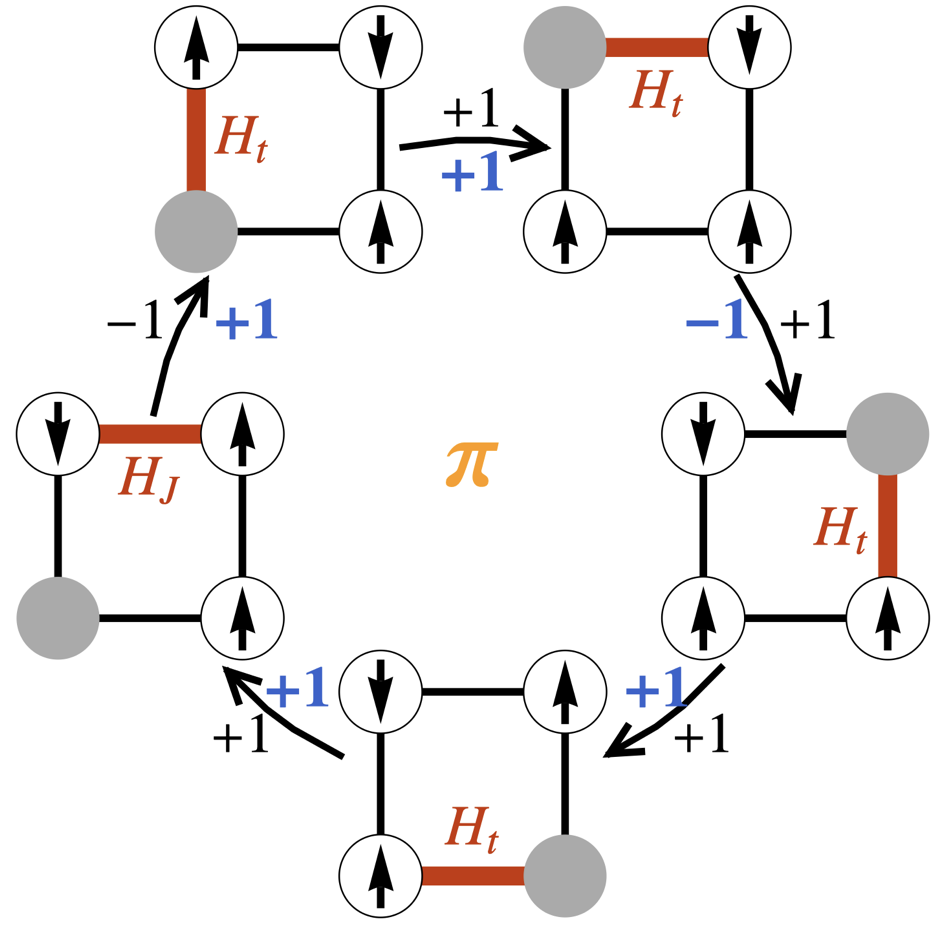}
    \caption{Illustration of the ``gauge invariance'' in an evolution loop for the $t$-$J$ model on a four-site lattice. The grey circles represent holes, and the arrows represent spins. The Hamiltonian term corresponding to the coming evolution process is labeled by thick red lines. The black (blue) $\pm 1$ marks the sign of the matrix elements of $(-H)$ before (after) the Marshall sign basis transformation. Both evolution loops result in $\Theta_C = \pi$.}
    \label{figR:4site_5step_evolution}
\end{figure}

Secondly, if the basis transformation is restricted to phase transformations, e.g., the Marshall sign transformation, the sign problem is basis-independent because the basis vectors always appear in pairs with their conjugates in each evolution path of the partition function. Specifically, for an $n$-step evolution loop $C$, the corresponding phase $\Theta_C$ in the phase factor $\tau_C = e^{i\Theta_C}$ can be expressed as
\begin{equation}
  \Theta_C = \sum_{k=0}^{n-1} \opr{Im} \ln \left\langle \alpha_{k+1} \right| (-H) \left| \alpha_{k} \right\rangle,
\end{equation}
where the temporal periodic boundary condition is imposed by $\left| \alpha_0 \right\rangle = \left| \alpha_n \right\rangle$.
One can set all diagonal elements positive by default since they can be arbitrarily shifted by adding a constant to the Hamiltonian. If a basis transformation changes $\left| \alpha_{k} \right\rangle \rightarrow e^{i\theta} \left| \alpha_{k} \right\rangle$, the resulting phase change in $\Theta_C$ will cancel with its conjugate, i.e., 
\begin{equation}
    e^{i\theta}\left| \alpha_{k} \right\rangle \left\langle \alpha_{k} \right| e^{-i\theta} = \left| \alpha_{k} \right\rangle \left\langle \alpha_{k}\right|.
\end{equation}
That is to say, although such a basis transformation may alter individual matrix elements of the Hamiltonian, the overall phase $\Theta_C$ remains unchanged for a closed evolution loop. Fig.\,\ref{figR:4site_5step_evolution} depicts a specific example to further clarify the above statement. Under the original Ising basis, each spin flip step carries a negative sign as
\begin{equation}
    \left\langle\up_i \dn_j\right| \left(-S_i^{+} S_j^{-}\right) \left| \dn_i \up_j\right\rangle < 0.
\end{equation}
In the presence of hole hopping, an odd number of spin flip steps can occur in an evolution loop, as marked by the black $\pm 1$ in Fig.\,\ref{figR:4site_5step_evolution}, and the accumulated phase in this example is $\Theta_C=\pi$. On the other hand, after performing the Marshall sign basis transformation $U_{\text{Ms}}$, the matrix elements of the spin flip steps become positive, i.e.,
\begin{equation}
    U_{\text{Ms}}^\dagger \left(-S_i^{+} S_j^{-}\right) U_{\text{Ms}} = S_i^{+} S_j^{-}.
\end{equation}
Instead, the exchange between a hole and a down spin results in a negative sign, as marked by the blue $\pm 1$ in Fig.\,\ref{figR:4site_5step_evolution}, and the accumulated phase is again $\Theta_C=\pi$. This concrete example demonstrates that although the two matrix representations differ by a phase transformation, the accumulated frustration phase after completing an evolution loop remains identical.

In summary, the quantum frustration/sign problem is not only a computational challenge but also encodes the intrinsic physical natures of corresponding quantum many-body systems, and it is basis-independent if the basis transformation is restricted to phase transformations. In this bosonic $t$-$J$ model, it appears as a ``gauge-invariant Berry phase'' in the physical real-space local hole-spin basis, dictating the quantum interference effect among individual holes and spins that entangles them into a non-trivial many-body ground state.

\subsection{Relation with the continuous Berry phase}
In the main text, we mention that the frustration phase can be seen as a generalized discrete version of the Berry phase action in the continuous path integral formalism of conventional models, in the sense that it is an ``adiabatic'' phase independent of the evolution speed, in contrast to the dynamical phase stemming from the exponentiated imaginary unit $i$ in the real-time evolution. Here, we provide another perspective to justify the close relationship between the frustration phase and the conventional continuous Berry phase.

For a parametrized quantum state $\ket{\alpha(R)}$, the conventional continuous Berry phase takes the form of
\begin{equation}
    \gamma_C = i \oint_{C} \braoprket{\alpha(R)}{\nabla_R}{\alpha(R)} \cdot d R,
\end{equation}
where $C$ represents a closed path in the parameter space. If one replaces the continuous path with a discrete series of quantum states $\{ \ket{\alpha_0}, \ket{\alpha_1}, \cdots, \ket{\alpha_{n-1}}\}$, the Berry phase becomes
\begin{equation}
    \gamma_C = \opr{Im}\ln \braket{\alpha_{n-1}}{\alpha_{n-2}} \cdots \braket{\alpha_{2}}{\alpha_{1}} \braket{\alpha_{1}}{\alpha_{0}}.
\end{equation}
That is to say, $\gamma_C$ counts the phase differences between each pair of adjacent states. On the other hand, as mentioned in the main text, the frustration phase $\Theta_C$ in $\tau_C=e^{i\Theta_C}$ takes the form of
\begin{equation}
    \Theta_C = \opr{Im}\ln \braoprket{\alpha_{n-1}}{(-H)}{\alpha_{n-2}} \cdots \braoprket{\alpha_{2}}{(-H)}{\alpha_{1}} \braoprket{\alpha_{1}}{(-H)}{\alpha_{0}}.
\end{equation}
Hence, the frustration phase can be seen exactly as a Berry phase if one absorbs the operator $(-H)$ into the original series of quantum states. We emphasize that the discreteness stems from the expansion scheme where a discrete complete basis is inserted. It is possible to realize a continuous form of the frustration phase if some appropriate continuous complete basis is applied like in the spin coherent state path integral.

\section{Analysis on the phase of pairing order parameter}

\subsection{Pairing in bosonic systems}
In this section, we discuss the basic properties of possible pairing states allowed in bosonic models. Here, we first remark that a singlet pair of spinful bosons is directed, i.e.,
\begin{equation}
\begin{aligned}
    \Delta_{ji} &= \frac{1}{\sqrt{2}}\left( B_{j\up} B_{i\dn}-B_{j\dn} B_{i\up} \right) = \frac{1}{\sqrt{2}} \left( B_{i\dn} B_{j\up} - B_{i\up} B_{j\dn} \right) \\
    & = \frac{1}{\sqrt{2}} \left( - B_{i\up} B_{j\dn} + B_{i\dn} B_{j\up} \right) = -\Delta_{ij},
\end{aligned}
\end{equation}
distinct from a singlet pair of fermions, which is undirected, i.e.,
\begin{equation}
\begin{aligned}
    \Delta_{ji} &= \frac{1}{\sqrt{2}}\left( c_{j\up} c_{i\dn}-c_{j\dn} c_{i\up} \right) = - \frac{1}{\sqrt{2}} \left( c_{i\dn} c_{j\up}- c_{i\up} c_{j\dn}\right) \\
    &= - \frac{1}{\sqrt{2}} \left( - c_{i\up} c_{j\dn} + c_{i\dn} c_{j\up}\right) = \Delta_{ij},
\end{aligned}
\end{equation}
where the negative signs from the permutation of singlets and the permutation of fermions cancel out. Please note the distinction between the notations $\Delta_{i\alpha}$ and $\Delta_{ij}$. $\Delta_{i\alpha}$ can be seen as $\Delta_{ij}$ with $j=i+\alpha$ for short.

The generic two-body wavefunction of a ``bosonic Cooper pair'' with the decoupled spatial and spin sectors can be expressed as
\begin{equation}\label{eq:twobody_wf}
    \psi\left(R, r ; \sigma_1, \sigma_2\right)=\phi(R, r) \chi\left(\sigma_1, \sigma_2\right),
\end{equation}
where $\phi(R, r)$ and $\chi\left(\sigma_1, \sigma_2\right)$ represents the spatial and spin sectors. $R$ denotes the center-of-mass position and $r$ denotes the relative position. It is important to note that the exchange of two bosons, i.e., $r \rightarrow -r$ and $\sigma_1 \rightarrow \sigma_2$, will not result in an additional minus sign, which differs from conventional fermionic systems. Therefore, if the spin sector is in an anti-symmetric singlet state $(\left|\up\dn\right\rangle - \left|\dn \up\right\rangle)/\sqrt{2}$, the corresponding spatial sector should exhibit odd parity $\phi(R, r) = -\phi(R, -r)$ leading to a $p$-wave symmetry, etc.

\begin{figure}
    \centering
    \includegraphics[width=0.86\linewidth]{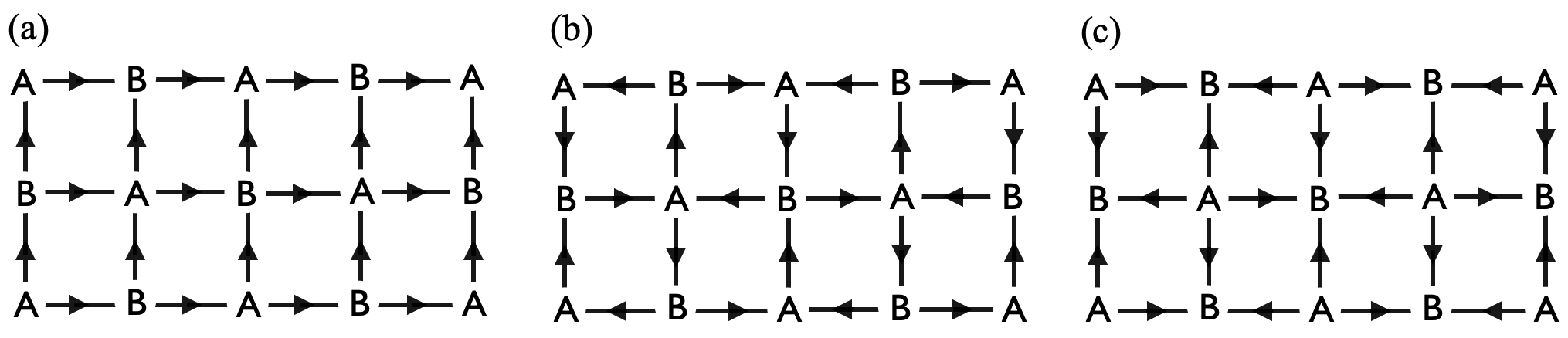}
    \caption{Three possible nearest-neighbor (NN) spin-singlet pairing sign patterns in spin-$1/2$ hard-core bosonic systems. The labels A and B denote the two sublattices of a square lattice. The arrow direction indicates the sign of the NN singlet pairing order parameter $\avg{\Delta_{i\alpha}}$. If $\alpha$ takes the same direction as the arrow, $\avg{\Delta_{i\alpha}}$ is positive, and vice versa.}
    \label{fig:sign_NN_arrow}
\end{figure}
% \hk{Other PDW possibilities.}

For a spatially uniform spin-singlet pairing order, as shown in Fig.\,\ref{fig:sign_NN_arrow}(a), it will naturally exhibit a $p$-wave rotational symmetry, i.e., a $\pi$ rotation yields a negative sign, because the bosonic spin-singlet is directed, i.e., $\Delta_{i\alpha} = -\Delta_{i+\alpha,-\alpha}$ with $\Delta_{i\alpha} = \sum_{\sigma}\sigma B_{i,\sigma} B_{i+\alpha,-\sigma} / \sqrt{2}$, in contrast to the fermionic case. Then we consider a spatially modulated singlet pairing order with a $(\pi,\pi)$ wavevector, i.e., it has staggered oscillations in both spatial directions. Figs.\,\ref{fig:sign_NN_arrow}(b) and (c) show two different possible nearest-neighbor (NN) spin-singlet pairing patterns under this condition, which can be characterized by the sign factors $(-1)^{i+\alpha_x}$ and $(-1)^{i+\alpha}$, respectively. Here the notation $(-1)^{i}$ takes $(-1)$ if $i\in \mathrm{A}$ and takes $(+1)$ if $i\in \mathrm{B}$, which can also be written as $(-1)^{x+y}$ with $i=(x,y)$ if the origin point belongs to sublattice B. $\alpha$ is the link vector in $\Delta_{i\alpha}$ and $\alpha_x$ represents its component along the $\hat{x}$ direction. These two NN pairing sign patterns exhibit a ``local'' $d$-wave symmetry and a ``local'' $s$-wave symmetry, respectively, in the sense that the four NN links around a single site follow the corresponding sign change rule. The DMRG results in this paper show that the actual pattern of the AFM+PDW phase in the bosonic $t$-$J$ model is Fig.\,\ref{fig:sign_NN_arrow}(c).

\begin{figure}
    \centering
    \includegraphics[width=1\linewidth]{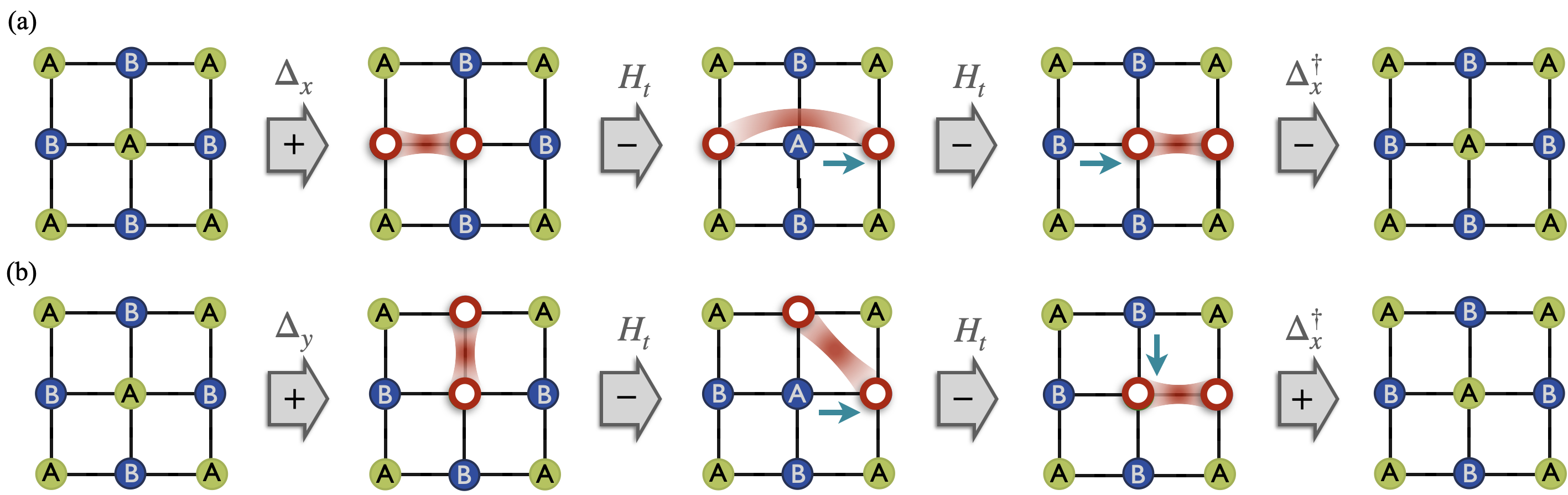}
    \caption{Schematic illustration of two representative imaginary-time evolution loops for the singlet pair-pair correlation under the Marshall basis. The labels A and B denote the two sublattices of a square lattice. The green and blue circles represent spins in $\left|\up\right\rangle$ and $\left|\dn\right\rangle$. The red hollow circles represent holes $\ket{0}$. The positive and negative signs in the grey arrow indicate the sign caused by the corresponding process. (a) A representative hopping procedure for a hole pair in the correlation function $\avg{\Delta_{i\hat{x}} \Delta_{i+\hat{x},\hat{x}}^\dagger }$. The negative sign created by $H_t$ arises from the matrix elements under the Marshall basis as in Eq.\,\eqref{eq:sign_tj}. (b) A representative hopping procedure for a hole pair in the correlation function $\avg{\Delta_{i\hat{y}} \Delta_{i\hat{x}}^\dagger }$. Here we use $\avg{\Delta_{j\beta} \Delta_{i\alpha}^\dagger }$ instead of $\avg{\Delta_{i\alpha}^\dagger \Delta_{j\beta}}$ just for the simplicity of illustration.}
    \label{fig:path_instances}
\end{figure}

\subsection{Pairing phase from the Marshall sign and constructive interference}
Here we provide a detailed argument to explain the origin of the $(\pi,\pi)$-PDW observed in DMRG simulations as depicted in Fig.\,\textcolor{darkblue1}{1}(b). Under the Marshall basis defined by Eq.\,\eqref{eq:marshall}, the singlet pair-pair correlation function can be expressed as
\begin{equation}\label{eq:pair_expansion}
\begin{aligned}
    \avg{\Delta_{i\alpha}^\dagger \Delta_{j\beta}} &= \frac{1}{Z} \opr{Tr}\left( e^{-\beta  H } \Delta_{i\alpha}^\dagger \Delta_{j\beta} \right) = \frac{1}{Z} \opr{Tr}\left( e^{-\beta U_{\text{Ms}} H U_{\text{Ms}}^\dagger} U_{\text{Ms}} \Delta_{i\alpha}^\dagger \Delta_{j\beta}  U_{\text{Ms}}^\dagger \right)\\
    &= \frac{1}{Z} \opr{Tr}\left( e^{-\beta U_{\text{Ms}} H U_{\text{Ms}}^\dagger} (-1)^{i +\alpha} \Delta_{i\alpha}^{t \dagger} (-1)^{j+\beta} \Delta_{j\beta}^t \right)\\
    &= (-1)^{(j-i)+(\beta-\alpha)} \frac{1}{Z} \opr{Tr}\left( \Delta_{j\beta}^t e^{-\beta U_{\text{Ms}} H U_{\text{Ms}}^\dagger} \Delta_{i\alpha}^{t \dagger} \right),
\end{aligned}
\end{equation}
where $\Delta_{i\alpha}^{t}=\sum_{\sigma} B_{i\sigma} B_{i+\alpha,\sigma}/\sqrt{2}$ is the $S^z=0$ spin-triplet pair annihilation operator. The sign factors $(-1)^{i+\alpha}$ and $(-1)^{j+\beta}$ emerges due to the Marshall sign transformation $U_{\text{Ms}}$ acting on the operators $\Delta_{i\alpha}^\dagger$ and $\Delta_{j\beta}$.

Similar to the series expansion of the partition function in Eq.\,\eqref{eq:expand_tj}, one can expand the pair-pair correlation function as
\begin{equation}\label{eq:expand_pair-pair}
\begin{aligned}
    \avg{\Delta_{i\alpha}^\dagger \Delta_{j\beta}} = (-1)^{(j-i)+(\beta-\alpha)} \frac{1}{Z} \sum_{n=0}^{\infty}  \frac{\beta^{n}}{n !} \opr{Tr}\left[\sum \Delta_{j\beta}^t \cdots \left(t T_{o \up}\right) \cdots \left(\frac{J}{2} T_{\up \dn}\right) \cdots \left(-t T_{o \dn}\right) \cdots \left(\frac{J}{2} V_{\up \dn} \right) \cdots \Delta_{i\alpha}^{t \dagger} \right]_{n}.
\end{aligned}
\end{equation}
Except the sign factor $(-1)^{(j-i)+(\beta-\alpha)}$ and the exchange of holes and $\dn$ spins $\left(-t T_{o \dn}\right)$, the remaining parts are positive-definite, including $\Delta_{j\beta}^t$ and $\Delta_{i\alpha}^{t \dagger}$. In general, one cannot determine the sign of the huge summation in Eq.\,\eqref{eq:expand_pair-pair} due to the existence of the negative ingredient $\left(-t T_{o \dn}\right)$. However, we know there exists a leading positive contribution from the ``in-tandem'' propagation of hole pairs, which always includes an even number of $\left(-t T_{o \dn}\right)$ resulting in an extensive summation of positive weights (``constructive interference''). If this contribution is dominant over the others (which is very likely as the other parts may cancel each other out due to the phase fluctuation), then the huge summation in Eq.\,\eqref{eq:expand_pair-pair} should be positive. Therefore, the overall sign of the NN singlet pair-pair correlation $\avg{\Delta_{i\alpha}^\dagger \Delta_{j\beta}}$ should be just determined by the factor
\begin{equation}
    (-1)^{(j-i)+(\beta-\alpha)},
\end{equation}
resulting from the Marshall sign. If we fix $i$ and $\alpha$ as a reference link, the relative sign of the correlation function should be determined by $(-1)^{j+\beta}$. Thus, the relative sign of the local pairing order parameter $\avg{\Delta_{i\alpha}}$ should have the form of
\begin{equation}
    (-1)^{i+\alpha}.
\end{equation}
Remember that the notation $(-1)^{i}$ takes $(\mp 1)$ for sublattice A and B respectively. The notation $\alpha$ means a unit vector $\{\pm\hat{x}, \pm\hat{y}\}$ on the square lattice. If $i\in A$, we have $i+\alpha\in B$, and vice versa. That is to say, up to an overall sign, we have $\avg{\Delta_{i\alpha}}<0$ for $i\in A$ and $\avg{\Delta_{i\alpha}}>0$ for $i\in B$. Combined with the fact that the singlet pair of spinful bosons satisfies
\begin{equation}
    \Delta_{i\alpha}=-\Delta_{i+\alpha,-\alpha},
\end{equation}
this is exactly the $(\pi,\pi)$-PDW pattern we observed in the DMRG simulations as well as the one depicted in Fig.\,\ref{fig:sign_NN_arrow}(c).

In summary, the strings of $\mathbb{Z}_2$ phases caused by the hopping of two holes under the Marshall basis can be canceled with each other by forming a tightly bound pair, resulting in constructive interference, so that the only remaining phase of the pairing order parameter comes from the transformed pair operators themselves, i.e., the Marshall sign. We further illustrate two representative evolution paths in Fig.\,\ref{fig:path_instances} as examples for clarity.

In addition, the above argument is based on the assumption that the ground state is not an FM state, i.e., the weights that mainly involve globally polarized spins are not dominant in the summation in Eq.\,\eqref{eq:expand_pair-pair}, where the sign factor $(-1)^{N^h_{\dn}}$ become trivial and no destructive interference occurs in the single-particle propagation. Otherwise, the contribution where holes form pairs would become sub-leading, and no pairing could be observed in the resulting ground state. 

Moreover, the above argument is not based on the specific N\'{e}el spin configuration. Instead, it is nontrivial for most of the components in a fluctuating AFM state that do not contain macroscopic FM domains. In other words, destructive interference of single-particle motion occurs for most of the components in an AFM-ordered state, including those containing defects, compared to a perfect classical N\'{e}el order. Here, a fluctuating AFM state means that the AFM background is not just a simple product state with spins in the N\'{e}el configuration; instead, it is a superposition of many configurations that can slightly deviate from the N\'{e}el configuration.

\section{Parton construction of the bosonic $\sigma t$-$J$ model} 
\subsection{Mean-field scheme}
\begin{figure}[t]
    \centering
    \includegraphics[width=0.55\linewidth]{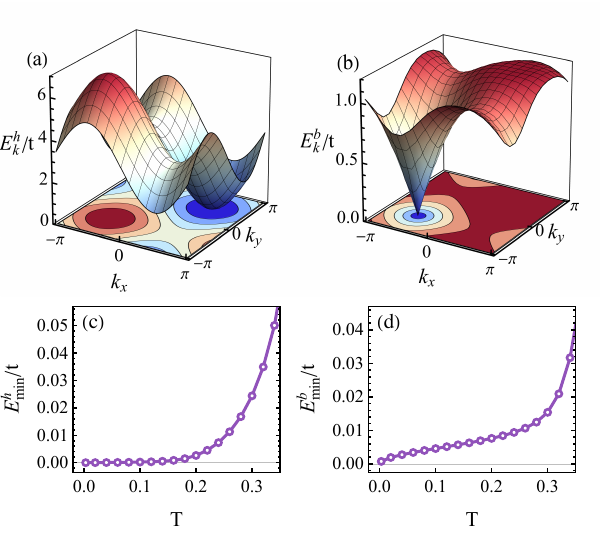}
    \caption{With parameters $J=t/3$ and doping density $\delta=1/12$. Dispersions for holons $E_k^h$ and spinons $E_k^b$ are shown in (a) and (b), respectively. The lowest excitation energy gaps for holons and spinons are shown in (c) and (d), respectively.}
    \label{fig:dis}
\end{figure}

In this section, we present more details of the parton construction approach used to describe the bosonic $\sigma t$-$J$ model. We decompose the original boson operator subject to the no-double-occupancy constraint into
\begin{equation}
    B_i=b_{i \sigma} h_i^{\dagger}
\end{equation}
where $b_{i \sigma}$ denotes the bosonic spinon with spin $\sigma$ and $h_i^{\dagger}$ denotes the bosonic holon. Then, the bosonic $\sigma t$-$J$ model in Eq.\,\eqref{eq:bosonic_sigma_tj} can be expressed as follows
\begin{eqnarray}\label{eq:HstJ}
    H_{\sigma t \text{-}J}=t \sum_{\langle i j\rangle, \sigma} \sigma \hat{\chi}_{i j, \sigma} \hat{\kappa}_{i j}^{\dagger}+\text{H.c.}-\frac{J}{2} \sum_{i j} \hat{\Delta}_{i j}^{\dagger} \hat{\Delta}_{ij},
\end{eqnarray}
where 
\begin{equation}
    \hat{\kappa}_{ij} = h_{i}^{\dagger} h_j,\quad
    \hat{\chi}_{ij, \sigma} = b_{i, \alpha}^{\dagger} b_{j, \sigma},\quad
    \hat{\Delta}_{ij}=\sum_\sigma \sigma b_{i, \sigma} b_{j,-\sigma},
\end{equation}
with the relation 
\begin{equation}
    \hat{\kappa}_{ji}= \hat{\kappa}_{i j}^{\dagger}, \quad \hat{\chi}_{j i, \sigma}= \hat{\chi}_{i j, \sigma}^{\dagger}, \quad \hat{\Delta}_{ji}=-\hat{\Delta}_{ij}.
\end{equation}
Based on the mean-field ansatz $\hat{\kappa}_{ij} = i\kappa$, $\hat{\chi}_{ij,\sigma} = i\sigma \chi$, and $\hat{\Delta}_{ij} = i\Delta$, the Hamiltonian~\eqref{eq:HstJ} at the mean-field level can be decomposed into the holon part $H_h$ and the spinon part $H_b$, with the form
\begin{eqnarray}\label{Hf}
    \begin{aligned}
    H_h =&  -4 t \chi\sum_k \mathcal{E}_k h_k^{\dagger}h_k -\mu \sum_k\left(h_k^{\dagger} h_k-\delta\right),
    \end{aligned}
\end{eqnarray}
and
\begin{eqnarray}\label{eq:Hb}
\begin{aligned}
    H_b =&-2t\kappa \sum_{k, \sigma} \sigma   \mathcal{E}_k b_{k, \sigma}^{\dagger} b_{k, \sigma} + J \Delta \sum_k \mathcal{E}_k \left( b_{-k, \dn} b_{k, \up} + \text{H.c.}\right) \\
    &+J \Delta^2 N -8 t N \chi \kappa+ \lambda \sum_{k}\left(\sum_{\sigma} b_{k, \sigma}^{\dagger} b_{k, \sigma} -1+\delta\right),
\end{aligned}
\end{eqnarray}
where $\mathcal{E}_k = \sin k_x + \sin k_y$, and $\kappa$, $\chi$, and $\Delta$ are all chosen to be real numbers. Here, $\mu$ and $\lambda$ are the chemical potentials controlling the holon number $\delta$ and spinon number $1-\delta$, respectively. From Eq.\,\eqref{Hf}, the dispersion for the holon is given by
\begin{equation}\label{eq:Ehk}
    E^h_k=-4 t \chi \mathcal{E}_k-\mu.
\end{equation}
Similarly, from Eq.\,\eqref{eq:Hb}, by introducing the following Bogoliubov transformation, we obtain the spinon dispersion
%\begin{equation}
%    H_b=\sum_{k,\sigma} E^b_k \gamma_{k,\sigma}^\dagger \gamma_{k,\sigma}
%\end{equation}
%with the dispersion 
\begin{equation}\label{eq:Ebk}
    E^b_k = \sqrt{\lambda_k^2-\Delta_k^2},
\end{equation}
where $\lambda_k=\lambda-2 t \kappa \mathcal{E}_k$ and $\Delta_k=J \Delta \mathcal{E}_k$.
%for the Bogoliubov spinon $\gamma_{k,\up}= u_k b_{k,\up}+ v_k b_{-k,\dn}^\dagger$ and $\gamma_{k,\dn}= u_k b_{-k,\dn}+ v_k b_{k,\up}^\dagger$.
The values of the mean-field ansatz and chemical potentials can be determined via self-consistent calculations as follows:
\begin{eqnarray}
\begin{aligned}
\frac{\partial F}{\partial \lambda} & =0 \Longrightarrow \frac{1}{N} \sum_k \frac{\left(\lambda-2 t \kappa \mathcal{E}_k\right)}{E_k^b} \opr{coth}\left(\frac{1}{2} \beta E_k^b\right)+(-2+\delta)=0, \\
\frac{\partial F}{\partial \mu} & =0 \Longrightarrow \frac{1}{N} \sum_k \frac{1}{e^{\beta E_k^h}-1}-\delta=0, \\
\frac{\partial F}{\partial \kappa} & =0 \Longrightarrow \frac{1}{N} \sum_k \frac{\mathcal{E}_k\left(\lambda-2 t \kappa \mathcal{E}_k\right)}{E_k^b} \opr{coth}\left(\frac{1}{2} \beta E_k^b\right)+4 \chi=0, \\
\frac{\partial F}{\partial \chi} & =0 \Longrightarrow \frac{1}{N} \sum_k \frac{\mathcal{E}_k}{e^{\beta E_k^h}-1}+2 \kappa=0, \\
\frac{\partial F}{\partial \Delta} & =0 \Longrightarrow 2-\frac{1}{N} \sum_k \frac{J \mathcal{E}_k^2}{E_k^b} \opr{coth}\left(\frac{1}{2} \beta E_k^b\right)=0.
\end{aligned}    
\end{eqnarray}
Based on the calculated mean-field parameters at $T \rightarrow 0$, the holon dispersion Eq.\,\eqref{eq:Ehk} is shown in Fig.\,\ref{fig:dis}(a), indicating that the band bottom is located at the momentum $(\pi/2, \pi/2)$. Similarly, the spinon dispersion Eq.\,\eqref{eq:Ebk} is shown in Fig.\,\ref{fig:dis}(b), exhibiting low-lying modes near the momentum $(-\pi/2, -\pi/2)$. The temperature dependence of the excitation gap for holons $E_{\text{min}}^h$ and for spinons $E_{\text{min}}^b$ are shown in Figs.\,\ref{fig:dis}(c) and (d), respectively. These figures show that both spinons and holons tend to be gapless in the zero temperature limit, indicating that both tend to undergo Bose-Einstein condensation as the temperature approaches zero. 

In addition, spinons with opposite spin directions occupy band minima at opposite momenta: the $\uparrow$-spinon is located at $(-\pi/2, -\pi/2)$ while the $\downarrow$-spinon is at $(\pi/2, \pi/2)$, as directly obtained from the corresponding Green's functions in Eqs.\,\eqref{Gb1}-\eqref{Gb2}. When combined with the condensation of holons at $(\pi/2, \pi/2)$, this results in actual bosons with $\uparrow$-spin condensing at $(0,0)$ momentum, while $\downarrow$-spin condenses at $(\pi,\pi)$ momentum. This finding is consistent with the $\sigma^r$ factor observed in the DMRG results [see Fig.\,\textcolor{darkblue1}{4}(a)].

\subsection{Spin response}
According to Eq.\,\eqref{eq:Hb}, the normal and anomalous Green's functions of spinons are given by:
\begin{eqnarray}
    & G_{\up \up}(k, \omega)=-\left\langle b_{k \up}(\omega) b_{k \up}^{\dagger}(\omega)\right\rangle=\frac{\lambda_k+\omega}{\omega^2-(E_k^b)^2}, \label{Gb1}\\
    &G_{\dn \dn}(k, \omega)=-\left\langle b_{k \dn}(\omega) b_{k \dn}^{\dagger}(\omega)\right\rangle=\frac{\lambda_{-k}+\omega}{\omega^2-(E_{-k}^b)^2}, \label{Gb2}\\
    & G_{* \dn \dn}(-k,-\omega)=-\left\langle b_{-k \dn}^{\dagger}(-\omega) b_{-k \dn}(-\omega)\right\rangle=\frac{\lambda_k-\omega}{\omega^2-(E_k^b)^2}, \\
    & G_{* \up \up}(-k,-\omega)=-\left\langle b_{-k \up}^{\dagger}(-\omega) b_{-k \up}(\omega)\right\rangle=\frac{\lambda_{-k}-\omega}{\omega^2-(E_{-k}^b)^2}, \\
    & F_{\up \dn}(k, \omega)=-\left\langle b_{k \up}(\omega) b_{-k \dn}(-\omega)\right\rangle=-\frac{\Delta_k}{\omega^2-(E_k^b)^2}, \\
    & F_{\dn \up}(k, \omega)=-\left\langle b_{k \dn}(\omega) b_{-k \up}(-\omega)\right\rangle=\frac{\Delta_k}{\omega^2-(E_{-k}^b)^2}, \\
    & F_{* \dn \up}(-k,-\omega)=-\left\langle b_{-k \dn}^{\dagger}(-\omega) b_{k \up}^{\dagger}(\omega)\right\rangle=-\frac{\Delta_k}{\omega^2-(E_k^b)^2}, \\
    & F_{* \up \dn}(-k,-\omega)=-\left\langle b_{-k \up}^{\dagger}(-\omega) b_{k \dn}^{\dagger}(\omega)\right\rangle=\frac{\Delta_k}{\omega^2-(E_{-k}^b)^2}.
\end{eqnarray}
Then, by the relation $S_{i}^{z}=\frac{1}{2} \sum_{\sigma} \sigma b_{i \sigma}^{\dagger}b_{i \sigma}$, the Matsubara spin-spin correlation function along the $z$-spin direction can be expressed as
\begin{eqnarray}
\begin{aligned}
    \chi^{zz}(i, j, \tau) &\equiv\left\langle\hat{T} S_j^z(\tau) S_i^z(0)\right\rangle_0=\frac{1}{4} \sum_{\sigma, \sigma^{\prime}} \sigma \sigma^{\prime}\left\langle\hat{T} b_{j, \sigma}^{\dagger}(\tau) b_{j, \sigma}(\tau) b_{i, \sigma^{\prime}}^{\dagger}(0) b_{i, \sigma^{\prime}}(0)\right\rangle_0\\
    &=\frac{1}{4}\left[G_{* \up \up}(r, \tau) G_{\up \up}(r, \tau)+G_{* \dn \dn}(r, \tau) G_{\dn \dn}(r, \tau)-F_{* \dn \up}(r, \tau) F_{\dn \up}(r, \tau)-F_{* \up \dn}(r, \tau) F_{\up \dn}(r, \tau)\right],
\end{aligned}
\end{eqnarray}
where $\langle \cdot \rangle_0$ denotes the expectation value under the mean-field state, and Wick's theorem is applied in the last line. Then, the spin correlation function in momentum and frequency space at $T=0$ can be expressed as
\begin{eqnarray}\label{eq:chizz}
\begin{aligned}
    \chi^{zz}\left(\boldsymbol{k}, i \omega_n\right)&=\int_0^\beta d \tau \sum_i e^{-i k \cdot r} e^{i \omega_n \tau} \chi^{zz}\left(r, \tau\right)\\
    =&-\frac{1}{16 N} \sum_q \frac{\left(E^b_q E^b_{k+q}-\lambda_q \lambda_{k+q}\right) + i\omega_n \left(-E^b_{k+q} \lambda_q+E^b_q \lambda_{k+q}\right) /\left(E^b_q+E^b_{k+q}\right) + \Delta_q \Delta_{k+q}}{E^b_q E^b_{k+q}}\\
    &\times\left(-\frac{1}{i\omega_n-E^b_q-E^b_{k+q}}+\frac{1}{i\omega_n+E_q^b+E^b_{k+q}}\right) + (k\leftrightarrow -k).
\end{aligned}
\end{eqnarray}
Similarly, by using the relation $S^{\pm}_i = \frac{1}{2}\left(S_i^x \pm i S_i^y\right)$, the Matsubara spin-spin correlation function along the $x$-$y$ plane can be expressed as
\begin{eqnarray}
\begin{aligned}
    \chi^{+-}(i, j, \tau) &\equiv\left\langle\hat{T} S_j^{+}(\tau) S_i^{-}(0)\right\rangle=\frac{1}{4}\left\langle\hat{T} b_{j, \up}^{\dagger}(\tau) b_{j, \dn}(\tau) b_{i, \dn}^{\dagger}(0) b_{i, \up}(0)\right\rangle_0\\
    & = \frac{1}{4} \left[G_{* \up \up}(r, \tau) G_{\dn \dn}(r, \tau)+F_{* \up \dn}(r, \tau) F_{\dn \up}(r, \tau)\right],
\end{aligned}
\end{eqnarray}
with the corresponding form in momentum and frequency space
\begin{eqnarray}\label{eq:chi+-}
\begin{aligned}
    &\chi^{+-}\left(\boldsymbol{k}, i \omega_n\right)=\int_0^\beta d \tau \sum_i e^{-i k \cdot r} e^{i \omega_n \tau} \chi^{+-}\left(r, \tau\right)\\
    =&-\frac{1}{N} \frac{1}{16}\sum_q \frac{\left(E^b_q E^b_{-k-q}-\lambda_q \lambda_{-k-q}\right)+i\omega_n \left(-E^b_{-k-q} \lambda_q+E^b_q \lambda_{-k-q}\right)/\left(E^b_q+E^b_{-k-q}\right) +\Delta_q \Delta_{-k-q}}{E^b_q E^b_{-k-q}}\\
    &\times\left(-\frac{1}{i\omega_n-E^b_q-E^b_{-k-q}}+\frac{1}{i\omega_n+E_q^b+E^b_{-k-q}}\right).
\end{aligned}
\end{eqnarray}
As a result, the imaginary part of the dynamic spin susceptibility Eqs.\,\eqref{eq:chi+-} and \eqref{eq:chizz}, after the analytic continuation $i \omega_n \rightarrow \omega + i \Gamma$, is depicted in Figs.\,\textcolor{darkblue1}{4}(c) and (d).

\section{Technical details and additional simulation results}
In this section, we offer technical information on the density matrix renormalization group (DMRG) simulations mentioned in the main text and additional numerical results to support our conclusions further.

\subsection{Numerical settings}
Extensive DMRG studies in the past few years of the Fermi-Hubbard and fermionic $t$-$J$ models have accumulated valuable experience and techniques on how to determine the ground-state properties of 2D systems~\cite{Jiang2018, Jiang2019, Jiang2020a, Qin2020, Gong2021, Lu2023, Jiang2023, Chen2024, Yang2024, Jiang2021, Chen2023}, providing helpful guidance for our current DMRG simulations of the bosonic $t$-$J$ model. We implement two DMRG schemes independently to obtain reliable results.
\begin{itemize}
    \item Grand canonical ensemble (GCE) calculations where spontaneous U(1) symmetry breaking is allowed, giving rise to non-zero order parameters of pairing or single-particle condensation~\cite{Jiang2021, Chen2023}. A chemical potential term $H_\mu=-\mu\sum_i n_i$ is introduced to control the doping level $\delta$ indirectly. The bond dimensions are kept up to $D=4000$ in the GCE simulations only to measure local properties on the natural ground states~\cite{Chen2023, Tasaki2019} within broken symmetry plateaus~\cite{Jiang2021}. This scheme can be considered as a certain interpolation between the exact ground states and the mean-field solutions, where the local properties can faithfully reflect the tendencies of long-range orders in the true ground state, despite that the long-distance correlations obtained by this scheme may not be accurate. We perform a variety of simulations with different initial states and temporary pinning fields to avoid being stuck in metastable states. The spin-U(1) symmetry is imposed in some of the GCE simulations in the AFM+PDW phase to speed up the computations and control the spin-rotation symmetry-breaking axis. 
    \item Canonical ensemble (CE) calculations where the charge-U(1) and spin-U(1) or SU(2) symmetries are imposed to fix the total particle number (hence the doping level) and total spin $S^z$. The phases of matter are determined by computing the decay behaviors of long-distance correlation functions~\cite{Jiang2018, Jiang2019, Jiang2020a, Qin2020, Gong2021, Lu2023, Jiang2023, Chen2024, Yang2024}. We keep up to $25000$ SU(2) multiplets or $D\approx 90000$ bond dimensions in CE simulations with truncation error $\epsilon \lesssim 5\times 10^{-6}$.
\end{itemize}
Our primary focus is the GCE simulations for width-$6,8$ systems and the CE simulations for width-$4,6,8$ systems, which yield consistent phase diagrams. Other scenarios, like width-$2$ systems, are also simulated in certain cases to support our conclusions further. We exploit either cylindrical boundary conditions (periodic in the $\hat{y}$ direction and open in the $\hat{x}$ direction) or fully open boundary conditions (open in both the $\hat{x}$ and $\hat{y}$ directions).

In particular, we remark that though spontaneous symmetry breaking can automatically occur in the GCE simulations even without any external pinning fields (which is checked by our simulations), sometimes we still apply very small pinning fields (around $10^{-2}$) on the lattice edges or bulks at the initial stage of DMRG optimization (when the bond dimension is relatively small) to induce symmetry breaking and stabilize the symmetry-breaking directions, such as the PDW pinning field $H_{\text{PDW}} = -h_{\text{PDW}} \sum_{i\alpha} (-1)^{x+y} (\Delta_{i\alpha} + \Delta_{i\alpha}^\dagger )$ with $i=(x,y)$, the AFM pinning field $H_{\text{AFM}} = -h_{\text{AFM}} \sum_{i} (-1)^{x+y} S_i^z$, the SF pinning field $H_{\text{SF}}= -h_{\text{SF}} \sum_{i\sigma} (B_{i\sigma} + B_{i\sigma}^\dagger)$, and the FM pinning field $H_{\text{FM}} = -h_{\text{FM}} \sum_{i} S_i^z$. The pinning fields are removed after a few sweeps and no pinning field is applied at the end of optimization. Combined with the consistent results from the CE calculations, one can reliably obtain the general phase diagram relevant to the 2D bosonic $t$-$J$ model.

% The consistency with the CE calculations also confirms the reliability of our GCE calculations.

In addition, we use the notation $\avg{\Delta_{i\alpha}}$ and $\avg{B_{i\sigma}}$ to denote the pairing and single-boson condensation local order parameters in the main text, which are real numbers in the numerical results. In principle, there is a phase arbitrariness for these U(1) symmetry-breaking order parameters. Here we fix this arbitrariness or say ``symmetry-breaking axis'' by code implementation with real-number wavefunctions such that $\avg{\Delta_{i\alpha}}$ and $\avg{B_{i\sigma}}$ are always real numbers. This implementation can also speed up the computations and improve the stability of the numerical results across different model parameters. Thus, the values of $\avg{\Delta_{i\alpha}}$ and $\avg{B_{i\sigma}}$ in the main text should be understood as the magnitudes along the symmetry-breaking axis.

\subsection{Additional simulation results}

\subsubsection{Correlation function analysis for the bosonic $t$-$J$ model}
\begin{figure}
    \centering
    \includegraphics[width=\linewidth]{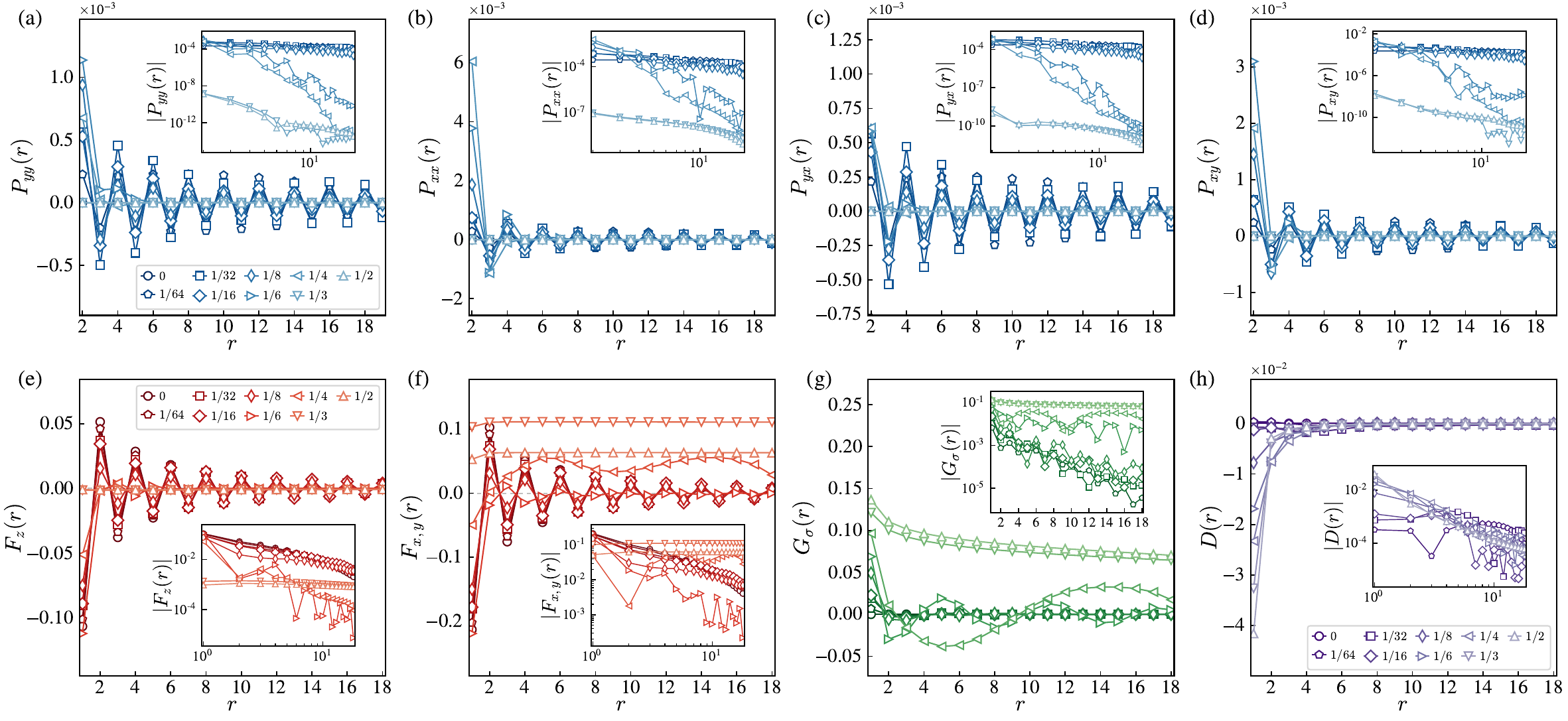}
    \caption{Correlation functions in the ground state of the bosonic $t$-$J$ model on cylinders of size $32\times 4$ or $36\times 4$ with $t/J=3$ and varying doping level $\delta$ labeled by different markers. (a-d) Spin-singlet pair-pair correlation between two NN links oriented in the $\{\hat{x},\hat{y}\}$ directions. (e,f) Spin-spin correlation along the $S^z$ and $S^{x,y}$ directions. (g) Single-particle Green's function. (h) Density-density correlation. The inset shows the logarithmic or semi-logarithmic plots.}
    \label{fig:bosonic_corr_sc_spin_green_nn}
\end{figure}

We first provide a detailed analysis of the correlation function data from the CE calculations for width-$4$ systems. Fig.\,\ref{fig:bosonic_corr_sc_spin_green_nn} shows the correlation functions in the ground state of the bosonic $t$-$J$ model on cylinders of size $32\times 4$ for the doping levels $\delta=0,1/64,1/32,1/16,1/8,1/4,1/2$ and size $36\times 4$ for the doping levels $\delta=1/6,1/3$. The spin-singlet pair-pair correlation is defined as
\begin{equation}
    P_{\alpha\beta}(r) = \langle \Delta_{i_0\alpha}^{\dagger} \Delta_{(i_0+r),\beta}\rangle,    
\end{equation}
where $\Delta_{i\alpha} = \frac{1}{\sqrt{2}}\sum_{\sigma}\sigma B_{i,\sigma} B_{i+\alpha,-\sigma}$ is the spin-singlet pair annihilation operator. $\alpha\in \{\hat{x},\hat{y}\}$ denotes the nearest-neighbor (NN) link. $\hat{x}$ and $\hat{y}$ denotes the two unit basis vectors of the square lattice. $\sigma$ takes $\{+1,-1\}$ for $\{\up,\dn\}$ when serving as a coefficient. $i_0=(x_0,y_0)$ is the reference site and $r$ is a real space vector connecting the created and annihilated particles. By default, we take $x_0=L_x/4$ and set the distance vector $r$ along the $\hat{x}$-direction. The spin-spin correlation is defined as
\begin{equation}
    F_\gamma(r) = \avg{S_{i_0}^\gamma S_{i_0+r}^\gamma},
\end{equation}
where $\gamma\in\{x,y,z\}$ denotes the spin component. The single-particle Green's function and the density-density correlation are defined as
\begin{eqnarray}
    G_{\sigma}(r) &=& \avg{B_{i_0,\sigma}^\dagger B_{(i_0+r),\sigma}}, \\
    D(r) &=& \avg{ n_{i_0} n_{i_0+r} } - \avg{ n_{i_0} } \avg{ n_{i_0+r} }.
\end{eqnarray}
The density average $\avg{n_i}$ is subtracted in $D(r)$ to represent the correlation of density fluctuations. We summarize the corresponding extracted power exponent $K$ and correlation length $\xi$ by fitting the power-law decay $\sim r^{-K}$ and the exponential decay $\sim e^{-r/\xi}$ respectively in Table.\,\ref{tab:btj}.

% \begingroup
\setlength{\tabcolsep}{8pt} % Default value: 6pt
\renewcommand{\arraystretch}{1.3} % Default value: 1

% \usepackage{multirow}
\begin{table}
\begin{tabular}{|c|c|c|c|c|c|c|}
\hline
Phase                              & Doping        & $G_{\sigma}(r)$  & $P_{\alpha\beta}(r)$ & $F_{z}(r)$                         & $F_{xy}(r)$                        & $D(r)$           \\ \hline
AFM$^*$                            & $\delta=0$    & $=0$             & $=0$                   & $K_F\approx 1.3$, $\xi_F\approx 5$ & $K_F\approx 1.3$, $\xi_F\approx 5$ & $=0$             \\ \hline
\multirow{3}{*}{AFM + $\pi$-PDW}   & $\delta=1/32$ & $\xi_G\approx 3$ & $K_P\approx 0.8$       & $K_F\approx 1.1$                   & $K_F\approx 1.1$                   & $K_D\approx 2$ \\ \cline{2-7} 
                                   & $\delta=1/16$ & $\xi_G\approx 3$ & $K_P\approx 1.0$       & $K_F\approx 0.9$                   & $K_F\approx 0.9$                   & $K_D\approx 2$ \\ \cline{2-7} 
                                   & $\delta=1/8$  & $\xi_G\approx 3$ & $K_P\approx 1.0$       & $K_F\approx 0.9$                   & $K_F\approx 0.9$                   & $K_D\approx 2$ \\ \hline
\multirow{2}{*}{Intermediate region} & $\delta=1/6$  & $K_G\approx 0.6$ & $\xi_P\approx 1$       & $K_F\approx 1.8$, $\xi_F\approx 4$ & $K_F\approx 0.7$                   & $K_D\approx 2$ \\ \cline{2-7} 
                                   & $\delta=1/4$  & $K_G\approx 0.2$ & $\xi_P\approx 1$       & $K_F\approx 1.8$, $\xi_F\approx 4$ & $K_F\approx 0$                     & $K_D\approx 2$ \\ \hline
\multirow{2}{*}{FM + SF}           & $\delta=1/3$  & $K_G\approx 0.2$ & $\approx 0$            & $\approx 10^{-3}$                  & $\approx 10^{-1}$                  & $K_D\approx 2$ \\ \cline{2-7} 
                                   & $\delta=1/2$  & $K_G\approx 0.2$ & $\approx 0$            & $\approx 10^{-3}$                  & $\approx 10^{-1}$                  & $K_D\approx 2$ \\ \hline
\end{tabular}
\caption{The estimated fitting exponents for the correlation functions in the ground states of the bosonic $t$-$J$ model at $t/J=3$ on $4$-leg cylinders. The system size is chosen as $32\times 4$ for $\delta=0,1/32,1/16,1/8,1/4,1/2$ and $36\times 4$ for $\delta=1/6,1/3$. $G_\sigma(r)$, $P_{\alpha\beta}(r)$, $F_z(r)$, $F_{xy}(r)$ and $D(r)$ represent the single-particle Green's function, singlet pair-pair correlation, spin-spin correlation along $z$ and $x,y$ direction, and density-density correlation. The singlet pair orientation in $P_{\alpha\beta}(r)$ takes values in $\alpha,\beta \in \{\hat{x}, \hat{y}\}$, the information of which are merged into one column as they have similar decaying behaviors. The spin orientation in $G_\sigma(r)$ takes values in $\sigma \in \{\up, \dn\}$, which are also merged into one column as they have the same decaying behaviors. The notations $K_G$, $K_P$, and $K_F$ stand for power-law fitting $r^{-K}$ while $\xi_G$, $\xi_P$, $\xi_F$ and $\xi_D$ stands for exponential fitting $e^{-r/\xi}$. The cases where two fitting exponents are both shown mean that considerable ambiguity exists in determining fitting schemes due to numerical uncertainty. Here ``AFM$^*$'' means that though a spin gap is observed due to the finite-size effects of even-leg ladders, a true AFM long-range order will be developed in the 2D limit.}
\label{tab:btj}
\end{table}

% \endgroup

\textit{Spin gap in half-filled finite-width systems.---}For narrow ladders with even legs such as the width-$4$ systems investigated here, it is believed that a spin gap exists at half-filling based on numerical experience and the argument that an even number of $S=1/2$ spins on a rung constitutes an integer spin, and the resulting integer spin chain is gapped in a similar spirit of Haldane's conjecture. This is indeed observed by our simulations for width-$4$ systems as shown by the spin-spin correlations at $\delta=0$ in Fig.\,\ref{fig:bosonic_corr_sc_spin_green_nn} and Table.\,\ref{tab:btj}. However, the spin gap, as a finite-width effect, will vanish in the 2D limit where a true long-range AFM order develops, as suggested by the spontaneous AFM patterns observed in our GCE simulations for width-$8$ systems. Upon doping, the AFM spin-spin correlation in the CE calculations is enhanced counter-intuitively to a quasi-long-range behavior. The spontaneous AFM order parameter in the GCE calculations is also enhanced upon doping, as shown in Fig.\,\textcolor{darkblue1}{1}(d). This can be understood by the fact that, in the low doping regime, the doped holes form tightly bound pairs propagating ``in tandem'', which respects the AFM spin background. As a result, the AFM order is not only undisturbed by the doped holes but can even be strengthened by the compatible $(\pi, \pi)$-PDW order.

\textit{Details on the pair-pair correlations.---}The PDW immediately emerges once doping holes, as shown in Fig.\,\ref{fig:bosonic_corr_sc_spin_green_nn} by the pair-pair correlation in the system with only two doped holes, i.e., of size $32\times 4$ and doping $\delta=1/64$. The PDW oscillation is also observed in our CE calculations for width-$6$ systems, albeit only in short-distance pair-pair correlations at the level of the bond dimension $D\leq 10000$ because the increased computational difficulty of width-$6$ systems makes it challenging to accurately calculate long-distance correlations. In the FM+SF phase, the singlet pair-pair correlations almost vanish (around $10^{-9}$), as shown in Table.\,\ref{tab:btj} and the insets of Fig.\,\ref{fig:bosonic_corr_sc_spin_green_nn}. The spin-triplet pair-pair correlations exhibit algebraic decay (not shown here), which yet does not signify true pairing but rather a secondary effect resulting from single-boson condensation.

\begin{figure}
    \centering
    \includegraphics[width=0.8\linewidth]{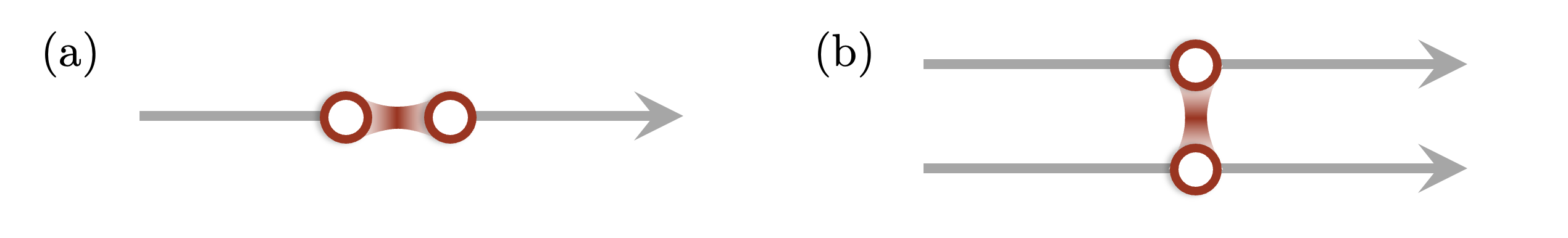}
    \caption{Illustration of two different modes of hole pair motion: (a) ``in-tandem'' vs (b) ``side-by-side''. The moving direction is always parallel to the bond direction of the hole pair in (a) while perpendicular to the bond direction in (b).}
    \label{fig:tandem_vs_side}
\end{figure}

\begin{figure}
    \centering
    \includegraphics[width=\linewidth]{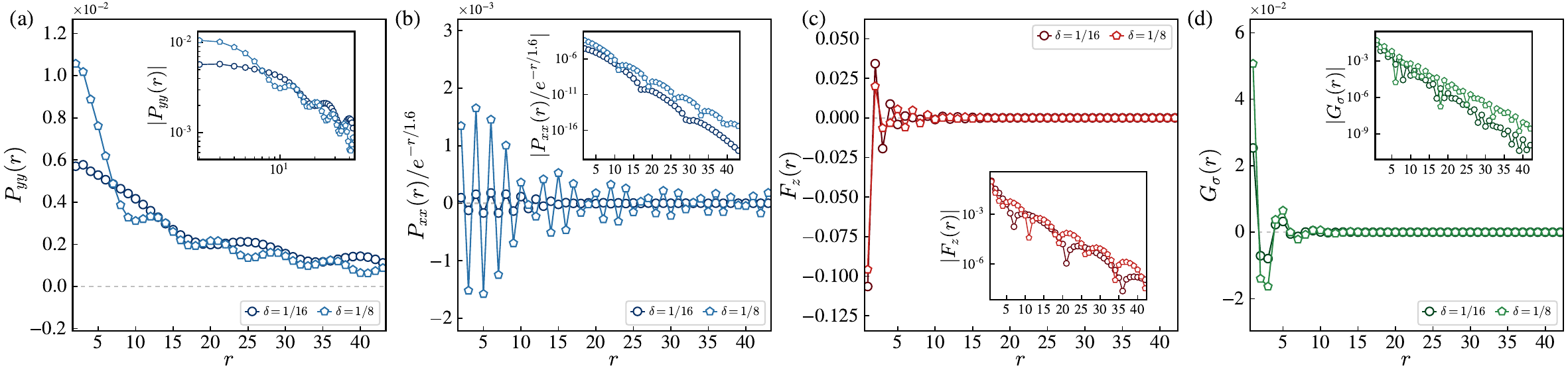}
    \caption{Correlation functions in the ground state of the bosonic $t$-$J$ model on $2$-leg ladders of size $64\times 2$ with $t_y=0$, $t_x/J=3$ and varying doping level $\delta$. (a,b) Spin-singlet pair-pair correlation between two NN links oriented in the $\{\hat{x}, \hat{y}\}$ directions. The correlation function in (b) is rescaled and divided by a factor $e^{-r/1.6}$ to show the oscillation pattern more clearly. (c) Spin-spin correlation along the $S^z$ direction. (d) Single-particle Green's function. The inset shows the logarithmic or semi-logarithmic plots.}
    \label{fig:2leg}
\end{figure}

\subsubsection{Probing the phase of pairing}
In the main text, we argue that the PDW results from the Marshall sign given that the fluctuating phases from imaginary time evolution under the Marshall basis are fully canceled by paired holes moving ``in tandem'', as depicted in Fig.\,\ref{fig:tandem_vs_side}(a). This argument can be further justified by measuring the pair-pair correlations in the absence of the inter-leg hopping $t_y$ (the hopping integral along the $\hat{y}$ direction). If we set $t_y=0$ while keeping $t_x/J=3$, the hole pair across different legs cannot move in tandem but can only move in two separate legs respectively, since the hole cannot hop among different legs. On the other hand, the hole pair within the same leg can still move in tandem. According to the argument in the main text, the former should have no PDW oscillation because an additional staggered sign arises from the ``side-by-side'' hopping process, as depicted in Fig.\,\ref{fig:tandem_vs_side}(b), while the latter should still have the same PDW oscillation as the cases with $t_y=t_x=3J$. Indeed, as shown by the DMRG results of $2$-leg systems with $t_y=0$ in Fig.\,\ref{fig:2leg}, the pair-pair correlation $P_{yy}(r)$, corresponding to the hole pair across different legs, has no PDW oscillation and turns out to be positive-definite. The pair-pair correlation $P_{xx}(r)$, corresponding to the hole pair within the same leg, still has a PDW oscillation at momentum $\pi$, though it decays exponentially as the physical nature of the ground states has also been changed by enforcing the 1D hopping restriction $t_y=0$~\cite{Zhang2022a, Zhang2023c}. Similar results have also been observed in width-$4$ systems with $t_y=0$.

\begin{figure}
    \centering
    \includegraphics[width=0.9\linewidth]{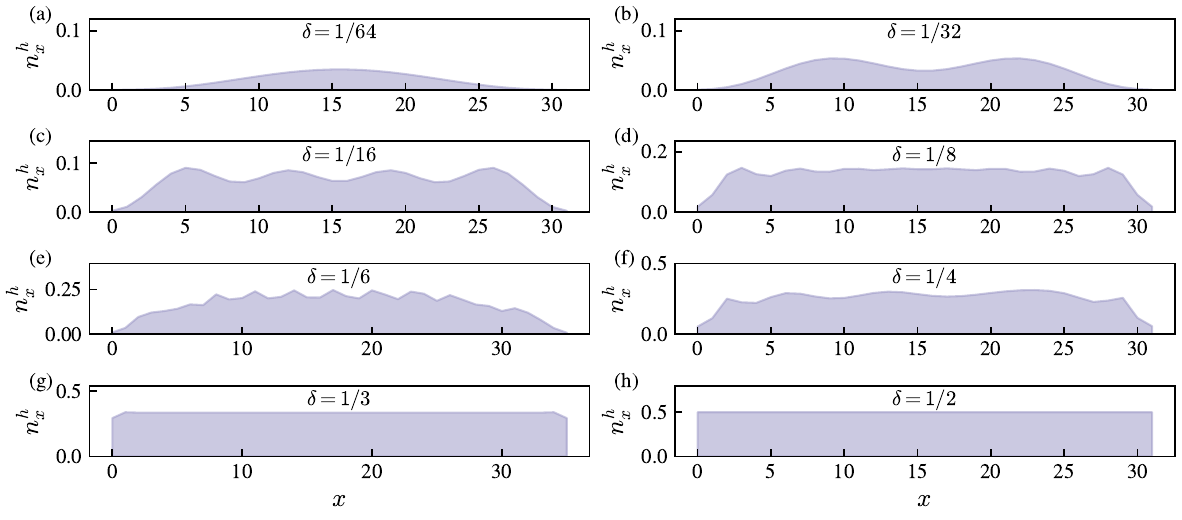}
    \caption{The hole density profiles in the ground state of the bosonic $t$-$J$ model on cylinders of size $32\times 4$ or $36\times 4$ with $t/J=3$ and varying doping level $\delta$.}
    \label{fig:4leg_nh_profile_subplot}
\end{figure}

\begin{figure}
    \centering
    \includegraphics[width=0.9\linewidth]{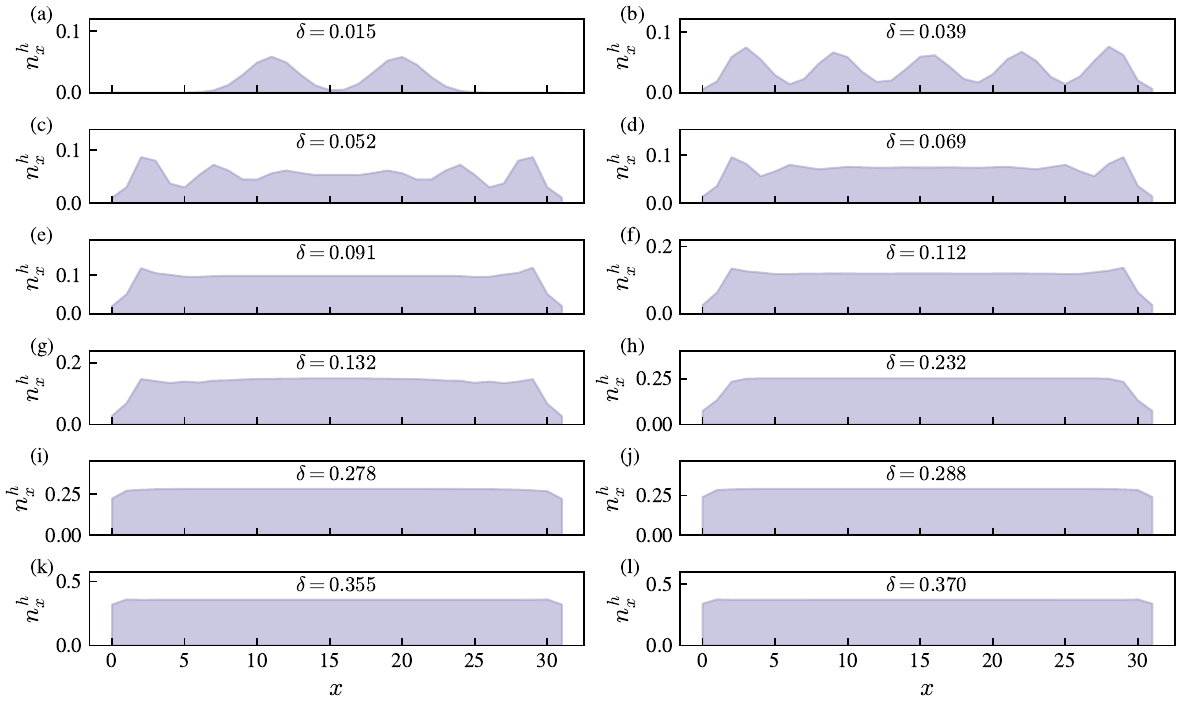}
    \caption{The hole density profiles in the ground state of the bosonic $t$-$J$ model on cylinders of size $32\times 8$ with $t/J=3$ and varying doping level $\delta$.}
    \label{fig:8leg_nh_profile_subplot}
\end{figure}

\subsubsection{Density distributions}
Figs.\,\ref{fig:4leg_nh_profile_subplot} and \ref{fig:8leg_nh_profile_subplot} show the hole density profiles $n_x^h = \sum_y n_{x,y}^h / L_y$ in width-$4$ and width-$8$ systems, where $n_{x,y}^h=n_i^h=1-n_i$. In the FM+SF phase, the density profiles are perfectly uniform. In the AFM+PDW phase, the profiles are almost uniform in the bulk at moderately low doping $\delta\gtrsim 0.05$, but exhibit some oscillations at very low doping $\delta\lesssim 0.05$, with each peak corresponding to around $2$ holes, which bears a certain resemblance with the inhomogeneity phenomena in real experiments. However, these density oscillations are different from the stripe orders commonly observed in the fermionic $t$-$J$ or Hubbard models because (i) the doping levels here are extremely low compared to the fermionic case; (ii) there is no AFM domain wall across the hole density profile peaks, i.e., the AFM order persists to the entire PDW phase instead of being modulated to spin density waves (SDW) at some momentum deviating from $\pi$~\cite{Xu2023}. Moreover, the density-density correlation $D(r)$, as shown in Fig.\,\ref{fig:bosonic_corr_sc_spin_green_nn} and Table.\,\ref{tab:btj}, decays much faster than those dominant correlations $P_{\alpha\beta}(r)$ or $G_{\sigma}(r)$. This feature also differs from the stripe orders in the fermionic case where the density-density correlations are dominant. From another perspective, this also suggests that the bosonic $t$-$J$ model might be computationally easier than the fermionic $t$-$J$ or Hubbard models due to the absence of intensely competing orders.

\begin{figure}
    \centering
    \includegraphics[width=\linewidth]{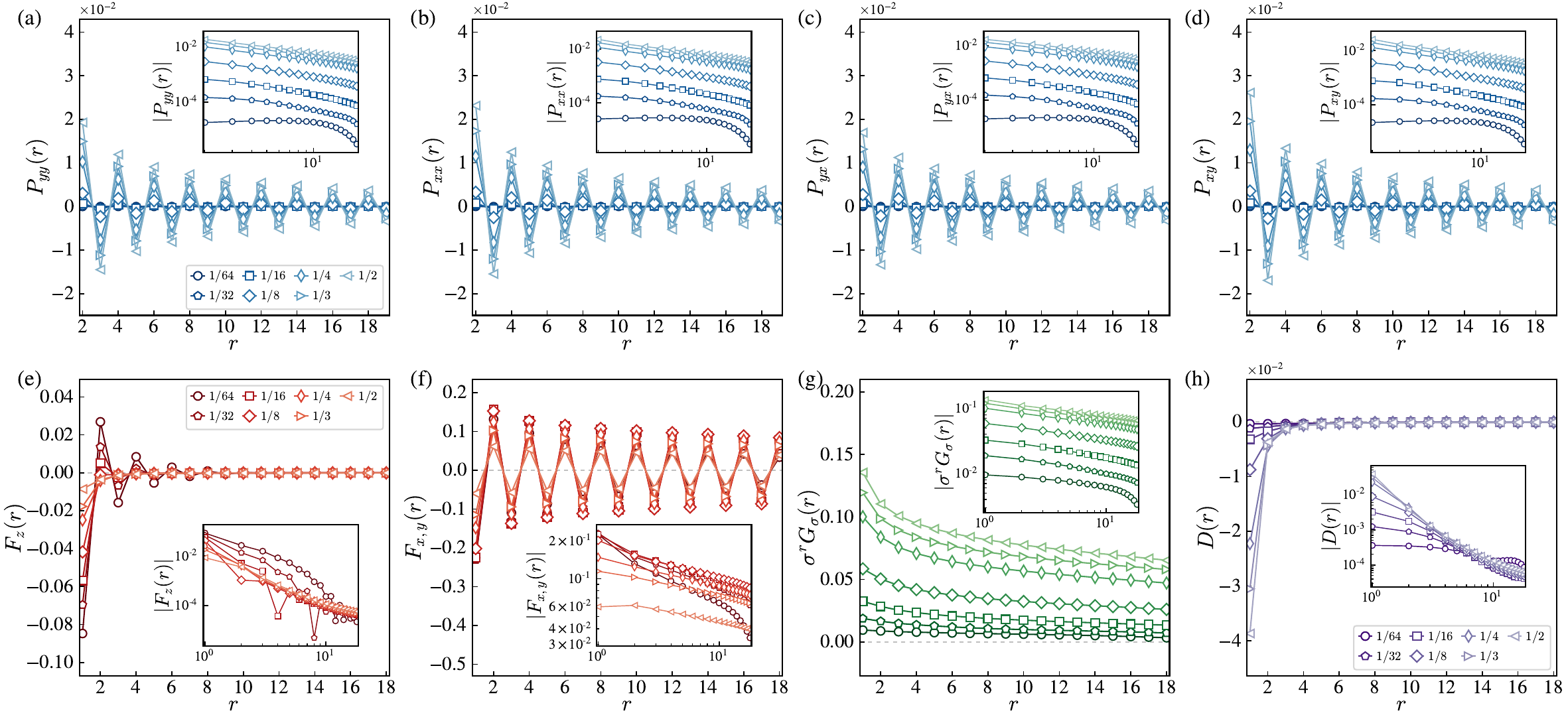}
    \caption{Correlation functions in the ground state of the bosonic $\sigma t$-$J$ model on cylinders of size $32\times 4$ or $36\times 4$ with $t/J=3$ and varying doping level $\delta$ labeled by different markers. (a-d) Spin-singlet pair-pair correlation between two NN links oriented in the $\{\hat{x},\hat{y}\}$ directions. (e,f) Spin-spin correlation along the $S^z$ and $S^{x,y}$ directions. (g) Single-particle Green's function. (h) Density-density correlation. The inset shows the logarithmic plots. The distinct behavior of $\delta=1/64$ is inclined to be a finite-size effect since there are only two holes in this $32\times 4$ system at doping $\delta=1/64$.}
    \label{fig:bosonic_sigma_corr_sc_spin_green_nn}
\end{figure}

% \usepackage{multirow}
\begin{table}
\begin{tabular}{|c|c|c|c|c|c|c|}
\hline
Phase                        & Doping        & $G_{\sigma}(r)$  & $P_{\alpha\beta}(r)$ & $F_{z}(r)$                         & $F_{xy}(r)$                        & $D(r)$           \\ \hline
AFM$^*$                      & $\delta=0$    & $=0$             & $=0$                   & $K_F\approx 1.3$, $\xi_F\approx 5$ & $K_F\approx 1.3$, $\xi_F\approx 5$ & $=0$             \\ \hline
\multirow{6}{*}{xy-AFM + SF} & $\delta=1/32$ & $K_G\approx 0.3$ & $K_P\approx 1$         & $K_F\approx 2$                   & $K_F\approx 0.3$                   & $K_D\approx 2$ \\ \cline{2-7} 
                             & $\delta=1/16$ & $K_G\approx 0.3$ & $K_P\approx 1$         & $K_F\approx 2$                   & $K_F\approx 0.3$                   & $K_D\approx 2$ \\ \cline{2-7} 
                             & $\delta=1/8$  & $K_G\approx 0.3$ & $K_P\approx 0.9$       & $K_F\approx 2$                   & $K_F\approx 0.3$                   & $K_D\approx 2$ \\ \cline{2-7} 
                             & $\delta=1/4$  & $K_G\approx 0.2$ & $K_P\approx 0.8$       & $K_F\approx 2$                   & $K_F\approx 0.2$                   & $K_D\approx 2$ \\ \cline{2-7} 
                             & $\delta=1/3$  & $K_G\approx 0.2$ & $K_P\approx 0.7$       & $K_F\approx 2$                   & $K_F\approx 0.2$                   & $K_D\approx 2$ \\ \cline{2-7} 
                             & $\delta=1/2$  & $K_G\approx 0.2$ & $K_P\approx 0.7$       & $K_F\approx 2$                   & $K_F\approx 0.2$                   & $K_D\approx 2$ \\ \hline
\end{tabular}
\caption{The estimated fitting exponents for the correlation functions in the ground states of the bosonic $\sigma t$-$J$ model at $t/J=3$ on $4$-leg cylinders. The system size is chosen as $32\times 4$ for $\delta=0,1/32,1/16,1/8,1/4,1/2$ and $36\times 4$ for $\delta=1/3$. $G_\sigma(r)$, $P_{\alpha\beta}(r)$, $F_z(r)$, $F_{xy}(r)$ and $D(r)$ represent the single-particle Green's function, singlet pair-pair correlation, spin-spin correlation along $z$ and $x,y$ direction, and density-density correlation. The singlet pair orientation in $P_{\alpha\beta}(r)$ takes values in $\alpha,\beta \in \{\hat{x}, \hat{y}\}$, the information of which are merged into one column as they have similar decaying behaviors. The spin orientation in $G_\sigma(r)$ takes values in $\sigma \in \{\up, \dn\}$, which are also merged into one column as they have the same decaying behaviors. The notations $K_G$, $K_P$ and $K_F$ stand for power-law fitting $r^{-K}$ while $\xi_G$, $\xi_P$, $\xi_F$ and $\xi_D$ stands for exponential fitting $e^{-r/\xi}$. The cases where two fitting exponents are both shown mean that considerable ambiguity exists in determining fitting schemes due to numerical uncertainty. Here ``AFM$^*$'' means that though a spin gap is observed due to the finite-size effects of even-leg ladders, a true AFM long-range order will be developed in the 2D limit.}
\label{tab:sigma_btj}
\end{table}

\subsubsection{Correlation function analysis for the bosonic $\sigma t$-$J$ model}
Here we provide a detailed analysis of the correlation function data from the CE calculations for the bosonic $\sigma t$-$J$ model on width-$4$ systems. Fig.\,\ref{fig:bosonic_sigma_corr_sc_spin_green_nn} shows the correlation functions in the ground state of the bosonic $\sigma t$-$J$ model on cylinders of size $32\times 4$ for the doping levels $\delta=1/64,1/32,1/16,1/8,1/4,1/2$ and size $36\times 4$ for the doping levels $\delta=1/3$. We summarize the corresponding extracted power exponent $K$ and correlation length $\xi$ in Table.\,\ref{tab:sigma_btj}. Since this bosonic $\sigma t$-$J$ model is a sign-problem-free model, previous work has actually already performed accurate quantum Monte Carlo (QMC) calculations~\cite{Boninsegni2001, Boninsegni2002, Boninsegni2008, Smakov2004} on the model in Eq.\,\eqref{eq:btj_minus} which is equivalent to this bosonic $\sigma t$-$J$ model. The conclusions there are consistent with ours, also indicating an SF ground state. One benefit from repeating the simulations with DMRG here is that the comparison with the QMC results allows us to validate the reliability of our DMRG simulations and our interpretation of the DMRG data throughout the paper, such as the extremely slow decay of $G_{\sigma}(r)$ in the CE calculations indeed indicates an SF phase. Moreover, similar to the FM+SF phase in the bosonic $t$-$J$ model, the algebraic decay of the singlet pair-pair correlations shown in Figs.\,\ref{fig:bosonic_sigma_corr_sc_spin_green_nn}(a)-(d) does not signify true pairing but rather a secondary effect resulting from single-boson condensation.

\subsubsection{Hole-hole correlation in two-hole-doped systems}

\begin{figure}
    \centering
    \includegraphics[width=0.6\linewidth]{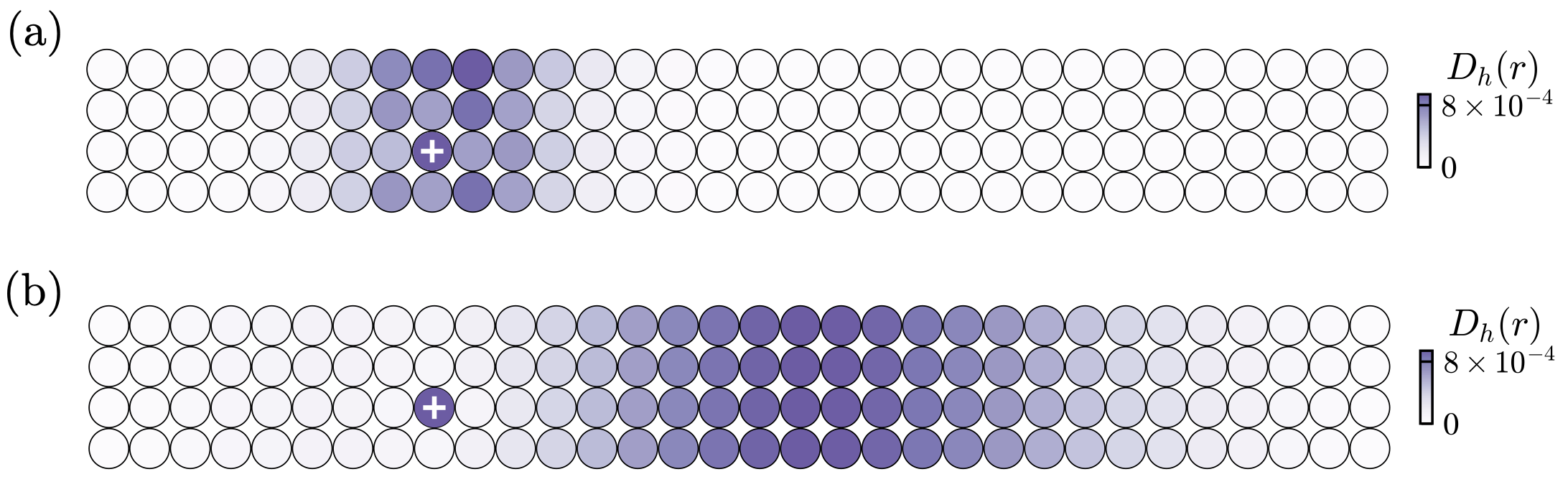}
    \caption{The hole-hole density correlation for the two-hole-doped (a) bosonic $t$-$J$ and (b) $\sigma t$-$J$ models with $t/J=3$ on cylinders of size $32\times 4$. The cross represents the reference site.}
    \label{fig:btj_sbtj_Nx=32_Ny=4_holehole_corr}
\end{figure}

\begin{figure}
    \centering
    \includegraphics[width=0.68\linewidth]{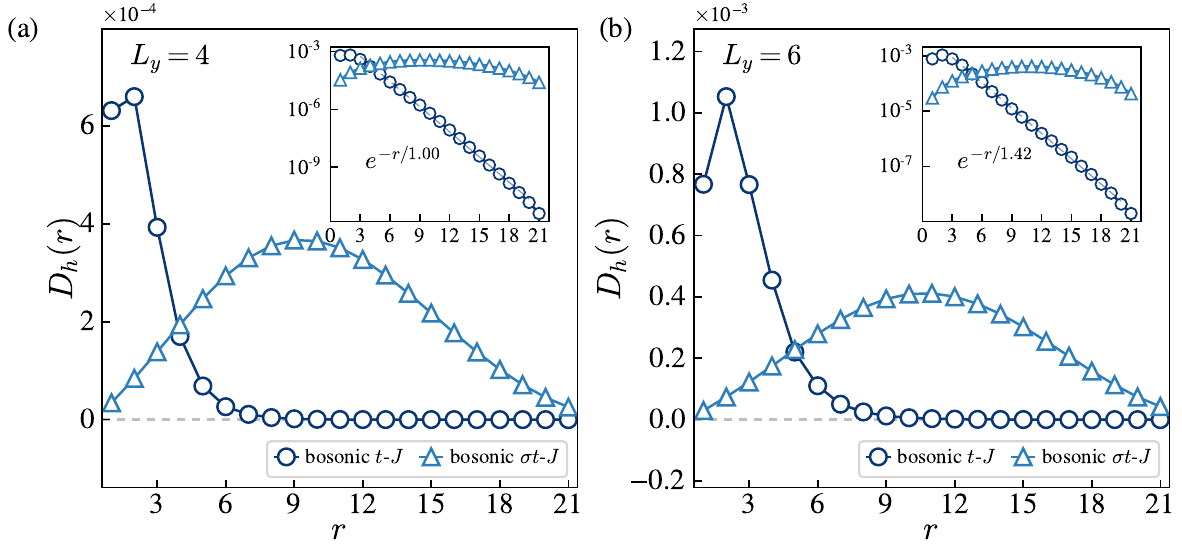}
    \caption{The profile of the hole-hole density correlation for the two-hole-doped bosonic $t$-$J$ and $\sigma t$-$J$ models with $t/J=3$ on cylinders of size $32\times 4$ and open square lattices of size $32\times 6$.}
    \label{fig:bosonic_Nx=32_Ny=4-6_nhnh}
\end{figure}

The hole-hole density correlation is defined as $D_h(r) = \avg{n_{i_0}^h n_{i_0+r}^h}$. For a two-hole-doped system, $D_h(r)$ can be interpreted as the spatial distribution of one hole over the lattice given another hole is pinned at site $i_0$. Hence, $D_h(r)$ can be used to detect the existence of BEC-type pairing besides the pair-pair correlations. In the main text, we have shown the hole-hole correlation in width-$6$ systems, where the fully open boundary condition is exploited, in contrast to Ref.\,\cite{Homeier2024}. Fig.\,\ref{fig:btj_sbtj_Nx=32_Ny=4_holehole_corr} shows the hole-hole correlation in width-$4$ cylinders, which yields consistent results with width-$6$ calculations. Namely, the two holes are very close to each other in space, forming a tightly bound pair in the bosonic $t$-$J$ model while exhibiting a mutually separated and loose distribution in the bosonic $\sigma t$-$J$ model, which aligns with the corresponding PDW order and SF order at the low doping regime, respectively. We further plot the profile of $D_h(r)$ with $r$ along the $\hat{x}$ direction in Fig.\,\ref{fig:bosonic_Nx=32_Ny=4-6_nhnh} and extract the correlation length of $D_h(r)$ for the bosonic $t$-$J$ model, which equals about one lattice constant for both width-$4$ and width-$6$ systems, implying an extremely small hole pair size in the AFM+PDW phase.

% Q: Why can $D(r)$ be smaller at the NN site than at the NNN site? A: Because the in-tandem motion should be a process in which the front hole moves first and then the back hole follows. So there is some time when the two holes are located at NNN sites.

\subsubsection{The intermediate phase and the evolution of orders}

Besides the prominent AFM+PDW and FM+SF phases discussed in the main text, Fig.\,\ref{fig:bosonic_corr_sc_spin_green_nn} reveals an intermediate region between the two typical phases, including $\delta=1/6,1/4$ at least at the level of $L_y=4$. In the following, we will show that together with this intermediate phase, the evolution of various orders with doping can also be well understood based on the interference effect.

\begin{figure}
    \centering
    \includegraphics[width=0.84\linewidth]{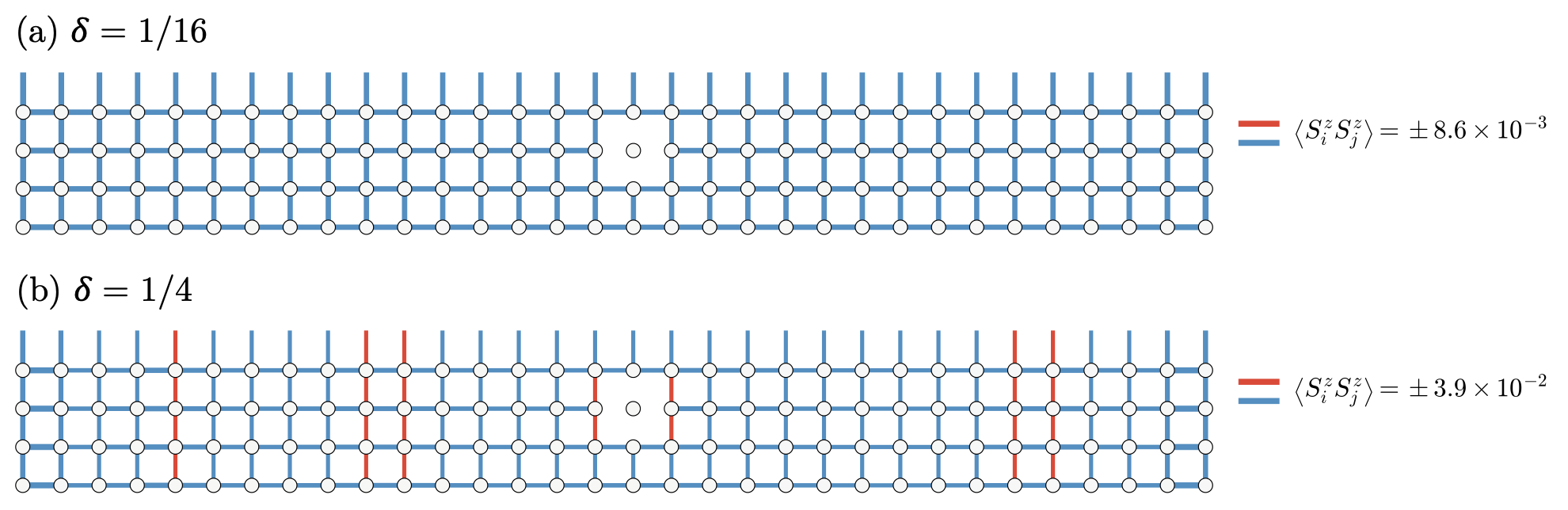}
    \caption{The spin-spin correlation function $\avg{n^h_i S^z_j S^z_k}$ for all nearest neighbor sites $\avg{jk}$ given a hole projected at site $i$, as marked by the vacancy square at the center of the system, in (a) the AFM+PDW phase at doping $\delta=1/16$ and (b) the intermediate phase at doping $\delta=1/4$, obtained by DMRG simulations for the bosonic $t$-$J$ model on a cylinder of size $32\times 4$. Blue links represent AFM correlations, while red links represent FM correlations. Polaron-like patterns can be observed from the FM spin-spin correlations around the projected hole in the intermediate phase.}
    \label{figR:nhSzSz}
\end{figure}

\begin{figure}
    \centering
    \includegraphics[width=0.32\linewidth]{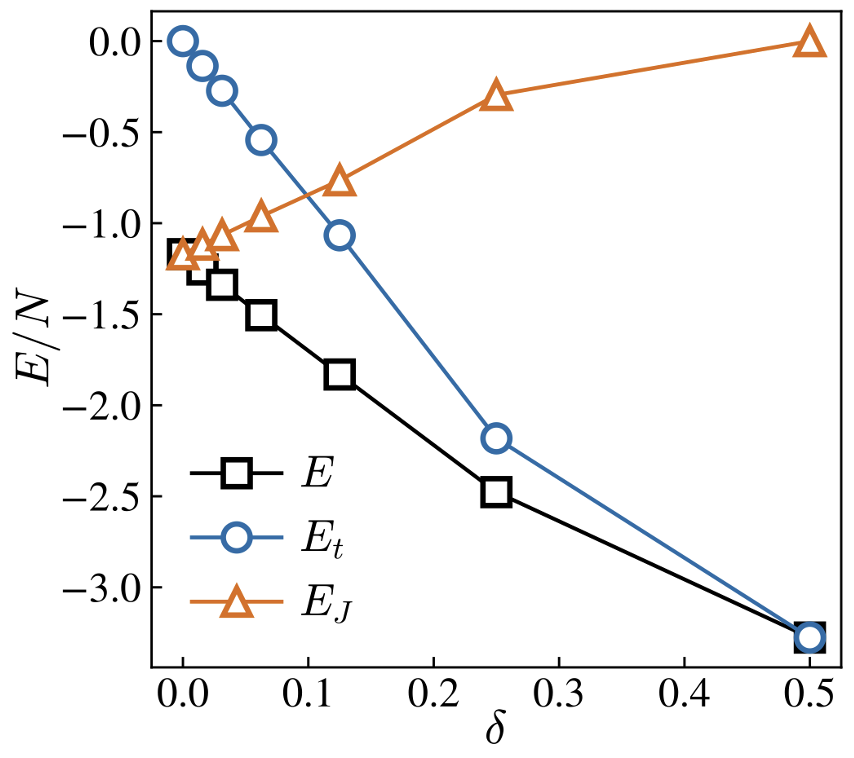}
    \caption{The ground state energy $E$ together with its components from the hopping term $E_t$ and the spin exchange term $E_J$ as a function of the hole doping level $\delta$, obtained from DMRG simulations for the bosonic $t$-$J$ model on a cylinder of size $32\times 4$.}
    \label{figR:energy_doping}
\end{figure}

Upon doping the antiferromagnet, the pairing mechanism immediately activates and generates a PDW order that initially becomes stronger with increasing hole doping. However, as the hole doping increases, the hopping term, which involves holes, will gradually dominate the total energy over the spin exchange term, which only involves spins. Since single-hole motion favors an Nagaoka FM spin background where no phase frustration occurs [$(-1)^{N^h_{\dn}}\rightarrow 1$], further hole doping will in turn weaken the AFM and PDW orders with a transition tendency to the FM and single-particle condensation. Fig.\,\ref{figR:nhSzSz} shows that there are indeed the FM spin-spin correlations around a hole at intermediate doping levels. Therefore, the PDW order will first increase with doping, then decrease and vanish when the hopping energy dominates over the spin exchange energy. This can be evidenced by our calculation of the separated energy evolution in Fig.\,\ref{figR:energy_doping}, where the cross point of the hopping energy and the spin exchange energy is exactly around the end point of the AFM+PDW phase $\delta_c\approx 0.1$.

% This intermediate region also shows signatures in the simulations for systems of $L_y=8$, where the doping level $\delta$ can hardly fall into the interval near $\delta_c\approx 0.2$ by tuning the chemical potential $\mu$, suggesting that it is an unstable region with relatively high entanglement that is hard to be captured by DMRG simulations, compared to the stable phases on both sides. Thus, we consider the region as a transition region instead of a new well-defined phase, as indicated by the grey area in Fig.\,\textcolor{darkblue1}{1}(a). In the 2D thermodynamic limit, this transition region could either shrink into a true quantum critical point or persist as a finite region, a question we leave for future research.

% The partition function can be seen as the trace of the Gibbs density matrix. Each term in the tracial summation is proportional to the fidelity between the Gibbs density matrix and the chosen basis state vector, i.e., the amplitude of the component coefficient of the chosen basis state vector in the Gibbs state.

\subsubsection{The effect of the density-density term and the CE simulations for $L_y=6,8$}

We note that Ref.\,\cite{Harris2024} observes a stripe order at low doping levels and width $L_y=6$ in a similar bosonic $t$-$J$ model using DMRG calculations. Here, we briefly explain this difference with additional DMRG calculations for different model parameters and system widths. We consider an extended bosonic $t$-$J$ model
\begin{equation}\label{eq:extended_bosonic_tj}
\begin{aligned}
    H_{t\text{-}J} &= \mathcal{P}_s (H_t + H_J) \mathcal{P}_s, \\
    H_t &= - t\sum_{\avg{ij} \sigma} \left( B^{\dagger}_{i\sigma} B_{j\sigma} + \text{H.c.}\right), \\
    H_J &= J \sum_{\avg{ij}} \mathbf{S}_i\cdot \mathbf{S}_j + V \sum_{\avg{ij}} n_i n_j,
\end{aligned}
\end{equation}
which takes the coefficient of the density-density interaction term as a tunable parameter $V$, compared to the original model. We set $J=1$ as the energy unit and fix $t/J=3$. If $V=-\frac{1}{4} J$, this extended model reduces to the standard bosonic $t$-$J$ model. If $V=\frac{3}{4} J$, this extended model becomes the same as the one used in Ref.\,\cite{Harris2024}. We remark that the case of $V=-\frac{1}{4} J$ is a standard choice in the fermionic $t$-$J$ model as the strong coupling limit of the Fermi-Hubbard model, while the case of $V=\frac{3}{4}J$ results from the spectrum-reversed version of the FM bosonic $t$-$J$ model as the strong coupling limit of the spin-$1/2$ Bose-Hubbard model as considered by the experimental realization in Ref.\,\cite{Harris2024}.

We conducted additional DMRG calculations for density-density interaction parameter $V=-\frac{1}{4},\frac{3}{4}$ and system width $L_y=4,6,8$, implementing the global charge-U(1) symmetry and spin-SU(2) symmetry with up to $M=25000$ SU(2) multiplets (equivalent to $D\approx 90000$ $U(1)$ bond dimensions) or spin-U(1) symmetry with up to $D= 18000$ bond dimensions. We set the doping level $\delta=\frac{1}{16}$ for system size $32\times 4$, $\delta=\frac{1}{24}$ for $32\times 6$, and $\delta=\frac{1}{24},\frac{1}{16}$ for $24\times 8$, taking into account the divisibility between the number of holes and the number of lattice sites. We summarize our additional results in Table.\,\ref{tab:0.75nn}.

% \begingroup
\setlength{\tabcolsep}{8pt} % Default value: 6pt
\renewcommand{\arraystretch}{1.3} % Default value: 1

\begin{table}[]
\begin{tabular}{|c|c|c|c|}
\hline
Width                    & Doping                         & Model Parameter   & Phase    \\ \hline
\multirow{2}{*}{$L_y=4$} & \multirow{2}{*}{$\delta=\frac{1}{16}$} & $V=-\frac{1}{4}J$ & AFM+PDW  \\ \cline{3-4} 
                         &                                & $V=\frac{3}{4}J$  & AFM+PDW  \\ \hline
\multirow{2}{*}{$L_y=6$} & \multirow{2}{*}{$\delta=\frac{1}{24}$} & $V=-\frac{1}{4}J$ & AFM+PDW* \\ \cline{3-4} 
                         &                                & $V=\frac{3}{4}J$  & SDW+PDW* \\ \hline
\multirow{3}{*}{$L_y=8$} & \multirow{2}{*}{$\delta=\frac{1}{24}$} & $V=-\frac{1}{4}J$ & AFM+PDW  \\ \cline{3-4} 
                         &                                & $V=\frac{3}{4}J$  & AFM+PDW  \\ \cline{2-4} 
                         & $\delta=\frac{1}{16}$                  & $V=-\frac{1}{4}J$ & AFM+PDW  \\ \hline
\end{tabular}
\caption{Summary of DMRG results on the ground state phases of the extended bosonic $t$-$J$ model with different system width $L_y$ and density-density interaction parameter $V$. Here, ``SDW'' refers to a ``spin density wave'' with wave vector deviating from $\pi$, and ``PDW*'' refers to a variant of the original $(\pi,\pi)$-PDW order.}
\label{tab:0.75nn}
\end{table}
% \endgroup

\begin{figure}
    \centering
    \includegraphics[width=0.8\linewidth]{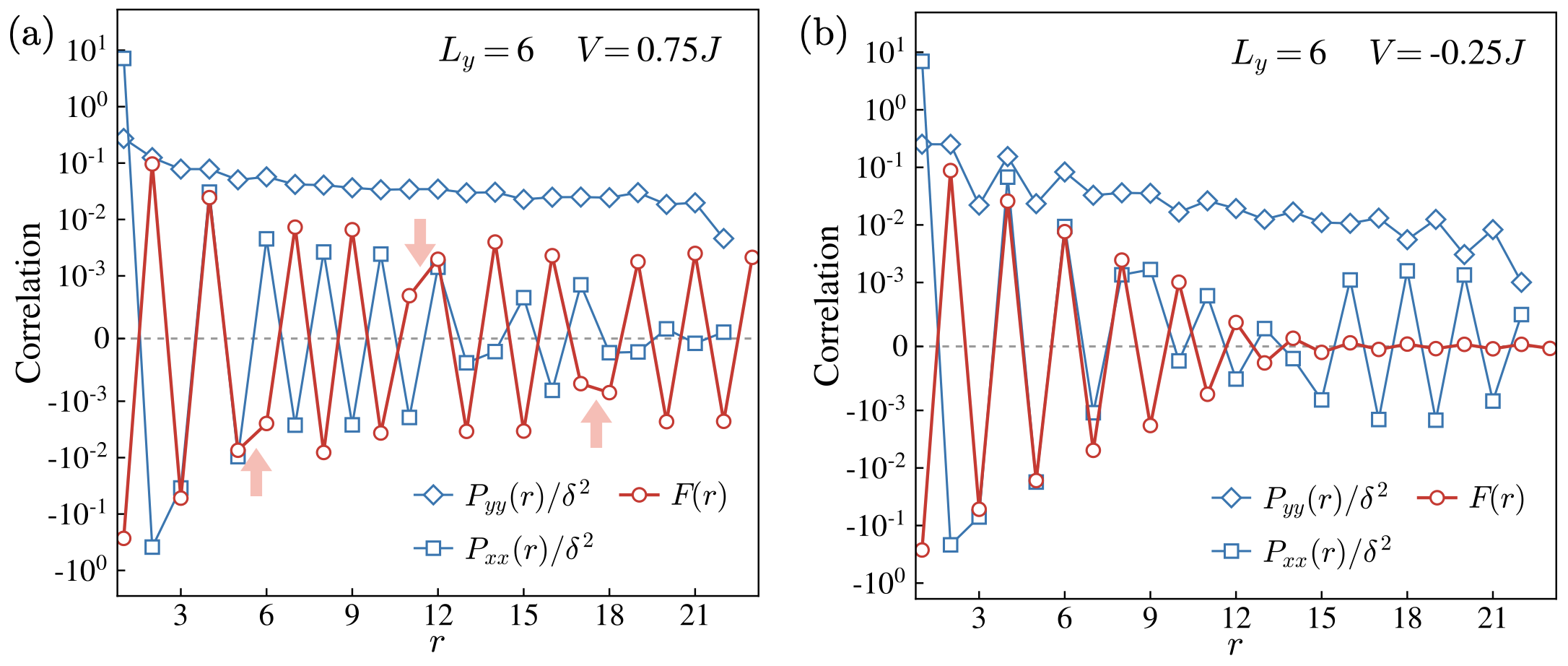}
    \caption{The singlet pair-pair correlation function $P_{\alpha\beta}(r)=\avg{\Delta_{i,\alpha}^\dagger \Delta_{i+r,\beta}}$ and the spin-spin correlation function $F(r)=\avg{\mathbf{S}_i\cdot \mathbf{S}_{i+r}}$ along the $\hat{x}$ direction in the ground state of the extended bosonic $t$-$J$ model on a cylinder of size $32\times 6$ with doping $\delta=\frac{1}{24}$ and density-density interaction (a) $V=\frac{3}{4}J$ and (b) $V=-\frac{1}{4}J$, obtained from DMRG simulations in CE with spin-SU(2) symmetry and up to $M=16000$ SU(2) multiplets (equivalent to $D\approx 56000$ effective U(1) bond dimensions). The correlations are rescaled using a symmetric logarithmic scale. The fitted power exponents are (a) $K_F\approx 1.4$ and $K_P\approx 0.7$; (b) $K_P\approx 1.6$. The red arrows mark the positions of the AFM domain walls. Here the pair-pair correlation keeps a uniform sign in $\hat{y}$-links along the $\hat{x}$ direction while exhibits sign oscillations in $\hat{x}$-links, see also Fig.\,\ref{figR:pdw_undirected_mu=5.5355}, which is slightly different from the standard $(\pi,\pi)$-PDW, and hence is denoted as ``PDW*'' in Table.\,\ref{tab:0.75nn}.}
    \label{figR:sc_spin_32_6_doping24_symlog}
\end{figure}

\begin{figure}
    \centering
    \includegraphics[width=0.7\linewidth]{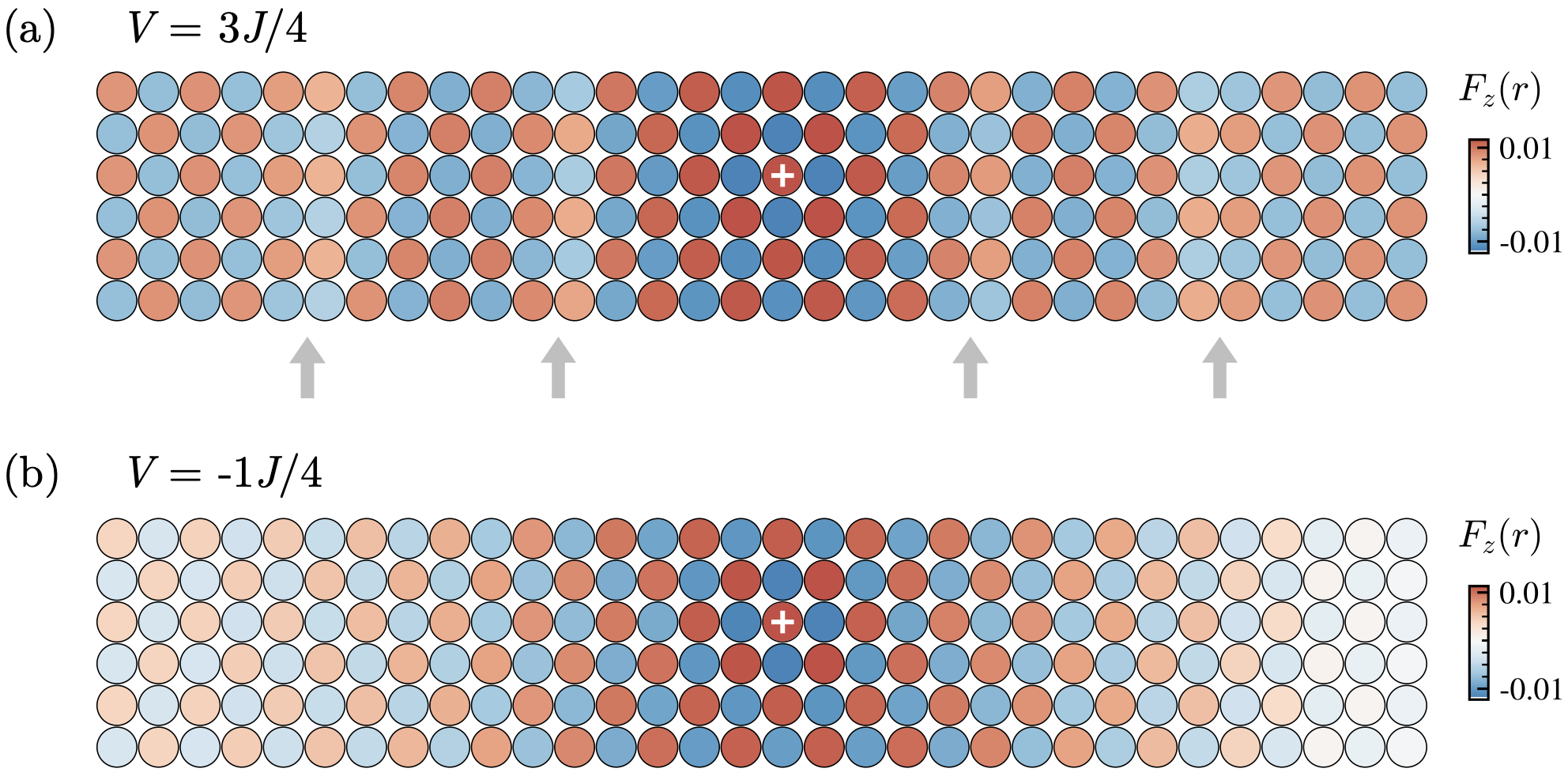}
    \caption{The 2D spin-spin correlation function $F_z(r)=\avg{S^z_i S^z_{i+r}}$ in the ground state of the extended bosonic $t$-$J$ model on a cylinder of size $32\times 6$ with doping $\delta=\frac{1}{24}$ and density-density interaction (a) $V=\frac{3}{4}J$ and (b) $V=-\frac{1}{4}J$, obtained from DMRG simulations in CE with spin-U(1) symmetry and bond dimension up to $D=18000$. The correlations are coded in color using a symmetric logarithmic scale. The cross represents the reference site. The grey arrows mark the positions of the AFM domain walls.}
    \label{figR:ref2all_szsz_32_6}
\end{figure}

\begin{figure}
    \centering
    \includegraphics[width=0.94\linewidth]{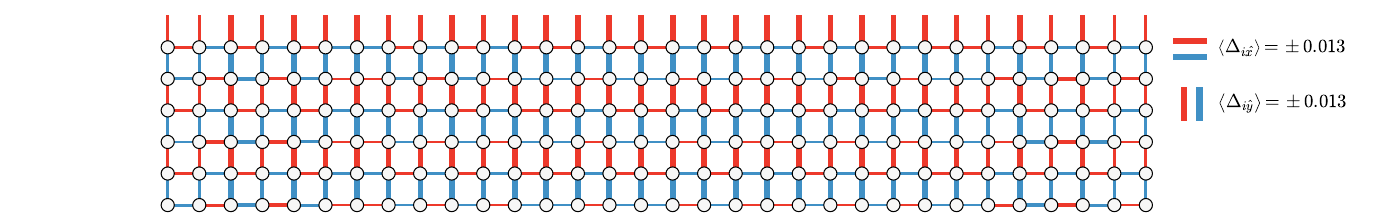}
    \caption{The typical PDW pattern on a cylinder of size $32\times 6$ with $V=-\frac{1}{4}J$ and $\delta\approx \frac{1}{24}$, denoted as ``PDW*'' in Table.\,\ref{tab:0.75nn}, obtained from DMRG simulations in GCE with bond dimension $D=4000$. The width of the edges is proportional to the magnitude of the spin-singlet pairing order parameter $\avg{\Delta_{i\alpha}}$.}
    \label{figR:pdw_undirected_mu=5.5355}
\end{figure}

First of all, in the same case as in Ref.\,\cite{Harris2024} where $L_y=6$, $V=\frac{3}{4}J$ and $\delta=\frac{1}{24}$, we obtain consistent results that there are indeed AFM domain walls in the spin-spin correlation function, i.e., spin ``stripes'', as shown in Figs.\,\ref{figR:sc_spin_32_6_doping24_symlog}(a) and \ref{figR:ref2all_szsz_32_6}(a), which we denote as ``spin density wave (SDW)'' in Table.\,\ref{tab:0.75nn}. However, we find that at the same time, there are coexisting quasi-long-range singlet pair-pair correlations that also exhibit PDW patterns, though they are slightly different from the standard $(\pi,\pi)$-PDW order, cf. Figs.\,\ref{figR:sc_spin_32_6_doping24_symlog} and \ref{figR:pdw_undirected_mu=5.5355}, which we denote as ``PDW*'' in Table.\,\ref{tab:0.75nn}. We note that similar PDW patterns have also been observed before in fermionic models~\cite{Yue2024, Zheng2024}. These strong pair-pair correlations are not included in Ref.\,\cite{Harris2024}, which mainly focuses on magnetic properties. Therefore, the general conclusion that the ground state of the bosonic $t$-$J$ model exhibits pairing orders at low doping regimes still holds, even in the case of $L_y=6$ and $V=\frac{3}{4}J$.

% they are slightly different from the standard $(\pi,\pi)$-PDW order by an extra uniform component on the $\hat{y}$ links

For $L_y=6$ and $V=-\frac{1}{4}J$, the pair-pair correlation remains the PDW* pattern as in the case of $L_y=6$ and $V=\frac{3}{4}J$, while the spin-spin correlation is weakened and presents no more stripe pattern, as shown in Figs.\,\ref{figR:sc_spin_32_6_doping24_symlog}(b) and \ref{figR:ref2all_szsz_32_6}. The presence of stripes in the case of $L_y=6$ and $V=\frac{3}{4}J$ can be understood by the fact that the additional repulsive density-density interaction $\left(+n_in_j\right)$ in the case of $V=\frac{3}{4}J$ makes holes tend to repel from each other and then facilitates the possible formation of separated hole-rich regions, which disarrange the normal AFM order and serve as AFM domain walls. Here we elaborate further on the PDW* order in Table.\,\ref{tab:0.75nn}. According to the argument above, the $(\pi,\pi)$-PDW is a natural result from the Marshall sign if the imaginary-time path integral is dominated by a leading positive contribution, which is highly likely since different paths of the ``in-tandem'' propagation of hole pairs can interfere constructively. However, if the contribution from the in-tandem propagation fails to dominate due to finite-size effects or geometric constraints, allowing alternative parts to contribute significantly, like the side-by-side motion illustrated in Fig.\,\ref{fig:tandem_vs_side}, the sign of the pairing order can no longer be determined solely by the Marshall sign. This ultimately leads to distinct outcomes such as the PDW* state.

\begin{figure}
    \centering
    \includegraphics[width=0.8\linewidth]{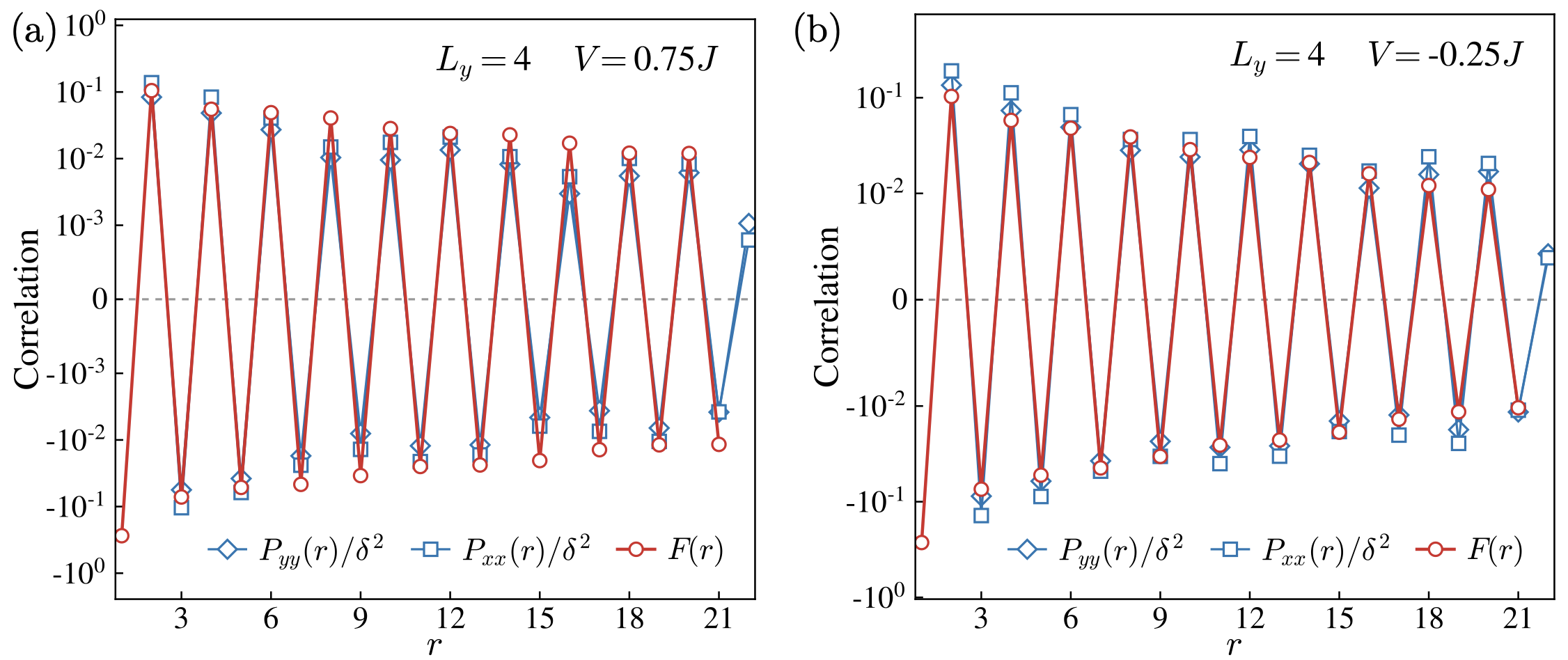}
    \caption{The singlet pair-pair correlation function $P_{\alpha\beta}(r)=\avg{\Delta_{i,\alpha}^\dagger \Delta_{i+r,\beta}}$ and the spin-spin correlation function $F(r)=\avg{\mathbf{S}_i\cdot \mathbf{S}_{i+r}}$ along the $\hat{x}$ direction in the ground state of the extended bosonic $t$-$J$ model on a cylinder of size $32\times 4$ with doping $\delta=\frac{1}{16}$ and density-density interaction (a) $V=\frac{3}{4}J$ and (b) $V=-\frac{1}{4}J$, obtained from DMRG simulations in CE with spin-U(1) symmetry and effective bond dimension up to $D= 10000$. The correlations are rescaled using a symmetric logarithmic scale. The fitted power exponents are $K_F\approx K_P\approx 1.0$ for both (a) and (b).}
    \label{figR:sc_spin_32_4_doping16_symlog}
\end{figure}

For systems of width $L_y=4$ with $\delta=\frac{1}{16}$, the ground state is in the ``AFM+PDW'' phase for both $V=-\frac{1}{4}J$ and $V=\frac{3}{4}J$, as shown in Fig.\,\ref{figR:sc_spin_32_4_doping16_symlog}. There is no AFM domain wall in the spin-spin correlation, not only for $V=-\frac{1}{4}J$ but also for $V=\frac{3}{4}J$. We note that this is consistent with Ref.\,\cite{Harris2024}, which also considers the case of $L_y=4$ and identifies the low-doping regime as an ``AFM/Stripes'' phase.

\begin{figure}
    \centering
    \includegraphics[width=0.8\linewidth]{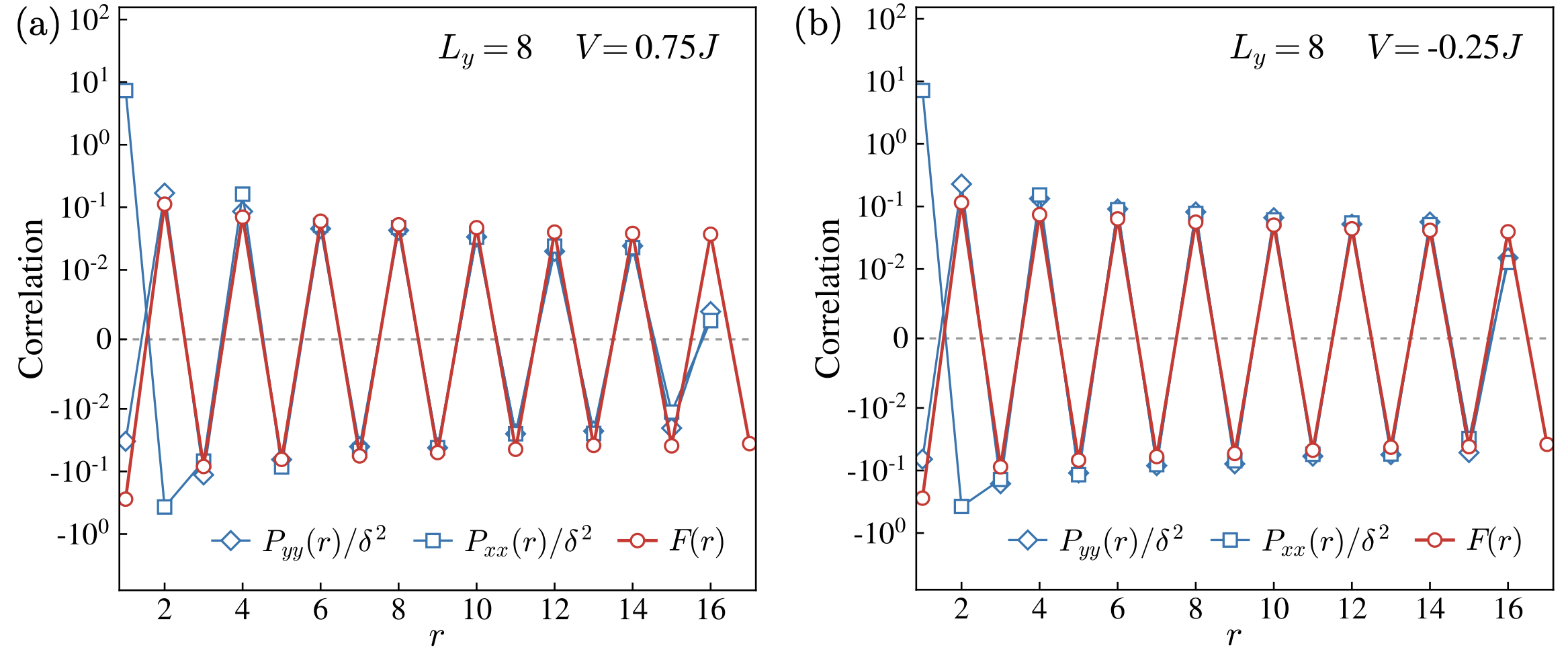}
    \caption{The singlet pair-pair correlation function $P_{\alpha\beta}(r)=\avg{\Delta_{i,\alpha}^\dagger \Delta_{i+r,\beta}}$ and the spin-spin correlation function $F(r)=\avg{\mathbf{S}_i\cdot \mathbf{S}_{i+r}}$ along the $\hat{x}$ direction in the ground state of the extended bosonic $t$-$J$ model on a cylinder of size $24\times 8$ with doping $\delta=\frac{1}{24}$ and density-density interaction (a) $V=\frac{3}{4}J$ and (b) $V=-\frac{1}{4}J$, obtained from DMRG simulations in CE with spin-SU(2) symmetry and up to $M=25000$ SU(2) multiplets (equivalent to $D\approx 90000$ U(1) bond dimensions). The correlations are rescaled using a symmetric logarithmic scale. The fitted power exponents are $K_F\approx K_P\approx 0.6$ for both (a) and (b).}
    \label{figR:sc_spin_24_8_doping24_symlog}
\end{figure}

\begin{figure}
    \centering
    \includegraphics[width=0.4\linewidth]{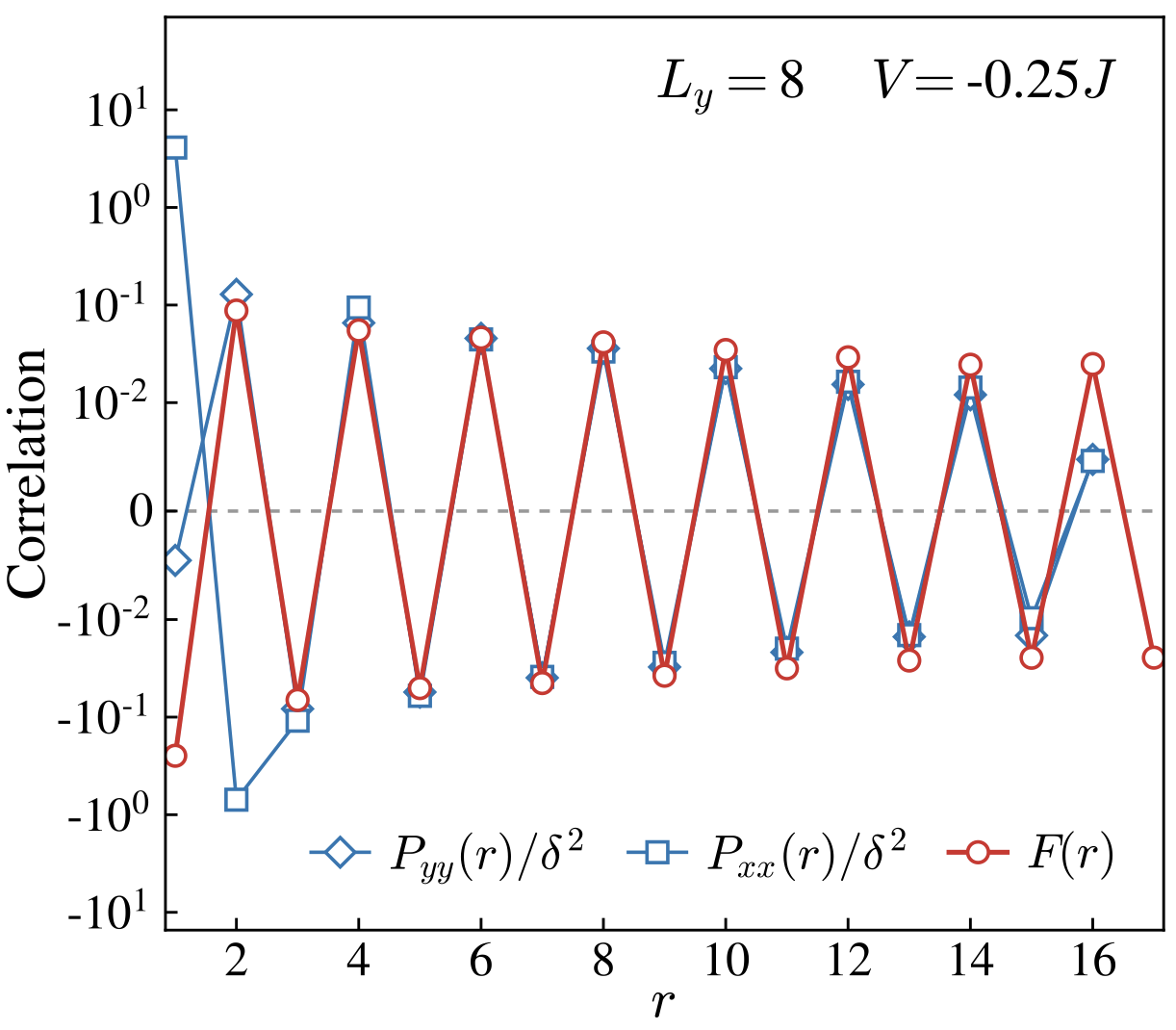}
    \caption{The same as in Fig.\,\ref{figR:sc_spin_24_8_doping24_symlog}(b) but for the doping level $\delta=\frac{1}{16}$. The data is obtained from DMRG simulations in CE with spin-SU(2) symmetry and up to $M=16000$ SU(2) multiplets (equivalent to $D\approx 60000$ U(1) bond dimensions). The fitted power exponents are $K_F\approx 0.6$ and $K_P\approx 1.0$.}
    \label{figR:sc_spin_24_8_doping=16_V=-0.25_symlog}
\end{figure}

For systems of width $L_y=8$ with $\delta=\frac{1}{24}$, the ground state is also in the ``AFM+PDW'' phase for both $V=-\frac{1}{4}J$ and $V=\frac{3}{4}J$, as shown in Fig.\,\ref{figR:sc_spin_24_8_doping24_symlog}, which exhibits the same behavior as the case of $L_y=4$. Moreover, we also compute the case of $\delta=\frac{1}{16}$ at $V=-\frac{1}{4}J$, which is also in the AFM+PDW phase, as depicted in Fig.\,\ref{figR:sc_spin_24_8_doping=16_V=-0.25_symlog}. This further corroborates the robustness of the PDW order at low doping levels. It is particularly worth emphasizing that, here we implement DMRG calculations in the CE scheme with both the charge-U(1) and spin-SU(2) symmetries for all $L_y=4,6,8$, the same scheme as in Ref.\,\cite{Harris2024}, and invest substantial computational resources to keep up to $M=25000$ SU(2) multiplets (equivalent to $D\approx 90000$ U(1) bond dimensions), to ensure the numerical reliability. As one can observe, we have obtained results that are consistent with both Ref.\,\cite{Harris2024} and our GCE simulations under the corresponding parameter settings. This not only validates the correctness of our main conclusion on the emergence of PDW order at low doping levels but also justifies the GCE computation scheme.

In summary, these results indicate that the striped spin correlation arises from the additional repulsive density-density interaction together with the special geometry of $L_y=6$, which is absent in the cases of $L_y=4,8$ with $V=-\frac{1}{4}J, \frac{3}{4}J$ and $L_y=6$ with $V=-\frac{1}{4}J$. More importantly, in all cases calculated above, there are always strong PDW correlations in the low-doping regime, even for the case in Ref.\,\cite{Harris2024}, firmly supporting our main conclusion.

\subsection{Comparison between two string effects in doped antiferromagnets}

We emphasize that the pairing mechanisms based on the phase-string effect and the spin-mismatch effect are two essentially different mechanisms, but they are cooperative and can contribute to the pairing orders simultaneously under certain conditions. The distinction can be explicitly seen from Eq.\,(\textcolor{blue}{4}) in the main text
\begin{equation}\label{eq:Z_tJ_supp}
    Z_{t\text{-}J}\equiv \opr{Tr} e^{-\beta H_{t\text{-}J}}=\sum_C \tau_C W_{t\text{-}J}[C].
\end{equation}
The spin-mismatch effect manifests itself on the non-negative weight $W[C]$ in Eq.\,\eqref{eq:Z_tJ_supp} where mismatched NN spins lead to an energy cost $\Delta E>0$ due to the spin interaction term $S^z_iS^z_j$, thereby reducing the weight $W[C]$ by a factor $e^{-\beta \Delta E}$. In contrast, the phase-string effect manifests itself on the phase factor $\tau_C = (-1)^{N_{\dn}^h}$ in Eq.\,\eqref{eq:Z_tJ_supp}, where $N_{\dn}^h$ counts the number of exchanges between holes and $\dn$ spins in an evolution loop.

It is important to note that the spin-mismatch effect can occur even in classical or sign-problem-free systems, while the phase string effect is an inherently quantum phenomenon. This distinction becomes particularly clear in the following two comparisons:
\begin{itemize}
    \item In the bosonic $\sigma t$-$J$ model, the phase-string effect is artificially switched off, i.e., $\tau_C = 1$ with all the weights $W[C]$ unchanged. Specifically, the spin-mismatch effect still exists because the $H_J$ term remains unchanged. According to the DMRG results in the main text, the pairing order disappears in this bosonic $\sigma t$-$J$ model, which highlights the significance of the phase-string effect in a non-trivial pairing phase.
    \item In the bosonic $t$-$J_z$ model~\cite{Smakov2004}, where the spin-flip terms are set to zero, preserving only the spin coupling along the $S^z$ direction $S^z_{i} S^z_{j}$, the system reduces to a sign-problem-free model which can be mapped to a classical model at the partition function level. The spin-mismatch effect remains in this $t$-$J_z$ model: the hopping of a single hole will disrupt the AFM spin configuration and incur an energy cost $\sim J_z$. Previous studies have shown that the ground state of the bosonic $t$-$J_z$ model is a single-particle superfluid~\cite{Smakov2004}, as opposed to the pairing ground state found here. This further underlines the indispensable role of the phase-string effect in the formation of pairing orders.
\end{itemize}
Therefore, the two effects are essentially different from each other. In fact, it was thought that the spin-mismatch effect alone could completely suppress the single-particle propagation due to the linearly growing energy costs. However, subsequent studies revealed that the single-particle mobility could be restored by following certain propagation paths or considering spin-flip processes to repair the mismatch links~\cite{Trugman1988}. The phase-string effect is exactly an emerging irreparable quantum effect after considering those spin-flip processes, which reveals that the propagation of a single hole will indeed be suppressed, yet due to the destructive interference among individual propagation paths in a spin background without macroscopic FM domains, thereby giving rise to the hole pairing through the cancellation of such phase frustration. 

In addition, the two effects can contribute to the pairing orders simultaneously because, as mentioned in the replies above, the in-tandem hole pair motion cancels the phase frustration and at the same time does not cause spin mismatches in the N\'{e}el spin configuration, so that no high-order spin-flip processes are needed to repair the spin background. Thus, the in-tandem hole pair motion not only realizes constructive interference but also gives rise to a summation over individual low-order large weights.

In summary, physically, a hole moves in an SU(2) AFM background will generally create a string defect composed of three non-commutative components under the hole-spin basis: (1) spin-mismatch effect or the $S^z$-string where $S^z$ is along the symmetry breaking axis; (2) the transverse $S^x$- and $S^y$-strings. The former is well known to be \textbf{reparable} by the spin-flip processes, which can be incorporated into the positive weight $W[C]$ in Eq.\,\eqref{eq:Z_tJ_supp}, while the latter is \textbf{irreparable} by the spin-flip processes, serving as the phase factor $\tau_C$ emerging in Eq.\,\eqref{eq:Z_tJ_supp}.

\subsection{Comparison with the fermionic $t$-$J$ model}

In cuprate superconductors and the fermionic $t$-$J$ model, the AFM order seems to have a competing relation with the pairing order~\cite{Keimer2015}, instead of the compatible coexistence observed in the bosonic $t$-$J$ model. This can be qualitatively understood from the different sign structures of the two models. In the bosonic case, the frustration of the single-hole propagation in an AFM background caused by the phase string effect can be perfectly canceled when holes form tightly binding pairs and move in tandem, which in turn respects the N\'{e}el spin configuration. This leads to the coexistence of the AFM and pairing order. By contrast, in the fermionic $t$-$J$ model, the key difference lies in the additional fermionic sign structure in the partition function, i.e., the sign factor in Eq.\,(\textcolor{blue}{3}) in the main text is replaced by
\begin{equation}\label{eq:fermionic_sign}
    \tau_C=(-1)^{N_{\dn}^h+N_{\mathrm{ex}}},
\end{equation}
where $N_{\mathrm{ex}}$ denotes the number of exchanges between holes in a closed loop $C$ of imaginary-time evolution. This implies that the hole-pair propagation, as discussed in the bosonic case, can experience an additional phase frustration in the fermionic case, hindering the formation of significant constructive interference and thereby suppressing the emergence of the long-range pairing order. Actually, the interference-based argument would become much more complex for the fermionic case due to the fermionic statistics, which is beyond the scope of semi-quantitative analysis. Especially, the fermionic statistics is expected to play an important role when involving the overlap of hole pairs, like in the BCS regime in the BEC-BCS crossover. However, a comprehensive understanding of the AFM and SC orders together with their phase transitions in the $t$-$J$ and related models requires more in-depth theoretical analysis and systematic numerical calculations.

\subsection{Justification for the GCE simulation method}

\begin{figure}[b]
    \centering
    \includegraphics[width=0.99\linewidth]{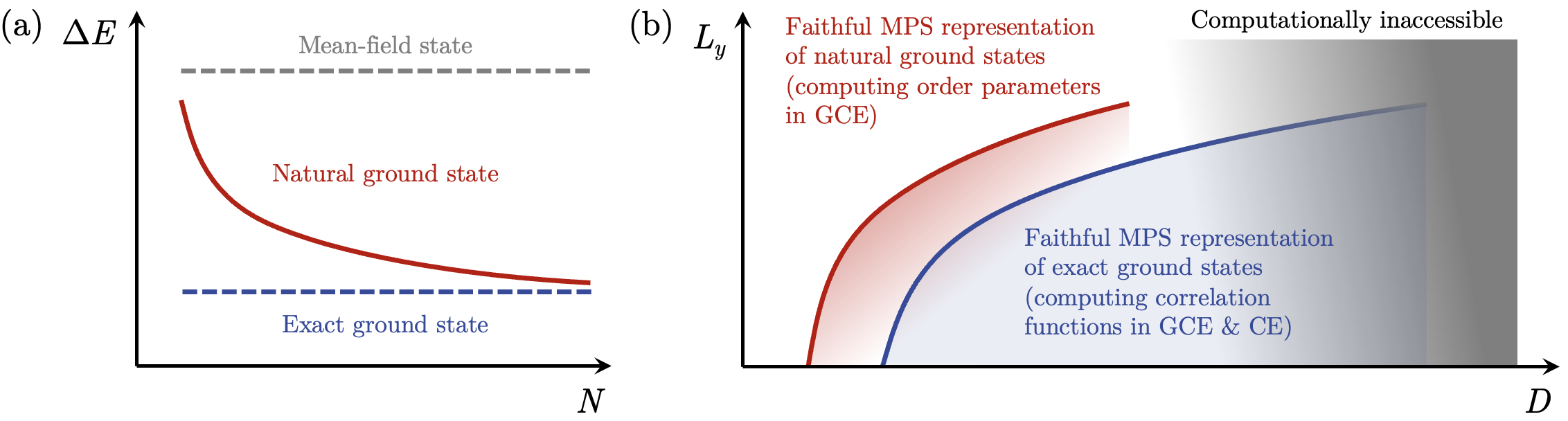}
    \caption{(a) Schematic of the energy gaps $\Delta E$ to the exact ground state of the natural ground state and the mean-field state for a given system with the tendency to SSB as a function of system size $N$. (b) The regions of faithful MPS representation of natural ground states (red) and exact symmetric ground states (blue) in the space spanned by the system width $L_y$ and bond dimension $D$. The grey area represents the computationally inaccessible region with overlarge bond dimensions and system sizes.}
    \label{figR:GCE}
\end{figure}

It is known that, for a quantum many-body system with the tendency to spontaneous symmetry breaking (SSB), there can be explicit SSB ground states in the thermodynamic limit with non-zero order parameters, e.g., the N\'{e}el order in Heisenberg antiferromagnets with 
\begin{equation}
    \avg{\mathbf{S}_i}\neq 0.
\end{equation}
However, for finite-size systems, due to symmetry constraints, a unique ground state has to respect all the symmetries of the Hamiltonian and only exhibits implicit SSB with zero order parameters and long-range correlations, e.g.,
\begin{equation}
    \avg{\mathbf{S}_i}= 0 \quad\text{and}\quad \lim_{|i-j|\rightarrow \infty} \avg{\mathbf{S}_i \cdot \mathbf{S}_j} \neq 0.
\end{equation}
Hence, it should be expected that as the system size becomes larger, there must exist a series of explicit SSB states whose energies become lower and lower till they are nearly degenerate with the exact implicit SSB ground state when approaching the thermodynamic limit [see Fig.\,\ref{figR:GCE}(a)]. These low-energy explicit SSB states are sometimes called ``natural ground states'' in the literature~\cite{Chen2023, Tasaki2019}, and it is known that a natural ground state is a superposition of low-energy eigenstates called the ``Anderson Tower of states''~\cite{Tasaki2019}.

In the context of DMRG or PEPS~\cite{Jiang2021, Chen2023, Liu2025}, taking the global charge-U(1) symmetry as an example, the CE simulations, where the charge-U(1) symmetry is imposed in the ansatz, directly target the exact symmetric ground state. On the other hand, the GCE simulations, where the charge-U(1) symmetry is not artificially imposed, in principle can target the natural ground states with non-zero pairing order parameters or single-particle condensations when the bond dimension is moderate, even though the GCE simulations will also converge to the exact symmetric ground state in the infinite bond dimension limit. This is because exact symmetric ground states need larger bond dimensions (or say entanglement) to establish long-range connected correlations. Moderate bond dimension effectively plays a similar role to external noisy perturbations that induce spontaneous symmetry breaking. Thus, the solutions from the GCE simulations with appropriate bond dimensions are not just above the mean-field level, because natural ground states are the true ground states in the thermodynamic limit, whereas naive mean-field solutions generally have finite energy gaps to the true ground state energy, as depicted in Fig.\,\ref{figR:GCE}(a).

To further illustrate the GCE scheme, in Fig.\,\ref{figR:GCE}(b), we depict the ranges of bond dimension and width where an MPS in the GCE or CE scheme can faithfully represent the natural or exact ground state of a given system with the tendency to SSB. Although the bond dimension required by the faithful MPS representation grows exponentially in $L_y$ for both natural and exact ground states, the former requires less due to the absence of long-range connected correlations. Moreover, order parameters only need local measurements and hence are easier to estimate than long-distance correlations. Therefore, the main advantage of the GCE simulations is that one can identify the potential SSB of the global charge-U(1) symmetry for larger systems with smaller computational costs, even for those beyond the scope of CE simulations, providing a more efficient way to detect pairing and superconductivity. On the contrary, admittedly, the disadvantage of the GCE-type simulations is that one might obtain small false positive signals for systems without the tendency to SSB, e.g., spin liquids.

In summary, we have argued that the GCE-type simulations can indeed provide correct order-parameter descriptions for systems with the tendency to SSB under appropriate bond dimensions and lattice sizes. Finally, we emphasize that we have also performed additional DMRG calculations in the traditional CE scheme for $L_y=6,8$, which show consistent results with the GCE simulations.

\subsection{Measuring pair-pair correlations in cold atom simulations}
To measure the correlators that are not products of Hermitian operators on quantum devices, e.g., the pair-pair correlators shown above, one needs to rewrite them as linear combinations of products of Hermitian operators. For qutrit systems where the bosonic $t$-$J$ model is defined, the Pauli matrices that are used in qubit systems as an orthogonal basis should be generalized, e.g., to the Gell-Mann matrices that span the Lie algebra of the $\mathrm{SU}(3)$ group. Here we exploit a more symmetric form by recombining the normal Gell-Mann matrices, i.e.,
\begin{equation}\label{eq:gell-mann_symmetric}
\begin{aligned}
   &\Lambda_{\up \dn}^{x} = \left|\up\right\rangle \! \left\langle \dn\right| + \left|\dn \right\rangle \! \left\langle \up\right|,\quad \Lambda_{\up \dn}^{y} = -i \left|\up\right\rangle \! \left\langle \dn\right| + i \left|\dn \right\rangle \! \left\langle \up\right|,\quad \Lambda_{\up \dn}^{z} = \left|\up\right\rangle \! \left\langle \up\right| - \left|\dn \right\rangle \! \left\langle \dn\right|, \\
   & \Lambda_{\up h}^{x} = \left|\up\right\rangle \! \left\langle h\right| + \left|h\right\rangle \! \left\langle \up\right|,\quad \Lambda_{\up h}^{y} = -i \left|\up\right\rangle \! \left\langle h\right| + i \left|h \right\rangle \! \left\langle \up\right|,\quad \Lambda_{\up h}^{z} = \left|\up\right\rangle \! \left\langle \up\right| - \left|h \right\rangle \! \left\langle h\right|, \\
    & \Lambda_{\dn h}^{x} = \left|\dn\right\rangle \! \left\langle h\right| + \left|h\right\rangle \! \left\langle \dn\right|,\quad \Lambda_{\dn h}^{y} = -i \left|\dn\right\rangle \! \left\langle h\right| + i \left|h \right\rangle \! \left\langle \dn\right|,\quad \Lambda_{\dn h}^{z} = \left|\dn\right\rangle \! \left\langle \dn\right| - \left|h \right\rangle \! \left\langle h\right|.
\end{aligned}
\end{equation}
Note that each of them is Hermitian and the three $\Lambda^z$ matrices are not independent of each other, e.g., $\Lambda_{\up \dn}^{z} = \Lambda_{\up h}^{z}-\Lambda_{\dn h}^{z}$. That is to say, there are eight independent matrices, consistent with the degrees of freedom in a $3\times 3$ unitary with determinant $1$. Using the $\Lambda$ matrices in Eq.\,\eqref{eq:gell-mann_symmetric}, the creation and annihilation operators of the hard-core bosons in the bosonic $t$-$J$ model can be expressed by
\begin{equation}\label{eq:b_lambda}
    B_{\sigma} = \frac{\Lambda_{\sigma h}^{x} - i\Lambda_{\sigma h}^{y}}{2},\quad B_{\sigma}^\dagger = \frac{\Lambda_{\sigma h}^{x} + i\Lambda_{\sigma h}^{y}}{2},
\end{equation}
where we omit the site indices for simplicity. Thus, by substituting Eq.\,\eqref{eq:b_lambda} into each creation and annihilation operator within a certain correlator, one can express any correlator as a linear combination of products of $\Lambda$ matrices. We remark that the correlators defined by the products of the $\Lambda$ matrices can be directly measured on quantum devices with single-qutrit rotation gates by rotating the measurement basis onto the eigenbasis of the operator to be measured. Alternatively, one can also apply randomized measurement schemes like the classical shadow tomography first and leave all the classical post-processing after measurements, which can further improve the efficiency when measuring multiple observables.

\section{Exact mapping to the $S=1$ spin model}
In the following, we provide a detailed discussion on the mapping between the spin-1/2 hard-core boson system and the pure spin-1 system. We can construct an explicit one-to-one correspondence between the local Hilbert space basis of the boson system and the spin-1 system
\begin{equation}
\begin{aligned}
    & \left |h\right\rangle \longleftrightarrow |0\rangle, \\
    & \left|\up\right\rangle \longleftrightarrow |1\rangle, \\
    & \left|\dn\right\rangle \longleftrightarrow \left|-1\right\rangle,
\end{aligned}
\end{equation}
where $\left| h \right\rangle$, $\left| \up \right\rangle$, and $\left| \dn \right\rangle$ denote the hole, spin-up, and spin-down states in the boson system, while $\left| 0 \right\rangle$, $\left| \pm 1 \right\rangle$ label the $S^z$ eigenstates of the pure spin-1 system. Then, the bosonic operators are mapped to
\begin{equation}\label{eq:boson_operator}
\begin{aligned}
    B_{\up}=\left|h\right \rangle\left\langle\up\right| & \longleftrightarrow |0\rangle\langle 1|=\left(1+\tilde{S}^z\right) \tilde{S}^{-},\\
    B_{\dn}=\left|h\right \rangle\left\langle\up\right| & \longleftrightarrow |0\rangle\langle -1|=\left(1-\tilde{S}^z\right) \tilde{S}^{+},\\
    n & \longleftrightarrow \tilde{S}^z \tilde{S}^z,
\end{aligned}
\end{equation}
and the spin-1/2 operators of bosons are mapped to
\begin{equation}\label{eq:spin_operator}
\begin{aligned}
    S^z & \longleftrightarrow \frac{1}{2} \tilde{S}^z, \\
    S^{+} & \longleftrightarrow \tilde{S}^{+} \tilde{S}^{+}, \\
    S^{-} & \longleftrightarrow \tilde{S}^{-} \tilde{S}^{-}. \\
\end{aligned}
\end{equation}
Here, the operators on the left-hand side act on the local Hilbert space of the spin-1/2 bosonic system, while those on the right-hand side (distinguished by the tilde symbols) act on the local Hilbert space of the pure spin-1 system. Under this mapping, the bosonic $t$–$J$ Hamiltonian $H_{t\text{-}J}=\mathcal{P}_s (H_t + H_J) \mathcal{P}_s$ is mapped to the following $S=1$ spin Hamiltonian
\begin{equation}\label{eq:SH}
\begin{aligned}
    \mathcal{P}_s H_t \mathcal{P}_s &\longleftrightarrow-t \sum_{\langle i j\rangle}\left[\tilde{S}_i^{+}\left(1+\tilde{S}_i^z\right)\left(1+\tilde{S}_j^z\right) \tilde{S}_j^{-}+\tilde{S}_i^{-}\left(1-\tilde{S}_i^z\right)\left(1-\tilde{S}_j^z\right) \tilde{S}_j^{+}+\text{H.c.}\right],\\
    \mathcal{P}_s H_J \mathcal{P}_s &\longleftrightarrow J \sum_{\langle i j\rangle}\left[\frac{1}{4} \tilde{S}_i^z \tilde{S}_j^z+\frac{1}{2} (\tilde{S}_i^{+}\tilde{S}_j^{-})^2+\frac{1}{2} (\tilde{S}_i^{-} \tilde{S}_j^{+})^2 -\frac{1}{4} (\tilde{S}_i^z \tilde{S}_j^z)^2\right],
\end{aligned}
\end{equation}
The pairing operator in the bosonic model is mapped to
\begin{equation}\label{eq:SO}
\begin{aligned}
    B_{i \up} B_{j \dn} &\longleftrightarrow \left(1+\tilde{S}_i^z\right) \tilde{S}_i^{-}\left(1-\tilde{S}_j^z\right) \tilde{S}_j^{+}. 
    %\\z\text{-spin order: } S_i^z = \sum_\sigma \sigma B_{i \sigma}^{\dagger} B_{i \sigma} &\longleftrightarrow \tilde{S}_i^{+}\left(1+\tilde{S}_i^z\right)\left(1+\tilde{S}_i^z\right) \tilde{S}_i^{-}-\tilde{S}_i^{-}\left(1-\tilde{S}_i^z\right)\left(1-\tilde{S}_i^z\right) \tilde{S}_i^{+}
\end{aligned}
\end{equation}
It can be found that both the pairing operator and the resulting Hamiltonian of the spin-1 model necessarily contain higher-order interactions, instead of purely two-spin interactions as in common spin models. These higher-order interactions can often be systematically reorganized in terms of the quadrupolar operators
\begin{equation}
    \tilde{Q}^{\alpha \beta}=\tilde{S}^\alpha \tilde{S}^\beta+\tilde{S}^\beta \tilde{S}^\alpha-\frac{4}{3} \delta^{\alpha \beta} \boldsymbol{1},
\end{equation}
with $\alpha,\beta\in\{x,y,z\}$. This allows the kinetic term in Eq.\,\eqref{eq:SH} to be simplified as
\begin{equation}\label{eq:SH2}
    \mathcal{P}_s H_t \mathcal{P}_s \longleftrightarrow -t \sum_{\langle i j\rangle}\left[ \frac{1}{2} \tilde{S}_i^{+} \tilde{S}_j^{-}+\frac{1}{4}\left(\tilde{Q}_i^{yz}-i \tilde{Q}_i^{z x}\right)\left(\tilde{Q}_j^{yz}+i \tilde{Q}_j^{z x}\right) +\text {H.c.} \right].
\end{equation}
Similarly, the spin exchange term in Eq.\,\eqref{eq:SH} can also be rewritten in terms of bilinear combinations of the eight generators of SU(3), i.e., $\tilde{S}^\alpha, \tilde{Q}^{\alpha\beta} $, which is not presented here due to the complex expressions. The pairing operator in Eq.\,\eqref{eq:SO} can be expressed as
\begin{equation}\label{eq:SO2}
\begin{aligned}
    B_{i \up} B_{j \dn} \longleftrightarrow
    \frac{1}{4} \tilde{S}_i^{-} \tilde{S}_j^{+}&-\frac{1}{8}\left(\tilde{Q}_i^{zx}-i \tilde{Q}_i^{yx}\right)\left(\tilde{Q}_j^{zx}+i \tilde{Q}_j^{yx}\right)\\&+\frac{1}{4 \sqrt{2}}\left(\tilde{Q}_i^{zx}-i \tilde{Q}_i^{yx}\right) \tilde{S}_j^{+}-\frac{1}{4 \sqrt{2}} \tilde{S}_i^{-}\left(\tilde{Q}_j^{zx}+i \tilde{Q}_j^{yx}\right)
    \end{aligned}
\end{equation}
which is also a bilinear combination of the dipolar spin operators and the quadrupolar operators. Although the bosonic $t$–$J$ model and the spin-1 model here are exactly mappable, they imply complementary perspectives on the underlying physics. The choice between the two representations should be guided by the specific physical interest and model simplicity.

\end{document}